\documentclass[11pt,a4paper]{article}
\usepackage{jheppub}       

\usepackage{bm}
\usepackage{xcolor}
\usepackage{tikz}          

\usetikzlibrary{calc,decorations.markings,decorations.pathmorphing,
                arrows.meta,positioning,fit,backgrounds,
                patterns,shapes.geometric}

\providecommand{\SGfill}{black!8}     
\providecommand{\Blobfill}{black!18}  
\providecommand{\HLcol}{black!22}     

\tikzset{
  prop/.style   = {line width=0.7pt},
  extl/.style   = {line width=0.7pt},
  dprop/.style  = {line width=0.7pt,
                   postaction={decorate,
                     decoration={markings,
                       mark=at position 0.5 with {
                         \fill[\Blobfill,draw=black,line width=0.4pt]
                           circle[radius=1.5pt];}}}},
  vtx/.style    = {circle,fill=black,inner sep=0pt,minimum size=5pt},
  blob/.style   = {circle,draw=black,fill=\Blobfill,line width=0.5pt,
                   inner sep=0pt,minimum size=9pt},
  subg/.style   = {draw=black,dashed,line width=0.6pt,rounded corners=3pt,
                   fill=\SGfill,inner sep=3pt},
  hl/.style     = {line width=4pt,\HLcol,line cap=round},
}

\newcommand{\extleg}[4]{%
  \draw[extl] #1 -- #2;%
  \ifx\relax#3\relax\else\node[font=\scriptsize,#4] at #2 {#3};\fi}

\tikzset{sqmark/.style={circle,draw=black,fill=white,line width=0.5pt,
                        inner sep=0pt,minimum size=3.4pt}}

\tikzset{
  gluon/.style = {decorate,line width=0.6pt,
                  decoration={coil,aspect=0.55,segment length=2.1mm,
                              amplitude=0.9mm}},
  ghost/.style = {line width=0.7pt,dash pattern=on 2.1pt off 1.7pt},
  quark/.style = {line width=0.8pt},
  ggvtx/.style = {circle,fill=black,inner sep=0pt,minimum size=5pt},
  ggarr/.style = {postaction={decorate,decoration={markings,
                   mark=at position 0.57 with {\arrow{Stealth[length=1.6mm]}}}}},
}

\usepackage[T2A]{fontenc}\usepackage[utf8]{inputenc}
\usepackage[english]{babel}
\usepackage{graphicx}

\usepackage{bm}
\usepackage{mathbbol}
\usepackage{amsthm}
\usepackage{mathrsfs}
\usepackage{latexsym}
\usepackage{commath} 

\usepackage{graphicx}
\usepackage{tikz}
\usetikzlibrary{babel,arrows.meta,positioning}
\usepackage{appendix}

\usepackage{subfig}
\usepackage{subcaption}
\usepackage{xcolor}

\graphicspath{{images/}}

\newcommand{\be}{\begin{equation}}
\newcommand{\ee}{\end{equation}}

\newcommand{\benn}{\begin{equation*}}
\newcommand{\eenn}{\end{equation*}}

\newcommand{\bea}{\begin{equation}\begin{aligned}}
\newcommand{\eea}{\end{aligned}\end{equation}}

\usepackage{mathtools}

\DeclarePairedDelimiter\bra{\langle}{\rvert}
\DeclarePairedDelimiter\ket{\lvert}{\rangle}
\DeclarePairedDelimiterX\braket[2]{\langle}{\rangle}{#1 \delimsize\vert #2}

\title{A tale of three points: mixed and tensor bispectra beyond leading order in Horndeski gravity}

\author[a,b,c,1]{Y.A.Ageeva}
\author[a,c,2]{and M.A.Stepanov}

\affiliation[a]{Institute for Nuclear Research of the Russian Academy of
  Sciences, 60th October Anniversary Prospect, 7a, 117312 Moscow, Russia}
\affiliation[b]{Institute for Theoretical and Mathematical Physics,
  M.V. Lomonosov Moscow State University, Leninskie Gory 1, 119991
  Moscow, Russia}
\affiliation[c]{Department of Particle Physics and Cosmology, Physics Faculty,
  M.V. Lomonosov Moscow State University, Leninskie Gory 1-2, 119991
  Moscow, Russia}
\emailAdd{$^1$ageeva@inr.ac.ru, $^2$stepanov.ma23@physics.msu.ru}

\abstract{
We derive the primordial
$\langle\zeta\zeta h\rangle$,
$\langle\zeta hh\rangle$, and
$\langle hhh\rangle$
bispectra in general Horndeski theory through first order in slow
variation, retaining their complete momentum and helicity dependence.
The calculation includes the time evolution of the cubic coefficients,
the correction to the scale factor, the first-order scalar and tensor
mode functions, and the corrected power-spectrum normalization. We verify the tensor-soft and scalar-soft Maldacena
consistency relations at the same order. We also study finite squeezing
and show that, when $c_S\neq c_T$, the relevant soft variable is the
ratio $c_Tq/(c_Sk)$, leading to an acoustic crossover not captured by
geometric squeezing alone. We construct the corresponding nonlinearity
parameters and apply the general results to a nonsingular contracting
model, in which the mixed bispectra retain sensitivity to the contraction
parameter through the scalar propagation speed.
}

\begin{document}
\maketitle

\section{Introduction}

Over the past century, advances in science have transformed the quest to understand the origin and history of everything that exists in our Universe from a subject confined to philosophy and religion into a domain of empirical science. We now know that the Universe contains an enormous number of galaxies and galaxy clusters, which together form its large-scale structure. Clearly, despite the approximate homogeneity and isotropy of the Universe on sufficiently large scales, small departures from exact homogeneity and isotropy must have been present at the early stages of its evolution; otherwise, the Universe would have remained perfectly homogeneous and stable structures such as our Galaxy could never have formed.

The most widely accepted explanation for the origin of these primordial inhomogeneities is a period of accelerated expansion in the early Universe, commonly referred to as inflation \cite{Starobinsky1980,Guth1981,Sato1981,Linde1982}. During this epoch, microscopic quantum fluctuations \cite{Chibisov:1982nx,Mukhanov:1981xt,Mukhanov:2005sc,Gorbunov:2011zzc} are stretched to cosmological scales and become effectively frozen. These primordial curvature perturbations subsequently seed the gravitational growth of matter inhomogeneities, eventually giving rise to the complex cosmic structure observed today. Understanding the origin and properties of these primordial perturbations is therefore one of the central problems of modern cosmology.

Direct observation of primordial quantum fluctuations is, however, impossible, just as the physical conditions of the early Universe cannot be reproduced experimentally. Nevertheless, their late-time imprints are observable, most notably in the anisotropies of the cosmic microwave background (CMB), the large-scale structure of the Universe, and, potentially, a primordial gravitational-wave background. Studying these observables therefore provides a unique window into the physics of the early Universe and quantum fields in curved spacetime.

For an accelerating cosmological background, the corresponding theoretical observables are late-time correlation functions. The two-point correlation function determines the primordial power spectrum, whose parameters are tightly constrained by CMB observations\cite{Planck:2019nip,BICEP:2021xfz}, whereas three-point correlation functions characterize primordial non-Gaussianity. With the continuing development of gravitational-wave cosmology, correlation functions involving tensor modes, i.e. gravitons, are also of particular interest.

Moreover, there is no unique model of inflation. Instead, there exists a wide variety of different realizations of the inflationary paradigm \cite{Starobinsky1980,Guth1981,Sato1981,MukhanovChibisov1981,Linde1982,AlbrechtSteinhardt1982,Hawking1982,GuthPi1982,Starobinsky1982,LindeFluctuations1982,Linde1983,BardeenSteinhardtTurner1983,MukhanovFeldmanBrandenberger1992,GarrigaMukhanov1999}, as well as an even larger number of alternative scenarios \cite{Novello:2008ra,Lehners:2008vx,Lehners:2011kr,Battefeld:2014uga,Qiu:2011cy,Easson:2011zy,Cai:2012va,Osipov:2013ssa,Qiu:2013eoa,Koehn:2013upa,Battarra:2014tga,Qiu:2015nha,Ijjas:2016tpn,Brandenberger:2016vhg,Ijjas:2018qbo,Kobayashi:2019hrl,Mironov:2019haz,Creminelli:2010ba,Creminelli:2012my,Hinterbichler:2012fr,Elder:2013gya,Pirtskhalava:2014esa,Nishi:2015pta,Kobayashi:2015gga}, many of which are compatible, to varying degrees, with current observational data. Moreover, explicit nonsingular constructions and subsequent developments in
Horndeski and beyond-Horndeski cosmology are discussed in
Refs. \cite{Mironov:2018Bounce,Mironov:2020Subluminal,
Mironov:2022Disformal,Mironov:2023Stable,
Mironov:2023Cartan,Mironov:2024Anisotropic,
Mironov:2024TorsionNoGo,Mironov:2024HealthyTorsion,
Mironov:2024Review,Volkova:2026Cuscuton,
Choi:2025Genesis,Choi:2025GenesisStarobinsky}. 

Constraints derived from the power spectrum alone are therefore insufficient to discriminate among models and rule out those that do not describe our Universe. To this end, higher-order correlation functions must be studied. With the expected improvement in the precision of measurements of primordial non-Gaussianity, characterized by three-point correlation functions, it should become possible to place significantly stronger constraints on many inflationary scenarios and their alternatives, and potentially to rule out some of them altogether.

In addition to the shape of the three-point correlation functions themselves, an important
characteristic of primordial non-Gaussianity is the nonlinearity parameter
$f_{\mathrm{NL}}$, which provides a convenient measure of the bispectrum amplitude
relative to the corresponding two-point correlation functions.
Constraints on primordial non-Gaussianity constitute one of the most sensitive probes
of the dynamics of the early Universe, since different inflationary scenarios may lead
to very similar power spectra while producing substantially different three-point
correlation functions. In standard single-field inflation, the expected level of
non-Gaussianity is small \cite{Maldacena:2002vr}, whereas more general scalar--tensor
dynamics, in particular within Horndeski theories, allows for a considerably richer
bispectrum structure \cite{Gao:2012ib,DeFelice:2013ar}. The scalar--scalar--tensor cubic action in quadratic DHOST theory was
recently derived in Ref. \cite{Mironov:2026SST}.

The purpose of the present work is to extend the calculation of
primordial bispectra containing tensor perturbations beyond the
leading slow-variation approximation. We derive the three-point
functions
$\langle\zeta\zeta h\rangle$,
$\langle\zeta hh\rangle$, and
$\langle hhh\rangle$
in general Horndeski theory, retaining the full momentum-triangle and
helicity dependence. The calculation includes the first-order
corrections to the scalar and tensor mode functions, the slow evolution
of the coefficients in the cubic actions, the correction to the scale
factor, and the conversion from leading reference spectra to corrected
power spectra. We then construct the corresponding nonlinearity
parameters. In the equilateral configuration, our expressions extend
the known leading-order Horndeski results of
Refs.\cite{Gao:2011vs,Gao:2012ib} to first order in slow variation. Moreover, the evaluations of three scalars bispectrum in the next-to-leading expansion by the slow-roll parameters are given in Refs. \cite{DeFelice:2011zh,DeFelice:2013ar}.

An important part of the analysis is the independent verification of
the mixed Maldacena consistency relations at the same order. In the
tensor-soft limit of $\langle h\zeta\zeta\rangle$, the result is fixed
by the momentum dependence of the scalar two-point function. In the
scalar-soft limit of $\langle\zeta hh\rangle$, the universal result
emerges only after the contributions of different cubic interactions,
their first-order corrections, and the corrected power-spectrum
normalization are combined. These limits therefore provide sensitive
checks of the complete calculation rather than additional assumptions
about the bispectra.

We also study the approach to the tensor-soft limit at finite momentum
ratio. For mixed modes with different propagation speeds, geometric
squeezing, $q/k\ll1$, does not necessarily imply a hierarchy of
acoustic frequencies. The relevant variable is
$c_Tq/(c_Sk)$, and the transition between the tensor-soft regime and
the region in which all three acoustic frequencies are comparable
occurs when this ratio is of order unity. This acoustic crossover
contains information that is not fixed by the strict consistency
relation.

Although inflation provides a successful mechanism for generating
primordial perturbations, it does not by itself determine the
pre-inflationary history or remove the issue of past geodesic
incompleteness \cite{Penrose:1964wq,Hawking:1966vg,Borde:1993xh,Borde:1996pt,Borde:2001nh,Yoshida:2018ndv}. Under appropriate causal, and curvature-convergence
assumptions, classical singularity theorems imply the existence of
incomplete causal geodesics
\cite{Penrose:1964wq,Hawking:1966vg,HawkingPenrose1970}. Consequently, the description of the earliest stage of the evolution of the Universe requires a separate analysis.

Inflation by itself therefore does not answer the question of what preceded the inflationary stage or what physical mechanism initiated inflation.
To understand how the problem of the initial singularity may be resolved, it is useful to examine the structure underlying the Hawking--Penrose singularity theorem in more detail.
One of the key ingredients in the proof is the Raychaudhuri equation \cite{Raychaudhuri:1955ta,Kar:2006ms} for a congruence of null geodesics,
\begin{equation}
	\frac{d\theta_\ell}{d\lambda}
	=
	-R_{\mu\nu}\ell^\mu\ell^\nu+\dots,
	\nonumber
\end{equation}
where omitted terms describe the twist, shear, and nonlinear self-focusing of the null geodesic congruence and are not essential for the present discussion. And $\lambda$ is an affine parameter along the null geodesic and $\ell^\mu$ is its tangent vector.

The evolution of a bundle of null geodesics is characterized by the expansion scalar $\theta_\ell$, which determines whether neighboring null geodesics diverge or converge. For $\theta_\ell>0$, the congruence expands, whereas for $\theta_\ell<0$ it contracts. Therefore, provided that
\begin{equation}\label{eq:2.2}
	R_{\mu\nu}\ell^\mu\ell^\nu\geq0,
\end{equation}
gravity leads to the focusing of null geodesics.
Under additional global assumptions, such as appropriate causality conditions and the existence of a trapped surface, geodesic focusing implies that the corresponding geodesics cannot be extended to arbitrarily large values of the affine parameter. Physically, this signals geodesic incompleteness and, in the context of the singularity theorems, the presence of a spacetime singularity.

In Einstein gravity, the null convergence condition
\eqref{eq:2.2} is equivalent to the null energy condition,
$T_{\mu\nu}\ell^\mu\ell^\nu\geq0$. For a spatially flat FLRW
background,
\begin{equation}
    \dot H=-\frac{\rho+p}{2M_{\rm pl}^2}.
    \nonumber
\end{equation}
A nonsingular bounce requires $\dot H>0$ near $H=0$ and hence
$\rho+p<0$, i.e. violation of the NEC. Modified gravity, or
relaxation of other assumptions entering the singularity theorems,
provides alternative possibilities.\footnote{
Other approaches to constructing nonsingular models employ, for
example, holographic or string-theoretic ingredients
\cite{Arefeva:2011qvf,Arefeva:2011yua,Silva:2015qna,
Silva:2020bnn,Silva:2023ieb}.
}

Let us also note that, when applied to the present-day Universe, the NEC prevents the Hubble parameter from increasing. Therefore, if the Hubble parameter were found to be increasing at the present epoch, this would indicate either a violation of the NEC by dark energy or a breakdown of General Relativity on cosmological scales\cite{Rubakov:2014jja}. This provides additional motivation for considering modified theories of gravity in the construction of nonsingular early-Universe scenarios.

Stable nonsingular evolution is difficult to realize within ordinary
two-derivative matter theories because violations of the null energy
condition are commonly accompanied by ghost or gradient
instabilities \cite{Tipler:1978zz}. In addition to the absence of such instabilities, the validity of the
classical description requires the relevant energy scales to remain
below the strong-coupling scale. Strong-coupling and
perturbative-unitarity constraints in related cosmological models were
studied in Refs. \cite{Ageeva:2020Classical,
Ageeva:2021StrongGravity,Ageeva:2022StrongCoupling,
Ageeva:2023Unitarity,Ageeva:2024KInflation,
Petrov:2026GWUnitarity}. Horndeski gravity provides a broad scalar--tensor
framework in which such questions can be studied systematically
\cite{Horndeski:1974wa,Kobayashi:2019hrl}.\footnote{There are extensions of Horndeski gravity such as recent higher-derivative scalar--tensor and
scalar--vector--tensor extensions, which  discussed in
Refs.\cite{Gorji:2026CovariantST,Gorji:2026HOMES}.} Its action is
\begin{align}
\mathcal S
&=
\int d^4x\,\sqrt{-g}
\left[
G_2(\phi,X)
-
G_3(\phi,X)\Box\phi
+
\mathcal L_4
+
\mathcal L_5
\right],
\label{action}
\end{align}
where
\begin{align}
\mathcal L_4
={}&
G_4(\phi,X)R
+
G_{4X}
\left[
(\Box\phi)^2
-
(\nabla_\mu\nabla_\nu\phi)
(\nabla^\mu\nabla^\nu\phi)
\right],
\nonumber\\
\mathcal L_5
=&
G_5(\phi,X)G_{\mu\nu}
\nabla^\mu\nabla^\nu\phi
\nonumber\\
&-
\frac16G_{5X}
\Big[
(\Box\phi)^3
-
3(\Box\phi)
(\nabla_\mu\nabla_\nu\phi)
(\nabla^\mu\nabla^\nu\phi)
\nonumber\\
&
+
2(\nabla^\mu\nabla_\alpha\phi)
(\nabla^\alpha\nabla_\beta\phi)
(\nabla^\beta\nabla_\mu\phi)
\Big],\nonumber
\end{align}
and
\begin{align*}
X
&=
-\frac12
g^{\mu\nu}
\partial_\mu\phi\,\partial_\nu\phi.
\end{align*}
Here $G_i(\phi,X)$ are arbitrary functions, and $G_4$ includes the
Einstein--Hilbert contribution; Einstein gravity is recovered for
$G_4=M_{\rm pl}^2/2$ and $G_5=0$. The general calculation is formulated directly for the Horndeski
action \eqref{action}, without assigning a distinguished Jordan- or
Einstein-frame interpretation to the metric $g_{\mu\nu}$. A definite
frame is fixed only in Subsec.\ref{sec:contracting_uni}: the
correlation functions are evaluated in the Einstein frame, whereas the
contracting background is physically interpreted in the Jordan frame.

As an application of the general calculation, we consider the
nonsingular cosmological model developed in
Refs.\cite{Ageeva:2020gti,Ageeva2023,Ageeva:2024knc}.
Its early-time evolution is a power-law contraction in the Jordan
frame, followed by a nonsingular transition. The Jordan-frame action
is
\begin{align}
S_J
=
\hat g
\int d^4x\,\sqrt{-g_J}\,\mathrm e^{2\mu\phi}
\Bigg\{
&\frac12R_J
+
\mathrm e^\phi
\left[
\bar c_2\mathrm e^\phi
+
\left(
d_2-\bar c_3(1+2\mu)
\right)
\sqrt{2X_J}
\right]
\nonumber\\
&+
2\mu(d_3+2\mu)X_J\ln X_J
\nonumber\\
&-
\frac{d_3+2\mu}{2}
(2+\ln X_J)\Box_J\phi
\Bigg\}.
\label{eq:j_frame}
\end{align}

The parameters entering this action have different roles. The
parameter $\chi$, introduced by the early-time solution
$a_J(t_J)\propto(-t_J)^\chi$, controls the contraction rate.
The parameter $\mu$ determines the time dependence of the conformal
factor and, together with $\chi$, fixes both the scalar and tensor
spectral tilts. The constants
$\bar c_2$, $d_2$, $\bar c_3$, and $d_3$
specify the Lagrangian. The parameters $d_2$ and $\bar c_3$ enter the
background and perturbation quantities relevant here only through the
combination
$\bar c_3(1+2\mu)-d_2$.
For the choice $d_3=-2$ used in our application, the background
equations determine $\bar c_2$ and this combination in terms of
$\mu$ and $\chi$.

The positive parameter $\hat g$ fixes the overall normalization of
the gravitational action and hence the amplitude of the primordial
power spectra. It can be determined from the measured scalar
amplitude, but it drops out of the dimensionless nonlinearity
parameters considered below. The spectral tilt can be used to replace
$\mu$ by
$\Delta\equiv1-n_s$. Consequently, after $n_s$ and the scalar
amplitude are fixed, the mixed nonlinearity parameters depend on one
remaining background parameter, which may be taken to be $\chi$.
Equivalently, this dependence can be expressed through the scalar
propagation speed
\begin{align*}
c_S
&=
(1-\chi)
\sqrt{\frac{1-n_s}{3}}.
\end{align*}
This explains why the model predictions in
Subsec.\ref{sec:contracting_uni} are presented as functions of
$n_s$ and $\chi$, or of $n_s$ and $c_S$.

The Jordan-frame action is related by the conformal transformation
\begin{align}
\Omega
&=
\frac{\mathrm e^{-\mu\phi}}{\sqrt{\hat g}},
\label{eq:conformal_factor}
\end{align}
or, precisely, $g_{\mu \nu}^J=\Omega^2 g_{\mu \nu}^E=\frac{\mathrm e^{-2 \mu \phi}}{\hat{g}} g_{\mu \nu}^E$,
to the Einstein-frame G-inflation action
\begin{align}
\mathcal S
&=
\int d^4x\,\sqrt{-g}
\left[
G_2(\phi,X)
-
G_3(\phi,X)\Box\phi
+
G_4R
\right],
\label{1.1.1}
\end{align}
where
\begin{align}
G_2
=&
\frac{\bar c_2\mathrm e^{-2(\mu-1)\phi}}{\hat g}
+
\frac{\sqrt{2X}\,d_2\mathrm e^{(1-\mu)\phi}}{\sqrt{\hat g}}
-
\frac{\sqrt{2X}\,\bar c_3(1+2\mu)
\mathrm e^{(1-\mu)\phi}}{\sqrt{\hat g}}
\nonumber\\
&-
6d_3\mu X
-
6\mu^2X,
\nonumber\\
G_3
=&
\frac12(d_3+2\mu)\ln X,
\nonumber\\
G_4
=&
\frac{M_{\rm pl}^2}{2}.\nonumber
\end{align}
In this representation $G_5=0$. Therefore, the model has
$\mu_5=\dot\phi XG_{5X}=0$, the additional Horndeski
three-graviton interaction is absent, and the tensor bispectrum is
generated by the universal general-relativistic interaction. The
tensor propagation speed is $c_T=1$. The mixed bispectra retain the
dependence on $\chi$ through $c_S$, while the purely tensor prediction
is fixed once the scalar tilt is specified.

The contracting interpretation is most transparent in the Jordan
frame, whereas the perturbation calculation is conveniently performed
in the conformally related Einstein frame. The conformal
transformation is therefore used as a calculational tool and does not
replace the physical interpretation of the solution as a contracting
cosmology followed by a nonsingular transition.\footnote{Although the conformally related Einstein-frame metric may be
geodesically incomplete, this is not by itself pathological in our
setup. By assumption, particles with time-independent masses are
minimally coupled to the Jordan-frame metric and therefore follow
Jordan-frame geodesics; in the Einstein frame, the corresponding
particle masses are generally field dependent. See
Ref.\cite{Rubakov:2022fqk} for a frame-independent discussion of
physical time and generalized geodesic completeness.} 

We describe scalar and tensor perturbations with the
Arnowitt--Deser--Misner decomposition
\cite{Arnowitt:1962hi},
\begin{align*}
ds^2
&=
-N^2dt^2
+
\gamma_{ij}
(dx^i+N^idt)
(dx^j+N^jdt).
\end{align*}
In unitary gauge, $\delta\phi=0$, and the propagating perturbations are
contained in
\begin{align*}
\gamma_{ij}
&=
a^2(t)\mathrm e^{2\zeta}(\mathrm e^h)_{ij},
&
\partial_i h_{ij}&=0,
&
\delta^{ij}h_{ij}&=0.
\end{align*}
The scalar perturbation is represented by $\zeta$, while $h_{ij}$\footnote{We also note, that $\zeta_J=\zeta_E, h_{i j}^J=h_{i j}^E$ under mentioned conformal transformations \eqref{eq:conformal_factor}.}
contains the two tensor helicities. The lapse and shift are fixed by
the Hamiltonian and momentum constraints and do not introduce
additional propagating degrees of freedom.

Our application to the nonsingular model provides primordial
predictions rather than direct observational bounds on $\chi$. Obtaining such bounds requires projecting the full momentum- and
helicity-dependent primordial bispectra onto the observable CMB
temperature and polarization bispectra and matching their
normalization to the templates used in data analyses. This
model-specific projection is left for future work.

The paper is organized as follows. In Sec.\ref{sec:res}, we give a
compact summary of the main results. In
Sec.\ref{sec:general_horn}, we derive the scalar and tensor mode
functions through first order in slow variation and introduce the
common acoustic pivot used in the three-point calculations. In
Sec.\ref{sec:mixed}, we evaluate the
$\zeta\zeta h$, $\zeta hh$, and $hhh$ bispectra for arbitrary
momentum triangles and tensor helicities. In Sec.\ref{sec:Maldacena},
we verify the tensor-soft and scalar-soft Maldacena consistency
relations at the same order. In Sec.\ref{sec:nonlinearity_nlo}, we
construct the corresponding nonlinearity parameters, give their
equilateral specialization, apply the general formulas to the
nonsingular contracting model in
Subsec.\ref{sec:contracting_uni}, and study the acoustic crossover
at finite squeezing. We summarize our conclusions in
Sec.\ref{sec:conclusion}. The explicit first-order temporal
integrals, polarization contractions, and analytic functions entering
the nonlinearity parameters are collected in the appendices.

\section{Results}
\label{sec:res}

We summarize the main results below. All background quantities are evaluated
at the common physical pivot unless stated otherwise, and
$\mathcal O(\varepsilon^2)$ terms are omitted.

\begin{enumerate}
\setlength{\itemsep}{0.6em}

\item \textbf{General Horndeski bispectra at first order.}

We compute $\langle\zeta\zeta h\rangle$, $\langle\zeta hh\rangle$, and
$\langle hhh\rangle$ for arbitrary momentum triangles and helicities. Our
spectral conventions are
\begin{subequations}
\label{eq:sg_spectra}
\begin{align}
\langle\zeta(\mathbf k)\zeta(\mathbf k')\rangle
&=(2\pi)^3\delta^{(3)}(\mathbf k+\mathbf k')P_\zeta(k),
&
\mathcal P_\zeta&=\frac{k^3}{2\pi^2}P_\zeta,
\\
\langle h^{(\lambda)}(\mathbf q)h^{(\lambda')}(\mathbf q')\rangle
&=(2\pi)^3\delta^{(3)}(\mathbf q+\mathbf q')
\delta_{\lambda\lambda'}P_h(q),
&
\mathcal P_{h,\mathrm{one\,helicity}}
&=\frac{q^3}{2\pi^2}P_h,
\\
\mathcal P_h&=2\mathcal P_{h,\mathrm{one\,helicity}}.
\end{align}
\end{subequations}
The three bispectrum amplitudes are defined by
\begin{subequations}
\begin{align}
\left\langle
\zeta(\bm k_1)
\zeta(\bm k_2)
h^{(\lambda)}(\bm k_3)
\right\rangle'
&=
\frac{(2\pi)^4\mathcal P_\zeta^2}
{k_1^3k_2^3k_3^3}
\mathcal A_\lambda^{(\zeta\zeta h)}
(k_1,k_2,k_3),
\label{eq:_zetazetah_norm}
\\
\left\langle
\zeta(\bm k_1)
h^{(\lambda_2)}(\bm k_2)
h^{(\lambda_3)}(\bm k_3)
\right\rangle'
&=
\frac{(2\pi)^4\mathcal P_\zeta^2}
{k_1^3k_2^3k_3^3}
\mathcal A_{\lambda_2\lambda_3}^{(\zeta hh)}
(k_1,k_2,k_3),
\label{eq:z_hh_bispectrum_definition}
\\
\left\langle
h^{(\lambda_1)}(\bm k_1)
h^{(\lambda_2)}(\bm k_2)
h^{(\lambda_3)}(\bm k_3)
\right\rangle'
&=
\frac{(2\pi)^4\mathcal P_h^2}
{k_1^3k_2^3k_3^3}
\mathcal A_{\lambda_1\lambda_2\lambda_3}^{(hhh)}
(k_1,k_2,k_3).
\label{eq:hhh_bispectrum_definition}
\end{align}
\end{subequations}
Throughout, the prime means that the complete momentum-conserving
factor
\begin{align*}
(2\pi)^3
\delta^{(3)}
\left(
\bm k_1+\bm k_2+\bm k_3
\right)
\end{align*}
has been removed.

The quantities on the left-hand sides are the primordial bispectra in
a definite helicity basis. Equivalently, the bispectrum $B_A$ used
below in Eq.\eqref{eq:fnl_definition} is the corresponding primed
three-point correlator, with the helicity labels understood. The
equations above define the reduced amplitudes $\mathcal A_A$ by
extracting from $B_A$ the common momentum factor
$k_1^{-3}k_2^{-3}k_3^{-3}$ and the chosen power-spectrum
normalization. Thus,
$\mathcal A_A$ and $B_A$ describe the same three-point function; they
differ only by the explicit normalization displayed above.

For the two mixed sectors, the convention adopted in this work uses
$\mathcal P_\zeta^2$, whereas the purely tensor sector is normalized by
the square of the total tensor spectrum $\mathcal P_h^2$. 
Consequently, the amplitudes above can be inserted directly into
Eq.\eqref{eq:fnl_definition} after applying the power-spectrum
conversion described in
Eq.\eqref{eq:nlf_denominator_conversion}.

We organize each bispectrum consistently to first order in the
collective slow-variation expansion. The small quantities are the
logarithmic rates of the Hubble parameter, the scalar and tensor
quadratic coefficients, and the corresponding propagation speeds,
together with the logarithmic rates of the coefficients in the cubic
actions. In particular, this set includes
$\epsilon$, $f_S$, $s$, $f_T$, and $s_T$. These quantities also
determine the scalar and tensor Hankel-index shifts
$\delta_S$ and $\delta_T$.

For every cubic interaction, two different expansions must be
distinguished. First, the value of its coefficient at the common
physical pivot is separated into its first nonvanishing contribution
in the adopted slow-variation counting and an additive first-order
correction. We denote these two parts by the superscript
``lead'' and by $\delta$, respectively. Here ``lead'' means the first
nonzero term in the model-dependent power counting; it does not
necessarily mean that every explicit time derivative has been removed.
All starred quantities are evaluated at the common physical pivot.

Second, the time-dependent part of the in-in calculation is expanded
around the same pivot. This expansion includes the logarithmic
evolution of the bare cubic coefficient, the first-order correction
to the scale factor, and the corrections to the three scalar or tensor
mode functions. These contributions modify the leading momentum- and
helicity-dependent shape. With this separation, the complete
amplitudes take the form
\begin{subequations}
\begin{align}
\mathcal A_\lambda^{(\zeta\zeta h)}
&=
\sum_{q=1}^{6}
\left(
\mathcal C_{q*}^{\rm lead}
+\delta\mathcal C_{q*}
\right)
\mathcal S_q^{(0)}
+
\sum_{q=1}^{6}
\mathcal C_{q*}^{\rm lead}\,
\delta\mathcal Q_q,
\label{eq:zeta_zeta_h_final_nlo_amplitude}
\\
\mathcal A_{\lambda_2\lambda_3}^{(\zeta hh)}
&=
\sum_{q=1}^{7}
\left(
\mathcal D_{q*}^{\rm lead}
+\delta\mathcal D_{q*}
\right)
\mathcal S_{q,\lambda_2\lambda_3}^{(0)}
+
\sum_{q=1}^{7}
\mathcal D_{q*}^{\rm lead}\,
\delta\mathcal Q_{q,\lambda_2\lambda_3}^{(\zeta hh)},
\label{eq:z_hh_final_nlo_amplitude}
\\
\mathcal A_{\lambda_1\lambda_2\lambda_3}^{(hhh)}
&=
\left(
\mathcal E_{\rm new,*}^{\rm lead}
+\delta\mathcal E_{\rm new,*}
\right)
\mathcal S_{\lambda_1\lambda_2\lambda_3,\rm new}^{(0)}
+
\mathcal S_{\lambda_1\lambda_2\lambda_3,\rm GR}^{(0)}
\nonumber\\
&\quad
+
\mathcal E_{\rm new,*}^{\rm lead}\,
\delta\mathcal Q_{\rm new}
+
\delta\mathcal Q_{\rm GR}.
\label{eq:hhh_final_nlo_amplitude}
\end{align}
\end{subequations}

In the first equation,
$\mathcal C_{q*}^{\rm lead}$ is the reduced coefficient multiplying
the leading shape of the $q$-th $\zeta\zeta h$ interaction, while
$\delta\mathcal C_{q*}$ is the additive first-order correction to its
value at the common pivot. The analogous quantities in the
$\zeta hh$ sector are
$\mathcal D_{q*}^{\rm lead}$ and
$\delta\mathcal D_{q*}$.

The functions $\mathcal S^{(0)}$ contain the complete leading
momentum and helicity dependence, including the leading temporal
integral. By contrast, the functions $\delta\mathcal Q$ contain the
first-order correction to this leading shape at fixed leading pivot
coefficient. They receive contributions from the logarithmic evolution
of the bare cubic coefficient, from the scale factor, and from the
first-order correction to each of the three mode functions. Only one
such correction is inserted at a time, because a product of two
first-order corrections is of $\mathcal O(\varepsilon^2)$.

The distinction between the two terms in each amplitude is therefore
important. The quantities
$\delta\mathcal C_{q*}$,
$\delta\mathcal D_{q*}$, and
$\delta\mathcal E_{\rm new,*}$ describe corrections to the coefficient
evaluated at the pivot. The logarithmic evolution of that coefficient
away from the pivot is not included there; it is part of the
corresponding $\delta\mathcal Q$. This separation prevents the same
slow time dependence from being counted twice.

In the tensor sector, the first shape is generated by the additional
Horndeski interaction proportional to $\mu_5$, while the second is the
universal GR contribution. The normalized leading coefficient of the
GR interaction is unity, so its first-order corrections are contained
entirely in $\delta\mathcal Q_{\rm GR}$. The precise reduced
normalizations are defined at their first use in
Eqs.\eqref{eq:reduced_c_coefficients},
\eqref{eq:reduced_b_coefficients}, and
\eqref{eq:reduced_tensor_coefficient}, while $\mu_5$ is defined in
Eq.\eqref{mu5_def}.

All temporal integrals entering the functions $\delta\mathcal Q$ are
evaluated analytically in the appendices. The final amplitudes
therefore contain no unevaluated time integrals.
All first-order functions are evaluated analytically in the appendices.

\item \textbf{NLO Maldacena tests.}

We describe each squeezed limit by a dimensionless coefficient after
factoring out the two relevant power spectra and the known angular or
helicity structure. In the tensor-soft $h\zeta\zeta$ limit,
$\mathcal R_{h\zeta\zeta}$ is the coefficient multiplying the
polarization-dependent angular factor in
$B_{h\zeta\zeta}/(P_hP_\zeta)$. It measures the modulation of the
short-scale scalar spectrum by a long-wavelength tensor mode. In the
scalar-soft $\zeta hh$ limit, $\mathcal R_{\zeta hh}$ is the coefficient
multiplying $\delta_{\lambda\lambda'}$ in
$B_{\zeta hh}/(P_\zeta P_h)$. It measures the modulation of the
short-scale tensor spectrum by a long-wavelength scalar mode. Their
precise definitions are given in
Eqs.\eqref{Eps_large} and \eqref{eq:ss_R_definition}, respectively.

The explicit soft limits reproduce the symmetry predictions,
\begin{align*}
\mathcal R_{h\zeta\zeta}
&=
\frac32+\epsilon+\frac12f_S+\frac12s,
&
\mathcal R_{\zeta hh}
&=
2\epsilon+f_T+s_T
=
-n_T.
\end{align*}
The first equality follows from a single surviving interaction, whereas
the second appears only after the leading time-derivative and
spatial-gradient contributions cancel and all first-order terms are
combined; see Eqs.\eqref{eq:sg_final_result} and
\eqref{eq:ss_final_response}. These agreements provide a sensitive
check of the relative normalization of the quadratic and cubic actions,
the first-order mode functions, and the corrected power spectra.

\item \textbf{Nonlinearity parameters.}

We define
\begin{align}
 f_{\rm NL}^{A}(k_1,k_2,k_3;k_0)
 &=\frac{10}{3}
 \frac{k_1^3k_2^3k_3^3B_A(k_1,k_2,k_3)}
 {(2\pi)^4\mathcal P_A^2(k_0)(k_1^3+k_2^3+k_3^3)}.
\label{eq:fnl_definition}
\end{align}
The arbitrary-triangle results are
\begin{subequations}
\begin{align}
 f_{\rm NL}^{(\zeta\zeta h),\lambda}
 &=\frac{10H^2}{3F_Tc_T\sum_i k_i^3}
 \sum_{q=1}^{6}
 \left[V_\lambda^{(q)}\eta_q\Lambda_qF_q\widehat c_q
 +(k_1\leftrightarrow k_2)\right],
\label{eq:fnl_zzh_general}
\\
 f_{\rm NL}^{(\zeta hh),\lambda_2\lambda_3}
 &=\frac{80H^2F_Sc_S}{3F_T^2c_T^2\sum_i k_i^3}
 \sum_{q=1}^{7}
 \left[V_{\lambda_2\lambda_3}^{(q)}\eta_q\Lambda_qF_q\widehat b_q
 +(k_2,\lambda_2\leftrightarrow k_3,\lambda_3)\right],
\label{eq:fnl_zhh_general}
\\
 f_{\rm NL}^{(hhh),\lambda_1\lambda_2\lambda_3}
 &=\frac{10}{3\sum_i k_i^3}
 \Bigg\{
 V_{\lambda_1\lambda_2\lambda_3}^{(\rm new)}
 \frac{k_1^2k_2^2k_3^2}{K^3}\frac{H}{4G_T}
 [\mu_{5*}^{\rm lead}+\delta\mu_{5*}
 +\mu_{5*}^{\rm lead}\rho_{\rm new}]
 \nonumber\\
 &\hspace{2.5cm}+
 V_{\lambda_1\lambda_2\lambda_3}^{(\rm GR)}
 \widetilde{\mathcal A}_{\rm GR}(k_i)(1+\rho_{\rm GR})
 \Bigg\}.
\label{eq:fnl_hhh_general}
\end{align}
\end{subequations}
The corrected coefficients are explicit analytic functions, so these
expressions can be used directly for arbitrary shapes or specialized to the
equilateral benchmarks below.

\item \textbf{Contracting model, finite squeezing, and observational prospects.}

For $d_3=-2$, one has $c_T=1$, $\mu_5=0$, and all first-order mode
corrections are proportional to $\epsilon=(1-n_s)/(3-n_s)$. The mixed
amplitudes retain $\chi$ through
$c_S=(1-\chi)\sqrt{(1-n_s)/3}$, while the tensor amplitude is fixed by
$n_s$; the factor-of-nine helicity relation in
Eq.\eqref{eq:bounce_helicity_ratio} also survives at first order. We quote
these as model predictions rather than observational bounds, since a bound
on $\chi$ requires a dedicated projection of
Eqs.\eqref{eq:fnl_zzh_general} and \eqref{eq:fnl_zhh_general} onto CMB
$T$-, $E$-, and $B$-mode bispectra \cite{Philcox:2024wqx}.

For $k_1=k_2=k$ and $k_3=q$, the true soft variable is
$u=c_Tq/(c_Sk)$. The NLO result approaches the Maldacena value for
$u\ll1$ and departs from it around the acoustic crossover $u\simeq1$,
which becomes $q/k\simeq c_S$ in the model. Thus a geometrically squeezed
triangle need not be acoustically soft when $c_S\ll1$. Low-noise CMB
polarization is the natural observational target: forecasts in
Ref.\cite{Philcox:2024wqx} identify LiteBIRD and the Simons Observatory
as particularly useful, because $B$ modes substantially improve tensor and
mixed-bispectrum constraints.

\end{enumerate}

\section{General slow-roll approximation for physical scalar and tensor perturbations}
\label{sec:general_horn}

In this section, we derive the general solutions for free scalar and tensor perturbations in Horndeski theory to first order in the slow-variation parameters. These results are subsequently used in the $\zeta\zeta h$, $\zeta hh$, and $hhh$ sectors. The corresponding next-to-leading analysis of the three-scalar sector was carried out in Ref.  \cite{DeFelice:2013ar}. 

\subsection{Scalar mode function}

Eliminating the lapse and shift at quadratic order gives the scalar
action of Refs. \cite{Kobayashi:2011nu,Gao:2012ib},
\begin{align}
S_{\zeta\zeta}^{(2)}
&=\int dt\,d^3x\,a^3
\left[G_S\dot\zeta^2-\frac{F_S}{a^2}(\partial_i\zeta)^2\right],
&c_S^2&=\frac{F_S}{G_S}.
\label{eq:quadratic_action_slowroll}
\end{align}
The coefficients entering the scalar and tensor quadratic actions are
determined by the Horndeski action's functions as follows
\cite{Kobayashi:2011nu,Gao:2012ib}:
\begin{subequations}
\label{eq:horndeski_quadratic_coefficients}
\begin{align}
F_T
&=
2\left[
G_4
-
X\left(
\ddot\phi\,G_{5X}+G_{5\phi}
\right)
\right],
\label{eq:horndeski_FT}
\\
G_T
&=
2\left[
G_4
-
2XG_{4X}
-
X\left(
H\dot\phi\,G_{5X}-G_{5\phi}
\right)
\right],
\label{eq:horndeski_GT}
\end{align}
as well as
\begin{align}
\Sigma
={}&
XG_{2X}
+
2X^2G_{2XX}
+
12H\dot\phi\,XG_{3X}
+
6H\dot\phi\,X^2G_{3XX}
-
2XG_{3\phi}
-
2X^2G_{3\phi X}
-
6H^2G_4
\nonumber\\
&+
6\Big[
H^2
\left(
7XG_{4X}
+
16X^2G_{4XX}
+
4X^3G_{4XXX}
\right)
\nonumber\\
&\hspace{2.3cm}
-
H\dot\phi
\left(
G_{4\phi}
+
5XG_{4\phi X}
+
2X^2G_{4\phi XX}
\right)
\Big]
\nonumber\\
&+
30H^3\dot\phi\,XG_{5X}
+
26H^3\dot\phi\,X^2G_{5XX}
+
4H^3\dot\phi\,X^3G_{5XXX}
\nonumber\\
&-
6H^2X
\left(
6G_{5\phi}
+
9XG_{5\phi X}
+
2X^2G_{5\phi XX}
\right),
\label{eq:horndeski_Sigma}
\\[1mm]
\Theta
={}&
-\dot\phi\,XG_{3X}
+
2HG_4
-
8HXG_{4X}
-
8HX^2G_{4XX}
+
\dot\phi\,G_{4\phi}
+
2X\dot\phi\,G_{4\phi X}
\nonumber\\
&-
H^2\dot\phi
\left(
5XG_{5X}
+
2X^2G_{5XX}
\right)
+
2HX
\left(
3G_{5\phi}
+
2XG_{5\phi X}
\right).
\label{eq:horndeski_Theta}
\end{align}
Finally,
\begin{align}
F_S
&=
\frac{1}{a}
\frac{d}{dt}
\left(
\frac{aG_T^2}{\Theta}
\right)
-
F_T,
\label{eq:ss_FS_identity}
\\
G_S
&=
\frac{\Sigma}{\Theta^2}G_T^2
+
3G_T.
\label{eq:horndeski_GS}
\end{align}
\end{subequations}
Here $G_{iX}\equiv\partial G_i/\partial X$ and
$G_{i\phi}\equiv\partial G_i/\partial\phi$, with repeated subscripts
denoting higher partial derivatives. Our $G_2(\phi,X)$ is denoted by
$K(\phi,X)$ in Refs. \cite{Kobayashi:2011nu,Gao:2012ib}.
The absence of ghost and gradient instabilities requires
\begin{equation}
F_S>0,
\qquad
G_S>0,
\qquad
F_T>0,
\qquad
G_T>0.
\label{eq:quadratic_stability_conditions}
\end{equation}
We also select the Bunch--Davies state in an adiabatic regime where the
fluctuation theory is perturbative. With $N=\ln a$ and
$d/dN=H^{-1}d/dt$, the rates entering the scalar modes are
\begin{align}
\epsilon&=-\frac{\dot H}{H^2},
&f_S&=\frac{\dot F_S}{HF_S},
&g_S&=\frac{\dot G_S}{HG_S},
&s&=\frac{\dot c_S}{Hc_S}=\frac12(f_S-g_S).
\label{eq:slowroll_parameters_general}
\end{align}
The slow-variation expansion assumes not only small instantaneous
rates but also their derivative hierarchy,
\begin{align}
\epsilon,\,f_S,\,s&=\mathcal O(\varepsilon),
&\frac{d\epsilon}{dN},\,\frac{df_S}{dN},\,\frac{ds}{dN}
&=\mathcal O(\varepsilon^2).
\label{eq:scalar_slow_hierarchy}
\end{align}
Here and throughout the paper, $\varepsilon$ denotes a formal
bookkeeping parameter that collectively characterizes the magnitude of
all slow-variation quantities; it should not be confused with the
Hubble slow-variation parameter
$\epsilon=-\dot H/H^2$. Accordingly, every logarithmic background or
interaction-coefficient rate introduced below is counted as
$\mathcal O(\varepsilon)$ whenever it is nonvanishing and well defined,
while one derivative with respect to $N$ raises the slow-variation
order by one.
Throughout, $G_4$ includes the Einstein--Hilbert contribution, so that
$G_4=M_{\rm pl}^2/2$ gives Einstein gravity and
$F\equiv2G_4/M_{\rm pl}^2$.\footnote{In what follows we use $M_{\mathrm{pl}}=1$.} When translating
Ref.\cite{DeFelice:2013ar}, its combinations
$M_{\rm pl}^2F\epsilon_s$ and $\eta_{sF}$ are therefore $F_S$ and
$f_S$, respectively. We use only $F_S$ and $f_S$ in the mode
calculation, without assuming that $F$ is constant or identifying
$f_S$ with $d\ln\epsilon_s/dN$.

For $\zeta_{\bm k}=u_k a_{\bm k}+u_k^*a_{-\bm k}^\dagger$, the
sound-horizon time and canonical normalization are
\begin{align}
dy_s&=c_S\,d\tau=\frac{c_S}{a}\,dt,
&q_s&=a\sqrt{\frac{2F_S}{c_S}},
&v_{s,k}&=q_su_k.
\label{eq:sound_horizon_time}
\end{align}
The first-order scalar mode and its corrections were obtained in
Ref.\cite{DeFelice:2013ar}, and we quote the result rather than
repeat its derivation. Canonical normalization and the use of $y_s$
reduce the mode equation to Hankel form after the rates are held
constant to first order, with
\begin{align}
\delta_S&\equiv\epsilon+\frac12f_S+\frac12s,
\label{eq:delta_s_definition}\\
\nu_S&=\frac32+\delta_S+\mathcal O(\varepsilon^2).
\label{eq:nu_s_slowroll}
\end{align}
Choosing a constant oscillator phase as specified below, the result is
\begin{align}
u_k(x_s)
={}&-\frac{\sqrt\pi H}{2\sqrt{2F_Sc_S}\,k^{3/2}}
 x_s^{3/2}\left(1+\frac{\epsilon+s}{2}\right)\mathrm e^{i\theta_{S*}}
 H_{\nu_S}^{(1)}\!\left[(1+\epsilon+s)x_s\right],
\label{eq:scalar_mode_slow_variation_x}
\end{align}
where $x_s\equiv\frac{c_Sk}{aH}$, $H_{\nu_S}^{(1)}$ is the Hankel function of the first kind and phase convention is discussed below.
Here the background functions are evaluated at the time associated
with $x_s$, while the rates can be evaluated at the common pivot to
the retained order. The Hankel index includes the slow variation of
the canonical normalization $q_s$ as well as the sound-horizon time;
$\nu_S=3/2$ alone does not imply that all background coefficients
are constant.
In the zero-rate limit $\epsilon=f_S=s=0$, with finite positive
$F_S$ and $c_S$ held fixed, the mode reduces to
\begin{align}
u_k^{(0)}(\tau)
&=\frac{iH}{2\sqrt{F_Sc_S}\,k^{3/2}}
 (1+ic_Sk\tau)\mathrm e^{-ic_Sk\tau}.
\label{eq:scalar_mode_leading}
\end{align}
This is the frozen-background de Sitter mode of
Ref.\cite{Gao:2012ib}, rather than a requirement that a particular
model admit an exact de Sitter limit with a nonzero scalar kinetic
coefficient.

For the interaction integrals we use one physical pivot for all
three legs, whose species are denoted by $X_i=S,T$. Its acoustic frequency and conformal-time
reference are
\begin{align}
K_*&\equiv\sum_{i=1}^3c_{X_i*}k_i=a_*H_*,
&\tau_*&\equiv-\frac1{K_*},
&L(\tau)&\equiv\ln(\tau/\tau_*)=\ln(-K_*\tau),
\label{eq:modes_common_pivot}\\
\omega_i&\equiv c_{X_i*}k_i,
&r_i&\equiv\frac{\omega_i}{K_*},
&x_i&\equiv-\omega_i\tau=r_i(-K_*\tau),
\label{eq:modes_frequency_variables}
\end{align}
For the further purposes we also introduce the dimensionless integration variable
\begin{align}
y
=
-K_*\tau.
\label{eq:app_zzh_y}
\end{align}
The scalar and tensor propagation speeds are denoted by $c_S$ and $c_T$,
respectively, as defined in Eqs.\eqref{eq:quadratic_action_slowroll} and
\eqref{eq:tensor_quadratic_action}.
Stars refer to the physical time $t_*$, whereas $\tau_*$ is the
leading conformal reference used inside first-order logarithms; its
difference from $\tau(t_*)$ affects these logarithms only at second
order. The three-tensor sector uses $K_*=K_{T*}=c_{T*}\sum_i k_i$,
while the species assignments reproduce the two mixed frequencies.
For a scalar mode we write $x_i=x_{si}$ and define
\begin{align}
A_{S,i}&\equiv\frac{H_*}{2\sqrt{F_{S*}c_{S*}}\,k_i^{3/2}},
&\theta_{S*}&\equiv\frac\pi2\left(\epsilon_*+\frac12f_{S*}\right),
&\widetilde u_{k_i}&\equiv\mathrm e^{-i\theta_{S*}}u_{k_i}.
\label{eq:scalar_phase_theta}
\end{align}
The phase convention differs from the exactly positive-frequency
canonical normalization by a constant rephasing, compensated in the
oscillators, and does not change the state or any propagator. The
additive correction is defined for the phase-stripped mode,
\begin{align}
\widetilde u_{k_i}^*(\tau)
&=u_{k_i}^{(0)*}(\tau)+\Delta u_{k_i}^*(\tau)
 +\mathcal O(\varepsilon^2A_{S,i}),
&u_{k_i}^{(0)*}&=-iA_{S,i}(1+ix_{si})\mathrm e^{-ix_{si}}.
\label{eq:scalar_mode_expansion}
\end{align}
To state the required results compactly, we also use the Hankel-order
function used throughout the  appendices. Its explicit
exponential-integral representation is given in
Eq.\eqref{eq:nlf_hankel_E1}, while the common regularization and
analytic evaluation of the corresponding integrals are developed in
Sec.\ref{app:zzh_hankel_integrals}:
\begin{align}
\mathfrak h^*(x)
&\equiv\sqrt{\frac\pi2}\,x^{3/2}
 \left.\partial_\nu H_\nu^{(2)}(x)\right|_{\nu=3/2},
&\mathfrak h^{*\prime}(x)&\equiv\frac{d\mathfrak h^*(x)}{dx}.
\label{eq:modes_hankel_order_function}
\end{align}
In the next two expressions $x=x_{si}$ and the stars on the rates
are suppressed, while $A_{S,i}$ and $c_{S*}k_i$ remain fixed:
\begin{align}
\Delta u_{k_i}^*(\tau)
={}&-A_{S,i}\mathrm e^{-ix}
 \Big[(\epsilon+s)(x-i)+isx^2
 +\{\delta_S(i-x)-isx^2\}L(\tau)
\nonumber\\
&\hspace{3.6cm}
 +\delta_S\mathrm e^{ix}\mathfrak h^*(x)\Big],
\label{eq:scalar_mode_correction}\\
\partial_\tau\Delta u_{k_i}^*(\tau)
={}&A_{S,i}c_{S*}k_i \mathrm e^{-ix}
 \Big[x(sx-i\epsilon)+\delta_S(i/x-1)
\nonumber\\
&\qquad
 +i(\delta_S-2s+isx)xL(\tau)
 +\delta_S\mathrm e^{ix}\mathfrak h^{*\prime}(x)\Big],
\label{eq:scalar_mode_derivative_correction}
\end{align}
where we also use \eqref{eq:delta_s_definition} and  \eqref{eq:modes_common_pivot}.
These are the first-order scalar results of
Ref.\cite{DeFelice:2013ar}, expressed at the mixed acoustic pivot.
The constant normalization phase is removed consistently from the
external and bulk mode functions entering the propagator.\footnote{In what follows, the boundary-to-bulk propagator contains a late-time external mode and a
mode evaluated at the interaction time. At first order, its correction
therefore contains the correction to each factor once.} The
inverse-power term in the derivative is combined with the
Hankel-order contribution before taking a late-time limit.

The external scalar mode at the same pivot is
\begin{align}
u_{k_i}(0)
&=i \mathrm e^{i\theta_{S*}}A_{S,i}
 \left[1+E_S(k_i)+\mathcal O(\varepsilon^2)\right],
\label{eq:scalar_superhorizon_amplitude}\\
E_S(k_i)&=-(\epsilon_*+s_*)
 -\delta_{S*}\left(\gamma_2+\ln r_i\right),
&\gamma_2&\equiv\gamma_E-2+\ln2.
\label{eq:gamma_2}
\end{align}
The logarithm accounts for transfer from the individual crossing
$c_S(t_{k_i})k_i=a(t_{k_i})H(t_{k_i})$ to $t_*$, while $u(0)$ denotes
the constant superhorizon solution after its decaying partner has
become negligible. The unresummed expansion is used only when the
logarithmic corrections are controlled, and a soft frequency fraction
is not set to zero inside an isolated mode insertion.


\subsection{Tensor mode function}

We quantize the transverse--traceless perturbation in a helicity basis,
\begin{align}
h_{ij}(\bm k,\tau)
={}&\sum_{\lambda=\pm2}
\Big[
 e_{ij}^{(\lambda)}(\bm k)h_k(\tau)a_{\bm k}^{(\lambda)}
 +e_{ij}^{(\lambda)*}(-\bm k)h_k^*(\tau)
  a_{-\bm k}^{\dagger(\lambda)}
\Big],
\label{eq:tensor_mode_expansion}\\
\big[a_{\bm k}^{(\lambda)},
 a_{\bm k'}^{\dagger(\lambda')}\big]
={}&(2\pi)^3\delta_{\lambda\lambda'}
\delta^{(3)}(\bm k-\bm k'),
\nonumber
\end{align}
with all remaining tensor-oscillator commutators equal to zero; the
expansion of $h'_{ij}$ is obtained from
Eq.\eqref{eq:tensor_mode_expansion} by differentiating the mode
functions and need not be written separately. We use a complex helicity basis of symmetric, transverse, traceless,
norm-one polarization tensors,
\begin{align}
e_{ij}^{(\lambda)}(\bm k)
&=e_{ji}^{(\lambda)}(\bm k),
&k_i e_{ij}^{(\lambda)}(\bm k)&=0,
&e_{ii}^{(\lambda)}(\bm k)&=0,
\nonumber\\
e_{ij}^{(\lambda)}(\bm k)
 e_{ij}^{(\lambda')*}(\bm k)
&=\delta_{\lambda\lambda'},
&e_{ij}^{(\lambda)}(\bm k)
 e_{ij}^{(\lambda)}(\bm k)&=0,
\nonumber\\
e_{ij}^{(\lambda)*}(\bm k)
&=e_{ij}^{(-\lambda)}(\bm k)
 =e_{ij}^{(\lambda)}(-\bm k).
\label{eq:tensor_polarization_conventions}
\end{align}
The final equality fixes the relative phases of the helicity bases at
opposite momenta and is therefore a convention rather than a
phase-independent identity. The helicity amplitude and the associated
tensor projector are defined once and for all by
\begin{align}
h^{(\lambda)}(\bm k,\tau)
&\equiv
 e_{ij}^{(\lambda)*}(\bm k)h_{ij}(\bm k,\tau),
\label{eq:helicity_extraction_convention}\\
\Pi_{ij,kl}(\bm k)
&\equiv
 \sum_{\lambda=\pm2}
 e_{ij}^{(\lambda)}(\bm k)
 e_{kl}^{(\lambda)*}(\bm k).
\label{eq:tensor_projector}
\end{align}
Accordingly, every external helicity projection below is performed
with the conjugate polarization tensor, whereas real or imaginary
projections in the in-in calculation act only on the temporal
integrals after the momentum and helicity tensors have been factored
out.

With this norm-one convention, the quadratic tensor action is
\begin{align}
S_T^{(2)}
&=\frac18\int dt\,d^3x\,a^3
 \left[G_T\dot h_{ij}^{\,2}
 -\frac{F_T}{a^2}(\partial_kh_{ij})^2\right],
&c_T^2&=\frac{F_T}{G_T}.
\label{eq:tensor_quadratic_action}
\end{align}
The coefficients $F_T$ and $G_T$ are given explicitly in
Eqs.\eqref{eq:horndeski_FT} and \eqref{eq:horndeski_GT},
respectively, while their positivity is included in
Eq.\eqref{eq:quadratic_stability_conditions}.
The canonical quantities follow the
Ref.\cite{Kobayashi:2011nu} read
\begin{align}
dy_T&=c_T\,d\tau,
&z_T&=\frac a2\sqrt{\frac{F_T}{c_T}},
&v_\lambda(\bm k,y_T)&=z_T h^{(\lambda)}(\bm k,y_T).
\label{eq:tensor_canonical_variables}
\end{align}
The tensor rates and index are
\begin{align}
f_T&\equiv\frac{\dot F_T}{HF_T},
&g_T&\equiv\frac{\dot G_T}{HG_T},
&s_T&\equiv\frac{\dot c_T}{Hc_T}=\frac12(f_T-g_T),
\label{eq:tensor_slow_rates}\\
\delta_T&\equiv\epsilon+\frac12f_T+\frac12s_T,
&\nu_T&=\frac32+\delta_T+\mathcal O(\varepsilon^2).
\label{eq:tensor_hankel_index}
\end{align}
The rates $f_T,g_T,s_T$ obey the same smallness and derivative
hierarchy as their scalar counterparts. Applying the method of
Ref.\cite{DeFelice:2013ar} to this tensor action gives the following
counterpart of Eq.\eqref{eq:scalar_mode_slow_variation_x}, with its
derivation collected in Appendix\ref{app:tensor_mode_functions}:
\begin{align}
h_k(y_h)
&=-\frac{\sqrt\pi H}{\sqrt{F_Tc_T}\,k^{3/2}}
 y_h^{3/2}\left(1+\frac{\epsilon+s_T}{2}\right)\mathrm e^{i\theta_{T*}} H_{\nu_T}^{(1)}\!\left[(1+\epsilon+s_T)y_h\right],
\label{eq:tensor_mode_slow_variation}\\
y_h&\equiv\frac{c_Tk}{aH},\quad
\theta_{T*}\equiv\frac\pi2\left(\epsilon_*+\frac12f_{T*}\right).
\nonumber
\end{align}
Both displayed Hankel solutions are understood through first order,
so their unexpanded form does not imply second-order accuracy.

At the common pivot, the tensor normalization and frozen mode are
\begin{align}
A_{T,i}&\equiv\frac{\sqrt2 H_*}{\sqrt{F_{T*}c_{T*}}\,k_i^{3/2}},
&x_{hi}&\equiv-c_{T*}k_i\tau,
\nonumber\\
h_{k_i}^{(0)}(\tau)
&=iA_{T,i}(1-ix_{hi})\mathrm e^{ix_{hi}}.
\label{eq:tensor_mode_leading}
\end{align}
The limit $\epsilon=f_T=s_T=0$ gives $\nu_T=3/2$ and
$y_h=x_{hi}$, thereby recovering this de Sitter expression without
requiring $c_T=1$. With
$\widetilde h_{k_i}=\mathrm e^{-i\theta_{T*}}h_{k_i}$, the corrections used
in the interaction integrals can be written without repeating the
scalar formulas:
\begin{align}
\widetilde h_{k_i}^*
&=h_{k_i}^{(0)*}+\Delta h_{k_i}^*
 +\mathcal O(\varepsilon^2A_{T,i}),
\nonumber\\
\Delta h_{k_i}^*
&=2\sqrt2\left.\Delta u_{k_i}^*\right|_{S\to T},
&\partial_\tau\Delta h_{k_i}^*
&=2\sqrt2\left.\partial_\tau\Delta u_{k_i}^*\right|_{S\to T}.
\label{eq:tensor_mode_insertions}
\end{align}
Here $S\to T$ means $F_S\to F_T$, $c_S\to c_T$, $f_S\to f_T$,
$s\to s_T$, $\delta_S\to\delta_T$, and $x_{si}\to x_{hi}$ at the
same sector pivot, while $\epsilon$ is unchanged. The factor
$2\sqrt2$ is fixed by the relative scalar and tensor canonical
normalizations, and no additional phase or factor of $i$ is inserted.
The external tensor mode is
\begin{align}
h_{k_i}(0)
&=i\mathrm e^{i\theta_{T*}}A_{T,i}
 \left[1+E_T(k_i)+\mathcal O(\varepsilon^2)\right],
\nonumber\\
E_T(k_i)&=-(\epsilon_*+s_{T*})
 -\delta_{T*}\left(\gamma_2+\ln r_i\right),
\label{eq:tensor_external_mode}
\end{align}
where $r_i$ is given by \eqref{eq:modes_frequency_variables}.
With these conventions, the corrected dimensionless spectra are
\begin{align}
\mathcal P_\zeta(k)
&=\frac{H_*^2}{8\pi^2F_{S*}c_{S*}}
 \left[1+2E_S(k)+\mathcal O(\varepsilon^2)\right],
\nonumber\\
\mathcal P_h(k)
&=\frac{2H_*^2}{\pi^2F_{T*}c_{T*}}
 \left[1+2E_T(k)+\mathcal O(\varepsilon^2)\right],
\label{eq:modes_corrected_spectra}
\end{align}
where $\mathcal P_h$ includes both tensor helicities, and the
constant phases cancel between the external and conjugate bulk
factors for either species, including when the latter is
differentiated. A first-order propagator insertion therefore
corrects its external and bulk endpoints exactly once, as
implemented in the subsequent bispectrum calculation.

With the scalar and tensor modes expanded through first order in slow variation, we now evaluate the $hhh$, $\zeta hh$, and $\zeta\zeta h$ bispectra in general Horndeski theory.

\section{Mixed and tensor bispectra at first order in slow variation}
\label{sec:mixed}

Using the general scalar and tensor mode functions obtained in the previous section, we compute the first-order slow-variation corrections to the three-point functions in general Horndeski theory.

\subsection{The in-in formalism and evaluation of bispectra}

To compute three-point correlation functions, we employ the perturbative Schwinger--Keldysh method, or equivalently the \(\text{in-in}\) formalism, based on the perturbative expansion of the time-evolution operator in a Dyson series \cite{Maldacena:2002vr}.
At first order in the interaction Hamiltonian, for an arbitrary operator \(Q\), we have
\begin{equation*}
	\bra{\Omega}Q(\tau)\ket{\Omega}
	\approx
	-i\int_{\tau_0}^{\tau}d\tilde\tau\,
	\bra{0}\big[Q(\tau),H_{\text{int}}(\tilde\tau)\big]\ket{0}
	=
	i\int_{\tau_0}^{\tau}d\tilde\tau\,
	\bra{0}\left[Q(\tau),\int d^3x\,\mathcal L_{\text{int}}(\tilde\tau)\right]\ket{0}
\end{equation*}
where in the second equality we use the relation \(H_{\mathrm{int}}^{(3)}=-\int d^3x\,\mathcal L_{\mathrm{int}}^{(3)}\) at the cubic order considered here. In what follows, \(Q\) denotes products of three scalar and/or tensor perturbations, including the $\zeta\zeta h$ and $\zeta hh$ combinations; the corresponding multi-indices are omitted for brevity.

The perturbation operators entering the interaction Lagrangian are initially written in coordinate space. Therefore, in the course of the calculation one must transform to momentum space and perform the integration over \(d^3x\). Integration over the spatial coordinates produces the standard factor expressing momentum conservation,
\begin{equation*}
(2\pi)^3\delta^{(3)}(\mathbf q_1+\mathbf q_2+\mathbf q_3).
\end{equation*}
In what follows, we omit this standard momentum-conserving factor from our calculations.

\subsection{Two scalars and one graviton}

The reduced cubic action for one tensor and two scalar modes is \cite{Gao:2012ib}
	\begin{align}
		S_{\zeta\zeta h}^{(3)}
		&=
		\int dt\,d^3x\,a^3
		\Bigg[
		\frac{c_1}{a^2}h_{ij}\zeta_{,i}\zeta_{,j}
		+
		\frac{c_2}{a^2}\dot h_{ij}\zeta_{,i}\zeta_{,j}
		+
		c_3 \dot h_{ij}\zeta_{,i}(\partial^{-2}\dot\zeta)_{,j}
		+
		\frac{c_4}{a^2}\partial^2 h_{ij}\zeta_{,i}(\partial^{-2}\dot\zeta)_{,j}
		\nonumber\\
		&
		+
		\frac{c_5}{a^4}\partial^2 h_{ij}\zeta_{,i}\zeta_{,j}
		+
		c_6 \partial^2 h_{ij}(\partial^{-2}\dot\zeta)_{,i}(\partial^{-2}\dot\zeta)_{,j}
				\Bigg]
		+
		S_{E,\zeta\zeta h}.
\label{eq:lagrangian_of_two_grav_and_one_scal}
	\end{align}
Here and below, $S_{E,A}$ denotes the integrated remainder associated
with the reduction of the mixed cubic action.\footnote{The remainder
collects terms proportional to the linear equations of motion and total
derivatives, whose standard treatment follows
Ref.\cite{Gao:2012ib}; in the regular expanding-attractor regime used
here, it does not modify the late-time tree-level correlators.} The six cubic coefficients are\footnote{Here $\mu_5$ has the same meaning as $\mu$ in the work \cite{DeFelice:2013ar}. The subscript is introduced to avoid confusion with the parameter $\mu$ entering the Lagrangian of our model.}
\begin{subequations}
\begin{align*}
	c_1
	&=
	F_S,
	\\
	c_2
	&=
	\frac{\Gamma}{4\Theta}
	\left(
	F_S-F_T
	\right)
	+
	\frac{G_T^2}{\Theta}
	\left[
	-\frac12
	+
	\frac{H\Gamma}{4\Theta}
	\left(
	3+
	\frac{\dot{G}_T}{HG_T}
	\right)
	-
	\frac14\frac{d}{dt}
	\left(
	\frac{\Gamma}{\Theta}
	\right)
	\right]
	\nonumber\\
	&\quad
	+
	\frac{\mu_5F_S}{G_T}
	+
	\frac{2HG_T\mu_5}{\Theta}
	-
	G_T
	\frac{d}{dt}
	\left(
	\frac{\mu_5}{\Theta}
	\right),
	\\
	c_3
	&=
	G_S
	\left[
	\frac32
	+
	\frac{d}{dt}
	\left(
	\frac{\Gamma}{2\Theta}
	+
	\frac{\mu_5}{G_T}
	\right)
	-
	\left(
	3H+
	\frac{\dot{G}_T}{G_T}
	\right)
	\left(
	\frac{\Gamma}{2\Theta}
	+
	\frac{\mu_5}{G_T}
	\right)
	\right],
	\\
	c_4
	&=
	G_S
	\left[
	-\frac{
		G_T^2-\Gamma F_T
	}{
		2\Theta G_T
	}
	-
	\frac{2H\mu_5}{\Theta}
	+
	\frac{d}{dt}
	\left(
	\frac{\mu_5}{\Theta}
	\right)
	+
	\frac{\mu_5}{G_T^2}
	\left(
	F_T-F_S
	\right)
	\right],
	\\
	c_5
	&=
	\frac{G_T^2}{2\Theta}
	\left[
	\frac{
		G_T^2-\Gamma F_T
	}{
		2\Theta G_T
	}
	+
	\frac{2H\mu_5}{\Theta}
	-
	\frac{d}{dt}
	\left(
	\frac{\mu_5}{\Theta}
	\right)
	-
	\frac{\mu_5}{G_T^2}
	\left(
	3F_T-F_S
	\right)
	\right],
	\\
	c_6
	&=
	\frac{G_S^2}{4G_T}
	\left[
	1
	+
	\frac{6H\mu_5}{G_T}
	-
	2G_T
	\frac{d}{dt}
	\left(
	\frac{\mu_5}{G_T^2}
	\right)
	\right],
\end{align*}
\end{subequations}
where $\Theta$ is defined in
Eq.\eqref{eq:horndeski_Theta} and
\begin{align}
\label{Gamma_big}
\Gamma&=
2G_4
-8XG_{4X}
-8X^2G_{4XX}
\nonumber\\
&-2H\dot{\phi}
\left(
5XG_{5X}+2X^2G_{5XX}
\right)
+2X
\left(
3G_{5\phi}+2XG_{5\phi X}
\right),
\end{align}
and where $\mu_5$ is given by 
\begin{align}
\mu_5
&\equiv
\dot\phi\,XG_{5X}.
\label{mu5_def}
\end{align}. 
In conformal time, the Lagrangian \eqref{eq:lagrangian_of_two_grav_and_one_scal} takes the form
\begin{align}
		S_{\zeta\zeta h}^{(3)}
		&=
		\int d\tau\,d^3x
		\Bigg[
		a^2 c_1\,
		h_{ij}\zeta_{,i}\zeta_{,j}
		+
		a c_2\,
		h'_{ij}\zeta_{,i}\zeta_{,j}
		+
		a^2 c_3\,
		h'_{ij}\zeta_{,i}\Psi_{,j}
		\nonumber\\
		&
		+
		a c_4\,
		\partial^2 h_{ij}\zeta_{,i}\Psi_{,j}
		+
		c_5\,
		\partial^2 h_{ij}\zeta_{,i}\zeta_{,j}
		+
		a^2 c_6\,
		\partial^2 h_{ij}\Psi_{,i}\Psi_{,j}
				\Bigg]
		+
		S_{E,\zeta\zeta h}.
\label{eq:zeta_zeta_h_conformal}
\end{align}
where $\Psi\equiv\partial^{-2}\zeta'$.

The common mixed pivot is defined through the instantaneous total
comoving acoustic frequency of the three modes by \eqref{eq:modes_common_pivot} and in this case reads
\begin{align}
	K_*
	&\equiv
	K_{\zeta\zeta h}(t_*)
	=
	c_{S*}(k_1+k_2)+c_{T*}k_3.\label{eq:zeta_zeta_h_conformal_reference}
\end{align}
All starred background quantities are evaluated at the common physical
interaction time $t_*$. The relation \eqref{eq:zeta_zeta_h_conformal_reference}
is used as the leading-order conformal-time reference point in the
logarithmic expansion, in direct analogy with the common acoustic pivot
used in the calculation of the three-scalar sector in Ref. \cite{DeFelice:2013ar}.

For constant $\epsilon_*$, the exact conformal time corresponding to the
physical crossing condition is
\begin{align*}
	\tau_{\times *}
	&=
	-\frac{1}{(1-\epsilon_*)K_*}
	=
	-\frac{1+\epsilon_*}{K_*}
	+
	\mathcal O(\varepsilon^2).
\end{align*}
Consequently,
\begin{align*}
	\ln\left(
	\frac{\tau}{\tau_{\times *}}
	\right)
	&=
	\ln\left(
	\frac{\tau}{\tau_*}
	\right)
	+
	\ln(1-\epsilon_*)
	\\
	&=
	\ln\left(
	\frac{\tau}{\tau_*}
	\right)
	-
	\epsilon_*
	+
	\mathcal O(\varepsilon^2).	
\end{align*}
Since in the present calculation this difference is always multiplied by a coefficient
that is first order in the slow-variation parameters, replacing $\tau_{\times *}$ with
$\tau_*$ changes the bispectrum only at order $\mathcal O(\varepsilon^2)$.

\paragraph{Power counting of the cubic coefficients.}
For any cubic coefficient $g_q$ (standing for $c_q$ or later -- for $b_q$ and 
$\mu_5$), ``lead'' denotes the first nonzero term in the specified
model-dependent power counting, not the result of deleting all
explicit time derivatives from the exact coefficient.
Relative statements such as
$\delta g_{q*}=\mathcal O(\varepsilon g_{q*}^{\mathrm{lead}})$ and
$\sigma_q=\dot g_q^{\mathrm{lead}}/(Hg_q^{\mathrm{lead}})$ are used
only when the leading coefficient is nonzero.
A non-singular form of the coefficient expansion is
\begin{equation}
g_q(\tau)=g_{q*}^{\mathrm{lead}}+\delta g_{q*}
-\frac{\dot g_{q*}^{\mathrm{lead}}}{H_*}
\ln(\tau/\tau_*)+\cdots.
\label{eq:cubic_coefficient_additive_convention}
\end{equation}

The reduced pivot coefficients entering
Eq.\eqref{eq:zeta_zeta_h_final_nlo_amplitude} are
\begin{align}
\mathcal C_{q*}^{\rm lead}
&=
\frac{H_*^2}{F_{T*}c_{T*}}\,
c_{q*}^{\rm lead},
&
\delta\mathcal C_{q*}
&=
\frac{H_*^2}{F_{T*}c_{T*}}\,
\delta c_{q*}.
\label{eq:reduced_c_coefficients}
\end{align}
Only the bare coefficient $c_q(\tau)$ is expanded inside the temporal
integral. The slow variation associated with the scalar and tensor
quadratic normalizations is already included in the corrected mode
functions and is not added a second time.

For a nonzero leading coefficient, we use the shorthand
\begin{equation}
\sigma_q\equiv
\frac{\dot g_q^{\mathrm{lead}}}{H g_q^{\mathrm{lead}}}.
\label{eq:cubic_running_rate}
\end{equation}
Every product $g_{q*}^{\mathrm{lead}}\sigma_{q*}$ in the following
formulas is understood as $\dot g_{q*}^{\mathrm{lead}}/H_*$.
At a zero of the leading coefficient the additive expression, rather
than an isolated logarithmic rate, is used. An identically absent
operator is removed before forming any rate. If an operator first
appears in the pivot correction, its retained contribution is
$\delta g_{q*}$ times the leading temporal function, without a
relative first-order mode correction to that term.

The truncation assumes
$dg_q^{\mathrm{lead}}/dN=\mathcal O(\varepsilon g_{q,\mathrm{char}})$,
$d^2g_q^{\mathrm{lead}}/dN^2=\mathcal O(\varepsilon^2g_{q,\mathrm{char}})$,
and $d\delta g_q/dN=\mathcal O(\varepsilon^2g_{q,\mathrm{char}})$, where
$g_{q,\mathrm{char}}$ is a nonzero characteristic coefficient scale.
The same convention applies in all three sectors and their appendices.
Relative first-order corrections per operator must be distinguished
from a fixed overall slow-variation order: when two coefficients start
at different orders, their leading and corrected contributions are
truncated according to that common overall counting.

The coefficients entering the cubic action are the bare coefficients
$c_q(t)$. Their pivot values and logarithmic time dependence follow
Eqs.\eqref{eq:cubic_coefficient_additive_convention} and
\eqref{eq:cubic_running_rate}, with $g_q=c_q$ and
$\sigma_q=\sigma_q^{(c)}$. The reduced coefficients are those of
Eq.\eqref{eq:reduced_c_coefficients}; only the bare coefficient is
expanded inside the time integral. The time dependence of
$H^2/(F_Tc_T)$ is already contained in the external and bulk tensor-mode
corrections and is not counted a second time.

The scale factor is expanded as
\begin{align}
	a(\tau)
	&=
	-\frac{1}{H_*\tau}
	\left[
	1+\epsilon_*
	-
	\epsilon_*
	\ln\!\left(
	\frac{\tau}{\tau_*}
	\right)
	\right]
	+
	\mathcal O(\varepsilon^2).
	\label{eq:zeta_zeta_h_scale_factor_expansion}
\end{align}
Equation\eqref{eq:zeta_zeta_h_scale_factor_expansion} follows by
expanding the constant-$\epsilon$ background about the physical pivot.
Its constant $\epsilon_*$ term is retained because
$\tau_*=-K_*^{-1}$ is the logarithmic reference time rather than the exact
conformal time of physical crossing.

The scalar--scalar--tensor bispectrum is defined in
Eq.\eqref{eq:_zetazetah_norm}. Its first-order organization is the one
stated in Eq.\eqref{eq:zeta_zeta_h_final_nlo_amplitude}. For later use,
we now specify the leading shapes and the separate sources of their
corrections.

The leading-order part is
\begin{equation}
\mathcal A_{\lambda,\mathrm{lead}}^{(\zeta\zeta h)}
=
\sum_{q=1}^{6}
\mathcal C_{q*}^{\mathrm{lead}}\,
\mathcal S_q^{(0)}.
\label{eq:zeta_zeta_h_leading_amplitude}
\end{equation}
where the six leading-order shapes are defined as
\begin{align*}
	\mathcal S_q^{(0)}(k_1,k_2,k_3)
	=
	V^{(q)}(k_1,k_2,k_3)\,
	J_*^{(q)}(k_1,k_2,k_3)
	+
	(k_1\leftrightarrow k_2),\quad
	q=1,\ldots,6,
\end{align*}
and the momentum structures are
\begin{align}
\label{eq:zzh_momentum_structures}
	&V^{(1)}
	=
	\frac{K}{8k_3^2}
	(k_1-k_2-k_3)
	(k_1+k_2-k_3)
	(k_1-k_2+k_3),
	\qquad
	K=k_1+k_2+k_3,
	\\
	&V^{(2)}
	=
	V^{(1)},
	\qquad
	V^{(3)}
	=
	\frac{1}{k_2^2}V^{(1)},
	\qquad
	V^{(4)}
	=
	\frac{k_3^2}{k_2^2}V^{(1)},
	\nonumber\\
	&V^{(5)}
	=
	k_3^2V^{(1)},
	\qquad
	V^{(6)}
	=
	\frac{k_3^2}{k_1^2k_2^2}V^{(1)},\nonumber
\end{align}
as well as
\begin{equation*}
    \begin{split}
	J_*^{(1)}
	&=
	-\frac{1}{H_*^2K_*^2}[c_{S*}^3(k_1+k_2)(k_1^2+k_1k_2+k_2^2)
		+
		2c_{S*}^2c_{T*}(k_1^2+k_1k_2+k_2^2)k_3
		\\&+
		2c_{S*}c_{T*}^2(k_1+k_2)k_3^2
		+
		c_{T*}^3k_3^3],
	\\
	J_*^{(2)}
	&=
	\frac{1}{H_*}
	\frac{
		c_{T*}^2k_3^2
		\left[
		2c_{S*}^2(k_1^2+3k_1k_2+k_2^2)
		+
		3c_{S*}c_{T*}(k_1+k_2)k_3
		+
		c_{T*}^2k_3^2
		\right]
	}{
		K_*^3
	},
	\\
	J_*^{(3)}
	&=
	\frac{1}{H_*^2}
	\frac{
		c_{S*}^2c_{T*}^2k_2^2k_3^2
		(c_{S*}k_1+K_*)
	}{
		K_*^2
	},
	\\
	J_*^{(4)}
	&=
	\frac{1}{H_*}
	\frac{
		c_{S*}^2k_2^2
		\left[
		c_{S*}^2(k_1+k_2)(2k_1+k_2)
		+
		3c_{S*}c_{T*}(2k_1+k_2)k_3
		+
		2c_{T*}^2k_3^2
		\right]
	}{
		K_*^3
	},
	\\
	J_*^{(5)}
	&=
	\frac{2}{K_*^4}
	\Big[
	c_{S*}^3(k_1+k_2)(k_1^2+3k_1k_2+k_2^2)
	+
	4c_{S*}^2c_{T*}(k_1^2+3k_1k_2+k_2^2)k_3
\\
	&
	+
	4c_{S*}c_{T*}^2(k_1+k_2)k_3^2
	+
	c_{T*}^3k_3^3
	\Big],
	\\
	J_*^{(6)}
	&=
	\frac{1}{H_*^2}
	\frac{
		c_{S*}^4k_1^2k_2^2
		(K_*+c_{T*}k_3)
	}{
		K_*^2
	}.
    \end{split}
\end{equation*}
The coefficient correction and the corrected temporal functions enter the
full amplitude exactly as in
Eq.\eqref{eq:zeta_zeta_h_final_nlo_amplitude}. The correction structures
are decomposed as
\begin{align*}
	\delta\mathcal Q_q
	&=
	\delta\mathcal Q_q^{[\ln c_q]}
	+
	\delta\mathcal Q_q^{[a]}
	+
	\delta\mathcal Q_q^{[\zeta_1]}
	+
	\delta\mathcal Q_q^{[\zeta_2]}
	+
	\delta\mathcal Q_q^{[h]},
	\quad
	q=1,\ldots,6.
\end{align*}
The logarithmic coefficient contribution follows from
Eqs.\eqref{eq:cubic_coefficient_additive_convention} and
\eqref{eq:cubic_running_rate}; the scale-factor contribution follows from
Eq.\eqref{eq:zeta_zeta_h_scale_factor_expansion}. The remaining terms are
obtained by replacing one scalar or tensor line by the corresponding
first-order mode correction in
Eqs.\eqref{eq:scalar_mode_correction},
\eqref{eq:scalar_mode_derivative_correction}, and
\eqref{eq:tensor_mode_insertions}.

In this organization, the correction $\delta c_{q*}$ is already included in the term $\delta\mathcal C_{q*}\mathcal S_q^{(0)}$, whereas $\delta\mathcal Q_q^{[\ln c_q]}$ contains only the logarithmic time dependence of the bare action coefficient. The evolution of $H^2/(F_Tc_T)$ is not included in this term, because the corresponding dependence is already contained in the corrected external tensor mode and the corrected bulk tensor mode. This prevents double counting of the tensor normalization.

Combining these contributions gives the amplitude already stated in
Eq.\eqref{eq:zeta_zeta_h_final_nlo_amplitude}. The explicit evaluation of the time integrals entering these
corrections is presented in Appendix\ref{app:zeta_zeta_h_slow_variation}.

\subsection{One scalar and two gravitons}

We now consider the most general cubic interaction involving one scalar
mode and two tensor modes. The scalar mode carries momentum $\bm{k}_1$,
while the two tensor modes carry momenta $\bm{k}_2$ and $\bm{k}_3$.
The cubic action in cosmic time is \cite{Gao:2012ib}
\begin{align}
		S_{\zeta hh}^{(3)}
		&=
		\int dt\,d^3x\,a^3
		\Bigg[
		b_1\zeta\,\dot h_{ij}\dot h_{ij}
		+
		\frac{b_2}{a^2}
		\zeta\,\partial_k h_{ij}\partial_k h_{ij}
		+
		b_3\partial_k(\partial^{-2}\dot\zeta)\,
		\dot h_{ij}\partial_k h_{ij}
		\\&
		+
		b_4\dot\zeta\,
		\dot h_{ij}\dot h_{ij}
		+
		\frac{b_5}{a^2}
		\partial^2\zeta\,
		\dot h_{ij}\dot h_{ij}
		+
		b_6\partial_i\partial_j(\partial^{-2}\dot\zeta)\,
		\dot h_{ik}\dot h_{jk}
		+
		\frac{b_7}{a^2}
		\partial_i\partial_j\zeta\,
		\dot h_{ik}\dot h_{jk}
				\Bigg]
		+
		S_{E,\zeta hh},
		\label{eq:general_z_hh_cubic_action_cosmic}
\end{align}
and
\begin{subequations}
\label{eq:z_hh_b_coefficients}
\begin{align}
	b_1
	&=
	\frac{3G_T}{8}
	\left[
	1
	-
	\frac{HG_T^2}
	{\Theta F_T}
	+
	\frac{G_T}{3}
	\frac{d}{dt}
	\left(
	\frac{G_T}
	{\Theta F_T}
	\right)
	\right],
	\\[1mm]
	b_2
	&=
	\frac{F_S}{8},
	\\[1mm]
	b_3
	&=
	-\frac{G_S}{4},
	\\[1mm]
	b_4
	&=
	\frac{G_T}
	{8\Theta F_T}
	\left(
	G_T^2-\Gamma F_T
	\right)
	+
	\frac{\mu_5}{4}
	\left[
	\frac{G_S}{G_T}
	-
	1
	-
	\frac{HG_T^2}
	{\Theta F_T}
	\left(
	6+
	\frac{\dot{G}_S}
	{HG_S}
	\right)
	\right]\\
	&
	+
	\frac{G_T^2}{4}
	\frac{d}{dt}
	\left(
	\frac{\mu_5}
	{\Theta F_T}
	\right),
	\\[1mm]
	b_5
	&=
	\frac{\mu_5G_T}{4\Theta}
	\left(
	\frac{F_SG_T}
	{F_TG_S}
	-
	1
	\right),
	\\[1mm]
	b_6
	&=
	-\frac{\mu_5}{2}
	\frac{G_S}{G_T},
	\\[1mm]
	b_7
	&=
	\frac{\mu_5}{2}
	\frac{G_T}{\Theta}.
    \end{align}
\end{subequations}
Passing to conformal time, we again use the conformal-time auxiliary
variable
$
\Psi
=
\partial^{-2}\zeta'
$.
The cubic action \eqref{eq:general_z_hh_cubic_action_cosmic} then becomes
\begin{align}
		S_{\zeta hh}^{(3)}
		&=
		\int d\tau\,d^3x
		\Bigg[
		a^2b_1\zeta\,
		h'_{ij}h'_{ij}
		+
		a^2b_2\zeta\,
		\partial_kh_{ij}\partial_kh_{ij}
		+
		a^2b_3\partial_k\Psi\,
		h'_{ij}\partial_kh_{ij}
		\label{eq:z_hh_conformal_action}\\&
		+
		ab_4\zeta'\,
		h'_{ij}h'_{ij}
		+
		b_5\partial^2\zeta\,
		h'_{ij}h'_{ij}
		+
		ab_6\partial_i\partial_j\Psi\,
		h'_{ik}h'_{jk}
		+
		b_7\partial_i\partial_j\zeta\,
		h'_{ik}h'_{jk}
				\Bigg]
		+
		S_{E,\zeta hh}.
		\nonumber
	\end{align}
The scalar and tensor free modes, together with their first-order
slow-variation corrections, were derived in the previous section.
In particular, the leading scalar mode and the corrections
$\Delta u_k^\star$ and $\partial_\tau\Delta u_k^\star$ are given by
\eqref{eq:scalar_mode_leading},
\eqref{eq:scalar_mode_correction}, and
\eqref{eq:scalar_mode_derivative_correction}.

For the present mixed correlator calculations, we use the notations \eqref{eq:modes_frequency_variables} and 
the common pivot scale as in \eqref{eq:modes_common_pivot}.
All scalar expressions apply with $k=k_1$ and the common variables of
Eqs.\eqref{eq:modes_common_pivot} and
\eqref{eq:modes_frequency_variables}. The conjugate tensor modes and their
conformal-time derivatives follow from
Eq.\eqref{eq:tensor_mode_leading}, with $a=2,3$, while their first-order
corrections are given by Eq.\eqref{eq:tensor_mode_insertions}. The
external corrections are those of
Eqs.\eqref{eq:gamma_2} and \eqref{eq:tensor_external_mode}, evaluated
with the frequency fractions of the present sector. The constant
normalization phases cancel between the two endpoints of each propagator.

The scalar and tensor boundary-to-bulk propagators have the same
first-order structure:
\begin{subequations}
\label{eq:complete_propagator_correction}
\begin{align}
G_\zeta^{(0)}(k;0,\tau)
&=
u_k^{(0)}(0)u_k^{(0)*}(\tau),
\\
\Delta G_\zeta(k;0,\tau)
&=
\Delta u_k(0)u_k^{(0)*}(\tau)
+
u_k^{(0)}(0)\Delta u_k^*(\tau),
\\[1mm]
G_h^{(0)}(k;0,\tau)
&=
h_k^{(0)}(0)h_k^{(0)*}(\tau),
\\
\Delta G_h(k;0,\tau)
&=
\Delta h_k(0)h_k^{(0)*}(\tau)
+
h_k^{(0)}(0)\Delta h_k^*(\tau).
\end{align}
\end{subequations}
For a differentiated field in the interaction, $\partial_\tau$
acts only on the mode evaluated at the interaction time $\tau$.
Each corrected propagator already includes the corrections to both
the late-time external mode and the mode at the interaction time,
so no additional external-mode factor is added.

We define the bispectrum in Eq.\eqref{eq:z_hh_bispectrum_definition}.
Throughout this subsection, the leading dimensionless reference spectrum is
the $E_S=0$ limit of Eq.\eqref{eq:modes_corrected_spectra}:
\begin{equation}
\mathcal P_{\zeta*}^{(0)}
=\frac{H_*^2}{8\pi^2F_{S*}c_{S*}}.
\label{P_lead}
\end{equation}
The corresponding dimensionful spectrum follows from
Eq.\eqref{eq:sg_spectra}.
The reduced coefficient multiplying the normalized contribution of the
$q$-th interaction is
\begin{align}
\mathcal D_q
&=
8H^2
\frac{F_Sc_S}{F_T^2c_T^2}\,
b_q.
\label{eq:reduced_b_coefficients}
\end{align}
Its leading part and its additive pivot correction are obtained by
replacing $b_q$ by $b_q^{\rm lead}$ and $\delta b_q$, respectively.
Inside the temporal integral, only the logarithmic evolution of the
bare coefficient $b_q(\tau)$ is expanded. Only the bare coefficients
$b_q(t)$ are expanded in the time integrals, as specified by
Eqs.\eqref{eq:cubic_coefficient_additive_convention} and
\eqref{eq:cubic_running_rate}.
The leading amplitude is then simply
\begin{align*}
	\mathcal A_{\lambda_2\lambda_3,\mathrm{lead}}^{(\zeta hh)}
	&=
	\sum_{q=1}^{7}
	\mathcal D_{q*}^{\mathrm{lead}}
	\mathcal S_{q,\lambda_2\lambda_3}^{(0)},
\end{align*}
where the seven leading shape functions are defined as
\begin{align*}
	\mathcal S_{q,\lambda_2\lambda_3}^{(0)}
	(k_1,k_2,k_3)
	&=
	V_{\lambda_2\lambda_3}^{(q)}
	(k_1,k_2,k_3)
	I_*^{(q)}(k_1,k_2,k_3)
	\\
	&+
	\left(
	k_2,\lambda_2
	\leftrightarrow
	k_3,\lambda_3
	\right),
	\qquad
	q=1,\ldots,7.
\end{align*}
The tensor projector is defined by \eqref{eq:tensor_projector}, so that
the seven tensor momentum structures are
\begin{subequations}
    \label{eq:z_hh_tensor_momentum_structures}
\begin{align}
	V_{ij,kl}^{(1)}
	&=
	\Pi_{ij,mn}(\bm{k}_2)
	\Pi_{kl,mn}(\bm{k}_3),
	\\
	V_{ij,kl}^{(2)}
	&=
	(\bm{k}_2\cdot\bm{k}_3)
	V_{ij,kl}^{(1)},
	\\
	V_{ij,kl}^{(3)}
	&=
	\frac{\bm{k}_1\cdot\bm{k}_3}{k_1^2}
	V_{ij,kl}^{(1)},
	\\
	V_{ij,kl}^{(4)}
	&=
	V_{ij,kl}^{(1)},
	\\
	V_{ij,kl}^{(5)}
	&=
	k_1^2V_{ij,kl}^{(1)},
	\\
	V_{ij,kl}^{(6)}
	&=
	\widehat{k}_{1m}\widehat{k}_{1n}
	\Pi_{ij,mm'}(\bm{k}_2)
	\Pi_{kl,nm'}(\bm{k}_3),
	\\
	V_{ij,kl}^{(7)}
	&=
	k_1^2V_{ij,kl}^{(6)}.
    \end{align}
    \end{subequations}
After projection onto definite helicities,
\begin{align*}
	V_{\lambda_2\lambda_3}^{(q)}
&=
e_{ij}^{(\lambda_2)*}(\bm{k}_2)
e_{kl}^{(\lambda_3)*}(\bm{k}_3)
	V_{ij,kl}^{(q)},
	\qquad
	q=1,\ldots,7.
\end{align*}
Next, the leading time-integral structures are
\begin{subequations}
    \label{eq:z_hh_I1}
    \begin{align}
	I_*^{(1)}
	&=
	\frac{1}{H_*^2}
	\frac{
		c_{T*}^4k_2^2k_3^2
		\left(
		c_{S*}k_1+K_*
		\right)
	}{
		K_*^2
	},
	\\[1mm]
	I_*^{(2)}
	&=
	-\frac{1}{H_*^2K_*^2}
	\Bigg[
	c_{S*}^3k_1^3
	+
	2c_{S*}^2c_{T*}k_1^2(k_2+k_3)
\\
	&
	+
	2c_{S*}c_{T*}^2k_1
	\left(
	k_2^2+k_2k_3+k_3^2
	\right)
	+
	c_{T*}^3(k_2+k_3)
	\left(
	k_2^2+k_2k_3+k_3^2
	\right)
	\Bigg],
	\\[1mm]
	I_*^{(3)}
	&=
	\frac{1}{H_*^2}
	\frac{
		c_{S*}^2c_{T*}^2
		k_1^2k_2^2
		\left(
		K_*+c_{T*}k_3
		\right)
	}{
		K_*^2
	},
	\\[1mm]
	I_*^{(4)}
	&=
	\frac{2}{H_*}
	\frac{
		c_{S*}^2c_{T*}^4
		k_1^2k_2^2k_3^2
	}{
		K_*^3
	},
	\\[1mm]
	I_*^{(5)}
	&=
	2c_{T*}^4k_2^2k_3^2
	\frac{
		3c_{S*}k_1+K_*
	}{
		K_*^4
	},
	\\[1mm]
	I_*^{(6)}
	&=
	I_*^{(4)},
	\\[1mm]
	I_*^{(7)}
	&=
	I_*^{(5)}.
\end{align}
\end{subequations}
The equalities $I_*^{(6)}=I_*^{(4)}$ and $I_*^{(7)}=I_*^{(5)}$ follow from the fact that the corresponding pairs of operators have the same time dependence, whereas their tensor momentum contractions are different.

The coefficients entering the cubic action
\eqref{eq:z_hh_conformal_action} are the bare coefficients $b_q(t)$. Their expansion is
Eq.\eqref{eq:cubic_coefficient_additive_convention}, with $g_q=b_q$ and
$\sigma_q=\sigma_q^{(b)}$ as in Eq.\eqref{eq:cubic_running_rate}. Their
reduced pivot values are those of
Eq.\eqref{eq:reduced_b_coefficients}. Only $b_q(t)$ is expanded inside
the time integral; the slow variation of the normalization multiplying it
is already included in the corrected scalar and tensor modes.

The expansion of the scale factor was obtained above in \eqref{eq:zeta_zeta_h_scale_factor_expansion}. For the seven vertices considered here, it can be written as
\begin{align}
	a^{p_q}(\tau)
	&=
	a_0^{p_q}(\tau)
	\left[
	1+
	p_q\epsilon_*
	\left(
	1-
	\ln\left(
	\frac{\tau}{\tau_*}
	\right)
	\right)
	\right]
	+
	\mathcal O(\varepsilon^2),
	\qquad
	a_0(\tau)
	=
	-\frac{1}{H_*\tau},
	\label{eq:z_hh_scale_factor_power_expansion}
\end{align}
with
$(p_1,p_2,p_3,p_4,p_5,p_6,p_7)=(2,2,2,1,0,1,0)$.

Suppressing purely momentum-dependent and tensorial factors, the leading
temporal functions for a fixed ordering of the tensor legs are
\begin{equation}\label{eq:z_hh_leading_temporal_functions}
    \begin{split}
	\mathcal I_1^{(0)}(\tau)
	&=
	a_0^2
	G_\zeta^{(0)}
	\partial_\tau G_{h,2}^{(0)}
	\partial_\tau G_{h,3}^{(0)},
	\\
	\mathcal I_2^{(0)}(\tau)
	&=
	a_0^2
	G_\zeta^{(0)}
	G_{h,2}^{(0)}
	G_{h,3}^{(0)},
	\\
	\mathcal I_3^{(0)}(\tau)
	&=
	a_0^2
	\partial_\tau G_\zeta^{(0)}
	\partial_\tau G_{h,2}^{(0)}
	G_{h,3}^{(0)},
	\\
	\mathcal I_4^{(0)}(\tau)
	&=
	a_0
	\partial_\tau G_\zeta^{(0)}
	\partial_\tau G_{h,2}^{(0)}
	\partial_\tau G_{h,3}^{(0)}=\mathcal I_6^{(0)}(\tau),
	\\
	\mathcal I_5^{(0)}(\tau)
	&=
	G_\zeta^{(0)}
	\partial_\tau G_{h,2}^{(0)}
	\partial_\tau G_{h,3}^{(0)}=\mathcal I_7^{(0)}(\tau).
    \end{split}
\end{equation}
Here
\begin{align*}
	G_\zeta^{(0)}
	\equiv
	G_\zeta^{(0)}(k_1;0,\tau),
	\qquad
	G_{h,a}^{(0)}
	\equiv
	G_h^{(0)}(k_a;0,\tau),
	\qquad
	a=2,3.
\end{align*}
For $q=3$, the expression above corresponds to the ordering in which
$\bm{k}_2$ is carried by the differentiated tensor mode. The second
ordering is generated only once, by
$(k_2,\lambda_2)\leftrightarrow(k_3,\lambda_3)$ in the final shape and
correction formulas below.

\paragraph{Projection and normalization of the temporal functions.}

The temporal functions in Eq.\eqref{eq:z_hh_leading_temporal_functions} retain
all leading external and bulk mode normalizations. Define
\begin{align*}
\mathcal M_*
\equiv
\left|u_{k_1}^{(0)}(0)\right|^2
\left|h_{k_2}^{(0)}(0)\right|^2
\left|h_{k_3}^{(0)}(0)\right|^2=
\frac{H_*^6}
{F_{S*}c_{S*}F_{T*}^2c_{T*}^2
 k_1^3k_2^3k_3^3}.
\end{align*}
Relative to the momentum structures in
Eq.\eqref{eq:z_hh_tensor_momentum_structures}, the spatial Fourier
signs of the seven operators are
$
(\varsigma_1,\ldots,\varsigma_7)
=(1,-1,1,1,-1,1,-1).
$
These signs do not include the signs of differentiated propagators,
which remain in the temporal functions themselves.

For a fixed ordering, introduce the normalized temporal projection
\begin{align}
\label{eq:zhh_temporal_projection}
\mathscr P_q[f]
\equiv{}&
\frac{\varsigma_q}{\mathcal M_*}
\lim_{\varepsilon_{\mathrm{BD}}\to0^+}
\operatorname{Re}
\left\{
\operatorname{FP}_{\alpha=0}
\left[
 i\int_{-\infty}^{0}d\tau\,
 (-K_*\tau)^\alpha
 \mathrm e^{\varepsilon_{\mathrm{BD}}K_*\tau}
 f(\tau)
\right]
\right\}.
\end{align}
The complex integral is evaluated in its initial convergence domain
and then analytically continued in $\alpha$. All terms belonging to
a complete mode insertion use the same regulators and are combined
before finite-part extraction and real projection.
The normalization $\mathcal M_*$ is fixed at the pivot.
The momentum and helicity structures $V^{(q)}$ are kept outside
$\mathscr P_q$.

The leading temporal functions satisfy\footnote{The temporal integrals generated by this projection, including
their logarithmic and mode-function corrections, are evaluated
explicitly in Appendix\ref{app:zeta_h_h_slow_variation}.}
\begin{equation*}
I_*^{(q)}=\mathscr P_q[\mathcal I_q^{(0)}].
\end{equation*}
Moreover,
\begin{equation*}
\frac{(2\pi)^4(\mathcal{P}_{\zeta*}^{(0)})^2}
{k_1^3k_2^3k_3^3}\,\mathcal D_{q*}
=2\mathcal M_*b_{q*}.
\end{equation*}
Thus the factor of two from the in-in commutator is already included
in the conversion to $\mathcal D_{q*}$ and is not inserted again in
$\mathscr P_q$. The second tensor ordering is added exactly once.

The complete first-order organization is already displayed in
Eq.\eqref{eq:z_hh_final_nlo_amplitude}. The pivot contribution
$\delta\mathcal D_{q*}$ is fixed by
Eq.\eqref{eq:reduced_b_coefficients}; below we decompose only the
corrected shape function that remains to be evaluated.

For each vertex
\begin{align*}
	\delta\mathcal Q_{q,\lambda_2\lambda_3}^{(\zeta hh)}
	=
	\delta\mathcal Q_q^{[\ln b_q]}
	+
	\delta\mathcal Q_q^{[a]}
	+
	\delta\mathcal Q_q^{[\zeta]}
	+
	\delta\mathcal Q_q^{[h_2]}
	+
	\delta\mathcal Q_q^{[h_3]},
	\qquad
	q=1,\ldots,7.
\end{align*}

The coefficient and scale-factor corrections use the same normalized
projection as the leading temporal functions:
\begin{align}
\delta\mathcal Q_q^{[\ln b_q]}
&=-\sigma_{q*}^{(b)}V_{\lambda_2\lambda_3}^{(q)}
  \mathscr P_q\!\left[L\,\mathcal I_q^{(0)}\right]
 +(k_2,\lambda_2\leftrightarrow k_3,\lambda_3),
\nonumber\\
\delta\mathcal Q_q^{[a]}
&=p_q\epsilon_*V_{\lambda_2\lambda_3}^{(q)}
  \mathscr P_q\!\left[(1-L)\mathcal I_q^{(0)}\right]
 +(k_2,\lambda_2\leftrightarrow k_3,\lambda_3).
\end{align}
Here $L=\ln(\tau/\tau_*)$ is defined in
Eq.\eqref{eq:modes_common_pivot}, and $\mathscr P_q$ in
Eq.\eqref{eq:zhh_temporal_projection}.

In particular,
\begin{align*}
	\delta\mathcal Q_5^{[a]}
	=
	0,\quad
	\delta\mathcal Q_7^{[a]}
	=
	0.
\end{align*}

The mode-function contributions are obtained from \(I_q^{(0)}\) by replacing one scalar or tensor propagator at a time by its first-order correction, \(G^{(0)}\to\Delta G\) (and analogously \(\partial_\tau G^{(0)}\to\Delta\partial_\tau G\)).

Let
$\Delta\mathcal I_q^{[\zeta]}$ denote the temporal function obtained from
$\mathcal I_q^{(0)}$ by replacing the scalar propagator by its first-order
correction, and let
$\Delta\mathcal I_q^{[h_a]}$ denote the temporal function obtained by correcting only
the $a$-th tensor propagator. Then
\begin{align*}
\delta\mathcal Q_q^{[\zeta]}
=&
V_{\lambda_2\lambda_3}^{(q)}
\mathscr P_q\!\left[\Delta\mathcal I_q^{[\zeta]}\right]
+(k_2,\lambda_2\leftrightarrow k_3,\lambda_3),
\\
\delta\mathcal Q_q^{[h_a]}
=&
V_{\lambda_2\lambda_3}^{(q)}
\mathscr P_q\!\left[\Delta\mathcal I_q^{[h_a]}\right]
+(k_2,\lambda_2\leftrightarrow k_3,\lambda_3),
\qquad a=2,3.
\end{align*}
Only one propagator is corrected at a time, since products of two
first-order corrections are of $\mathcal O(\varepsilon^2)$. Each corrected propagator includes both its external and bulk mode
correction. Hence no additional external-mode factor is added to
these expressions. Separating the two endpoints reproduces the
$E_{\mathrm{ext}}I_*^{(q)}$ and bulk-insertion terms in
Eq.\eqref{eq:app_zhh_delta_I_explicit}.

Combining the coefficient and temporal corrections gives the amplitude
already stated in Eq.\eqref{eq:z_hh_final_nlo_amplitude}. Explicit expressions for all seven corrected shape functions
$\delta\mathcal Q_{q,\lambda_2\lambda_3}^{(\zeta hh)}$
are given in Appendix\ref{app:zeta_h_h_slow_variation}.


\subsection{Three gravitons}

We now consider the most general cubic interaction of three tensor
perturbations in Horndeski theory. The tensor modes carry momenta
$\bm{k}_1$, $\bm{k}_2$, and $\bm{k}_3$ and have helicities
$\lambda_1$, $\lambda_2$, and $\lambda_3$, respectively.

The cubic Lagrangian for three tensor perturbations in cosmic time is
given by \cite{Gao:2012ib}
\begin{align}
	\mathcal L_{hhh}^{(3)}
	&=
	a^3
	\Bigg[
	\frac{\mu_5}{12}
	\dot h_{ij}\dot h_{jk}\dot h_{ki}
	+
	\frac{F_T}{4a^2}
	\left(
	h_{ik}h_{jl}
	-
	\frac12h_{ij}h_{kl}
	\right)
	\partial_k\partial_l h_{ij}
	\Bigg],
	\label{eq:hhh_cubic_lagrangian_cosmic}
\end{align}
where $\mu_5$ is given  by \eqref{mu5_def}. The first term contains one time derivative acting on each tensor
mode and is present only when $G_{5X}\neq0$. In what follows, we
denote this contribution by the index $\mathrm{new}$. The second term contains two spatial derivatives and has the same
operator structure as the cubic tensor vertex in general relativity.
We denote this contribution by the index $\mathrm{GR}$ in what follows.

Passing to conformal time via $dt
	=
	a\,d\tau$ and $\dot h_{ij}
	=
	\frac{h'_{ij}}{a}$
we obtain
\begin{align*}
	S_{hhh}^{(3)}
	=
	\int d\tau\,d^3x
	\Bigg[
	\frac{a\mu_5}{12}
	h'_{ij}h'_{jk}h'_{ki}
	+
	\frac{a^2F_T}{4}
	\left(
	h_{ik}h_{jl}
	-
	\frac12h_{ij}h_{kl}
	\right)
	\partial_k\partial_l h_{ij}
	\Bigg].
\end{align*}
The leading tensor modes and their conformal-time derivatives follow from
Eq.\eqref{eq:tensor_mode_leading}; their first-order corrections are given
by Eq.\eqref{eq:tensor_mode_insertions}. Since all three species are tensor
modes, the common definitions
\eqref{eq:modes_common_pivot},
\eqref{eq:modes_frequency_variables}, and \eqref{eq:app_zzh_y} give
$K_*=K_{T*}=c_{T*}K$ and $r_i=k_i/K$. The complete propagator corrections
are the $X=T$ case of Eq.\eqref{eq:complete_propagator_correction}; the logarithms in
the external corrections are therefore $\ln r_i=\ln(k_i/K)$.

For the new vertex, we introduce the tensor structure
\begin{align}
	\mathbb V_{i_1j_1i_2j_2i_3j_3}^{(\mathrm{new})}
	&=
	\Pi_{i_1j_1,lm}(\bm{k}_1)
	\Pi_{i_2j_2,mn}(\bm{k}_2)
	\Pi_{i_3j_3,nl}(\bm{k}_3).
	\label{eq:hhh_new_tensor_structure}
\end{align}
Projecting onto definite helicities gives
\begin{align*}
	V_{\lambda_1\lambda_2\lambda_3}^{(\mathrm{new})}
&=
e_{i_1j_1}^{(\lambda_1)*}(\bm{k}_1)
e_{i_2j_2}^{(\lambda_2)*}(\bm{k}_2)
e_{i_3j_3}^{(\lambda_3)*}(\bm{k}_3)
	\mathbb V_{i_1j_1i_2j_2i_3j_3}^{(\mathrm{new})}.
\end{align*}
For the universal GR vertex, we define
\begin{equation}
	\begin{split}
		\mathbb V_{i_1j_1i_2j_2i_3j_3}^{(\mathrm{GR})}
		&=
		\Bigg\{
		\Pi_{i_1j_1,ik}(\bm{k}_1)
		\Pi_{i_2j_2,jl}(\bm{k}_2)\\
		&\times
		\left[
		k_{3k}k_{3l}
		\Pi_{i_3j_3,ij}(\bm{k}_3)
		-
		\frac12
		k_{3i}k_{3k}
		\Pi_{i_3j_3,jl}(\bm{k}_3)
		\right]\\
		&
		+
		5\ \text{permutations of momenta and index pairs}
		\Bigg\}.
		\label{eq:hhh_GR_tensor_structure}
	\end{split}
\end{equation}
Its projection onto states of definite helicity is
\begin{align*}
	V_{\lambda_1\lambda_2\lambda_3}^{(\mathrm{GR})}
&=
e_{i_1j_1}^{(\lambda_1)*}(\bm{k}_1)
e_{i_2j_2}^{(\lambda_2)*}(\bm{k}_2)
e_{i_3j_3}^{(\lambda_3)*}(\bm{k}_3)
	\mathbb V_{i_1j_1i_2j_2i_3j_3}^{(\mathrm{GR})}.
\end{align*}
It is also convenient to define
\begin{align*}
	\widetilde{\mathcal A}_{\mathrm{GR}}
	(k_1,k_2,k_3)
	=
	-\frac{K}{16}
	\left[
	1
	-
	\frac{1}{K^3}
	\sum_{i\neq j}
	k_i^2k_j
	-
	\frac{4k_1k_2k_3}{K^3}
	\right].
\end{align*}
The leading shape functions associated with the two tensor vertices
are
\begin{align*}
	\mathcal S_{\lambda_1\lambda_2\lambda_3,
		\mathrm{new}}^{(0)}
	&=
	\frac{k_1^2k_2^2k_3^2}{K^3}
	V_{\lambda_1\lambda_2\lambda_3}^{(\mathrm{new})},
	\\
	\mathcal S_{\lambda_1\lambda_2\lambda_3,
		\mathrm{GR}}^{(0)}
	&=
	\widetilde{\mathcal A}_{\mathrm{GR}}
	(k_1,k_2,k_3)
	V_{\lambda_1\lambda_2\lambda_3}^{(\mathrm{GR})}.
\end{align*}
For the new vertex, the time-dependent factor can equivalently be written as
\begin{align*}
	\frac{k_1^2k_2^2k_3^2}{K^3}
=
	\frac{
		c_{T*}^3k_1^2k_2^2k_3^2
	}{
		K_{T*}^3
	}.
\end{align*}

The leading tensor reference spectrum is obtained from
Eq.\eqref{eq:modes_corrected_spectra} by setting $E_T=0$; its relation to
the one-helicity dimensionful spectrum is Eq.\eqref{eq:sg_spectra}. After normalization by the leading tensor spectrum, the reduced
coefficient of the additional tensor interaction is
\begin{align}
\mathcal E_{\rm new}
&=
\frac{H\mu_5}{4G_T}.
\label{eq:reduced_tensor_coefficient}
\end{align}
Its additive first-order correction at the common pivot is obtained by
replacing $\mu_{5*}^{\rm lead}$ with $\delta\mu_{5*}$. The normalized GR
coefficient is unity.
The leading amplitude therefore takes the form
\begin{align*}
	\mathcal A_{\lambda_1\lambda_2\lambda_3,
		\mathrm{lead}}^{(hhh)}
	=
	\mathcal E_{\mathrm{new},*}^{\mathrm{lead}}
	\mathcal S_{\lambda_1\lambda_2\lambda_3,
		\mathrm{new}}^{(0)}
	+
	\mathcal S_{\lambda_1\lambda_2\lambda_3,
		\mathrm{GR}}^{(0)}.
\end{align*}

We now turn to the first-order slow-variation corrections. The coefficient
$\mu_5(t)$ follows the additive expansion
Eq.\eqref{eq:cubic_coefficient_additive_convention}, with
$g_q=\mu_5$ and $\sigma_q=\sigma_{\mu_5}$ as in
Eq.\eqref{eq:cubic_running_rate}. Its reduced pivot correction is obtained
from Eq.\eqref{eq:reduced_tensor_coefficient} by replacing
$\mu_{5*}^{\mathrm{lead}}$ with $\delta\mu_{5*}$.
For the universal vertex, the corresponding time dependence is contained in $F_T$:
\begin{align*}
	F_T(\tau)
	=
	F_{T*}
	\left[
	1
	-
	f_{T*}
	\ln\left(
	\frac{\tau}{\tau_*}
	\right)
	\right]
	+
	\mathcal O(\varepsilon^2F_T).
\end{align*}
The pivot contribution proportional to $\delta\mu_{5*}$ is kept separate
from the logarithmic time dependence, exactly as specified by
Eqs.\eqref{eq:cubic_coefficient_additive_convention} and
\eqref{eq:reduced_tensor_coefficient}.

In the corrections to the time integrals, we separately account for the time dependence of the coefficients entering the cubic vertices, namely $\mu_5(\tau)$ and $F_T(\tau)$. The slow variation of the parameters entering the free tensor modes is already included in the corrections $\Delta G_h$ and $\Delta\partial_\tau G_h$ derived in the previous section.

For the scale-factor corrections we use \eqref{eq:z_hh_scale_factor_power_expansion}, with \(p_{\rm new}=1\) and \(p_{\rm GR}=2\).
Up to purely momentum-dependent and tensorial factors, the leading
temporal functions for the two interactions are
\begin{align*}
	\mathcal I_{\mathrm{new}}^{(0)}(\tau)
	&=
	a_0(\tau)
	\,
	\partial_\tau G_{h,1}^{(0)}
	\partial_\tau G_{h,2}^{(0)}
	\partial_\tau G_{h,3}^{(0)},
	\\
	\mathcal I_{\mathrm{GR}}^{(0)}(\tau)
	&=
	a_0^2(\tau)
	\,
	G_{h,1}^{(0)}
	G_{h,2}^{(0)}
	G_{h,3}^{(0)},
\end{align*}
where $G_{h,a}^{(0)}\equiv G_h^{(0)}(k_a;0,\tau)$, $a=1,2,3$.

For the new vertex, all three tensor modes are differentiated with
respect to conformal time. For the universal vertex, the spatial
derivatives are absorbed into the tensor momentum structure
$V_{\lambda_1\lambda_2\lambda_3}^{(\mathrm{GR})}$, so that its temporal
function contains three undifferentiated tensor propagators.

The complete temporal corrections are assembled in
Eq.\eqref{eq:app_hhh_delta_I} and converted into shape corrections through
Eq.\eqref{eq:app_hhh_delta_Q}. One tensor propagator is corrected at a
time: a differentiated propagator for the additional interaction and an
undifferentiated propagator for the GR interaction. The complete
propagator correction is Eq.\eqref{eq:complete_propagator_correction}; therefore no separate
external-mode term is added.
The difference between the two cases directly reflects the structure
of the corresponding cubic operators. For the new vertex,
\begin{align*}
	h'_{ij}h'_{jk}h'_{ki},
\end{align*}
all three tensor legs enter with conformal-time derivatives, and the
corresponding corrections therefore involve
$\Delta\partial_\tau G_h$. For the universal vertex, the temporal function
contains undifferentiated tensor modes, while the spatial derivatives
are already included in the momentum-space tensor structure.
Accordingly, the corrections are expressed in terms of
$\Delta G_h$.
Combining the pivot and temporal corrections gives the full amplitude in
Eq.\eqref{eq:hhh_final_nlo_amplitude}.

Collecting the leading contributions and all first-order slow-variation
corrections, the complete bispectrum amplitudes in the three sectors are
given by Eqs.\eqref{eq:zeta_zeta_h_final_nlo_amplitude},
\eqref{eq:z_hh_final_nlo_amplitude}, and
\eqref{eq:hhh_final_nlo_amplitude}. The time integrals entering the
corrected shape functions $\delta\mathcal Q_q$,
$\delta\mathcal Q_{q,\lambda_2\lambda_3}^{(\zeta hh)}$,
$\delta\mathcal Q_{\rm new}$, and $\delta\mathcal Q_{\rm GR}$ are
evaluated explicitly in
Appendices\ref{app:zeta_zeta_h_slow_variation},
\ref{app:zeta_h_h_slow_variation}, and
\ref{app:h_h_h_slow_variation}, respectively. In the following sections,
we use these full amplitudes first to verify the Maldacena consistency
relations in the mixed soft limits and then to construct the corresponding
nonlinearity parameters. 

\section{Maldacena consistency relation check in mixed sectors}
\label{sec:Maldacena}

\subsection{Soft graviton in \texorpdfstring{$h\zeta\zeta$}{hzeta zeta}}
\label{sec:soft_graviton}

The soft-graviton limit provides a useful test of the mixed bispectrum
beyond leading order. When a long tensor mode has frozen, its leading
effect on the short scalar modes is an anisotropic change of spatial
coordinates. The corresponding squeezed three-point function is
therefore fixed by the momentum dependence of the scalar two-point
function, independently of the detailed inflationary dynamics
\cite{Maldacena:2002vr,Hinterbichler:2013dpa}. In the present calculation,
this statement tests the relative normalization of the quadratic and
cubic actions, the correction to the scale factor, and the treatment of
both endpoints of each propagator. It is complementary to the
scalar-soft limit of $\langle\zeta hh\rangle$, where several cubic
operators can contribute at the same soft order and their sum must
satisfy the consistency relation. The latter is more demanding as a test
of cancellations between different vertices; the graviton-soft limit
isolates a single interaction and provides a particularly transparent
check of its complete first-order correction.

We work at tree level in the stable, expanding attractor regime \cite{Ageeva2026att} assumed
in the mode calculation, with the Bunch--Davies initial state and regular
background coefficients. In particular, the long tensor is the constant
adiabatic mode, and the short scalar modes are evaluated after their
decaying solutions have become negligible. The equivalence between the
reduced interaction variables and the late-time metric perturbations is
understood subject to the field-redefinition and boundary assumptions
of the preceding calculation. These conditions are essential: the
argument below is not an assertion about arbitrary non-attractor
solutions or about evolution through an arbitrary non-singular phase.

It is convenient to distinguish the dimensionful Fourier spectra from
the dimensionless reference spectra used to define the reduced
bispectra. Thus, we use earlier introduced notations \eqref{eq:sg_spectra}.
 With norm-one
polarizations, the helicity projection and the angular dependence of
the squeezed correlator are given by \eqref{eq:tensor_projector} and
\begin{equation*}
\mathcal E_\lambda(\hat{\mathbf q},\hat{\mathbf k})
\equiv e_{ij}^{(\lambda)*}(\hat{\mathbf q})\hat k_i\hat k_j.
\end{equation*}
For example, when $\hat{\mathbf q}=\hat{\mathbf z}$ and
$e_{ij}^{(\pm2)}=(\hat x_i\pm i\hat y_i)
(\hat x_j\pm i\hat y_j)/2$, the angular factor is
$\mathcal E_{\pm2}=\tfrac12\sin^2\vartheta\,
\mathrm e^{\mp2i\varphi}$. The scalar coefficient extracted below is independent
of this choice of polarization phase.

We take the tensor momentum soft while maintaining exact momentum
conservation,
\begin{equation*}
    \begin{split}
        \mathbf k_1
&=\mathbf k-\frac{\mathbf q}{2},
\qquad
\mathbf k_2=-\mathbf k-\frac{\mathbf q}{2},
\qquad
\mathbf k_3=\mathbf q,
\\
B_{h\zeta\zeta}^{\lambda}(\mathbf q,\mathbf k)
&\equiv
\left\langle h^{(\lambda)}(\mathbf q)
\zeta\!\left(\mathbf k-\frac{\mathbf q}{2}\right)
\zeta\!\left(-\mathbf k-\frac{\mathbf q}{2}\right)
\right\rangle'.
    \end{split}
\end{equation*}
Here $q=|\mathbf q|$, $k=|\mathbf k|$.
In addition to $q/k\ll1$, we require $c_{T*}q/(c_{S*}k)\ll1$, so that
the tensor is outside its propagation horizon while the scalar modes
interact. We define the dimensionless squeezed three-point coefficient
$\mathcal R_{h\zeta\zeta}$ by
\begin{align}
\label{Eps_large}
\frac{B_{h\zeta\zeta}^{\lambda}(\mathbf q,\mathbf k)}
{P_h(q)P_\zeta(k)}
=\mathcal E_\lambda(\hat{\mathbf q},\hat{\mathbf k})
\mathcal R_{h\zeta\zeta}(k)+o(1),
\qquad q/k\to0,
\end{align}
where $o(1)$ could contain terms such as
$\frac{q}{k}$, $\left(\frac{q}{k}\right)^2$, or $\frac{q}{k}\ln(q/k)$.
Both spectra in this definition are evaluated to the same
slow-variation order as the bispectrum. Defining the coefficient of
$\mathcal E_\lambda$, rather than dividing by it, also covers angular
configurations in which that factor vanishes. The quantity
$\mathcal R_{h\zeta\zeta}$ is distinct from the temporal projection
operator $\mathscr R$ used in the appendices.

Before evaluating the interaction integral, we determine the result
required by the adiabatic-mode argument. In the metric convention
$g_{ij}=a^2\mathrm e^{2\zeta}(\mathrm e^h)_{ij}$, a constant long tensor $h^L_{ij}$ is
removed locally by a homogeneous shear. The coordinate and momentum
transformations are
\begin{equation*}
    \begin{split}
        \widetilde x^i
=(\mathrm e^{h^L/2})^i{}_j x^j,
\quad
\widetilde k_i=(\mathrm e^{-h^L/2})_i{}^j k_j, \quad
\widetilde k^2
=k^2-h^L_{ij}k_i k_j+O((h^L)^2).
    \end{split}
\end{equation*}
The shear has unit determinant because $h^L_{ii}=0$. Neither the Fourier
measure nor its momentum-conserving delta function therefore supplies
an additional volume factor. The conditional short-mode spectrum is
$P_\zeta(\mathbf k\mid h^L)=P_\zeta(\widetilde k)$,
whose first variation is
\begin{align*}
\delta_{h^L}P_\zeta(\mathbf k)
=-h^L_{ij}k_i k_j
\frac{\partial P_\zeta(k)}{\partial k^2}
=-\frac12h^L_{ij}\hat k_i\hat k_j
\frac{dP_\zeta(k)}{d\ln k}.
\end{align*}
Contracting the long tensor with the external helicity mode gives the
mixed consistency relation of Ref. \cite{Maldacena:2002vr}, or,
equivalently, the anisotropic-scaling Ward identity of
Ref.\cite{Hinterbichler:2013dpa},
\begin{align}
        \lim_{q/k\to0}
\frac{B_{h\zeta\zeta}^{\lambda}(\mathbf q,\mathbf k)}
{P_h(q)}
&=-\frac12\mathcal E_\lambda
\frac{dP_\zeta(k)}{d\ln k},\\
\mathcal R_{h\zeta\zeta}(k)
&=-\frac12\frac{d\ln P_\zeta(k)}{d\ln k}
=\frac{3-[n_s(k)-1]}{2},
\\
n_s-1&\equiv\frac{d\ln\mathcal P_\zeta}{d\ln k}.
\label{eq:sg_tilt_identity}
\end{align}
The scale-invariant value is consequently $3/2$, not zero. Scale
invariance makes $\mathcal P_\zeta$ constant, whereas the dimensionful
Fourier spectrum still scales as $k^{-3}$; it is the latter that is
differentiated by the shear.

The quadratic calculation fixes the right-hand side independently of
any cubic time integral. At the individual scalar crossing time,
$c_S(t_k)k=a(t_k)H(t_k)$, the spectrum is
$\mathcal P_\zeta(k)=H_k^2/(8\pi^2F_{Sk}c_{Sk})
[1+\mathcal O(\varepsilon)]$, see \eqref{eq:modes_corrected_spectra}. The derivative of its first-order normalization
correction is second order under the slow-variation hierarchy. Since
\begin{equation*}
    \begin{split}
        \frac{d\ln k}{dN_k}
&=1-\epsilon_k-s_k,
\\
\frac{d}{dN_k}\ln\!\left(\frac{H_k^2}{F_{Sk}c_{Sk}}\right)
&=-2\epsilon_k-f_{Sk}-s_k,
\\
n_s-1
&=-2\epsilon_*-f_{S*}-s_*+\mathcal O(\varepsilon^2),
    \end{split}
\end{equation*}
we obtain the same first-order tilt as the general Horndeski
quadratic analysis of Ref. \cite{Kobayashi:2011nu}. The transfer of the
rates from the hard scalar crossing time to the common mixed pivot
changes this expression only at second order. The consistency
relation therefore predicts
\begin{align}
\mathcal R_{h\zeta\zeta}(k)
&=\frac32+\epsilon_*+\frac12f_{S*}+\frac12s_*
+\mathcal O(\varepsilon^2).
\label{eq:sg_symmetry_prediction}
\end{align}
Our task is to recover this coefficient directly from the NLO
bispectrum.

The reduction of the cubic action is especially simple in this limit.
Every scalar--scalar--tensor vertex other than the first contains a
time derivative or a Laplacian acting on the tensor. For the constant
growing tensor mode, $h_q(\tau)/h_q(0)=1+O(c_{T*}^2q^2\tau^2)$ and
$h_q'(\tau)/h_q(0)=O(c_{T*}^2q^2\tau)$ at leading order, with the same
soft-momentum suppression in the first-order corrections. The
remaining five vertices therefore give no contribution to the leading
soft coefficient. The surviving interaction is 
\begin{align*}
S_{c_1}^{(3)}
&=\int d\tau\,d^3x\,a^2F_S
h_{ij}\partial_i\zeta\partial_j\zeta,
\qquad c_1=F_S.
\end{align*}
see \eqref{eq:lagrangian_of_two_grav_and_one_scal} and the coefficients below this formula. It also follows by expanding the scalar gradient
term $-a^2F_S(\mathrm e^{-h})_{ij}\partial_i\zeta\partial_j\zeta$ to first
order in a constant tensor. The pivot value of this coefficient must
therefore be expanded consistently with the same $F_{S*}$ in the
quadratic normalization, rather than treated as an independent cubic
parameter.

To fix the normalization of the time-integral calculation, let
$p=c_{S*}k$. The leading propagator is
\begin{equation*}
        G_\zeta^{(0)}(k;0,\tau)
=P_\zeta^{(0)}(k)(1-ip\tau)\mathrm e^{ip\tau},
\end{equation*}
and the relevant two-point
normalizations are given by Eq.\eqref{P_lead} and by the leading limit of Eq.\eqref{eq:modes_corrected_spectra}.

The soft tensor propagator is independent of the integration time at
leading soft order. The spatial derivatives and the two scalar Wick
contractions give $2k^2\mathcal E_\lambda$. Taking the $q=1$
contribution of the general leading-order amplitude
\eqref{eq:zeta_zeta_h_leading_amplitude}, whose temporal dependence is
given by the first temporal function in Eq.\eqref{eq:app_zzh_T12}, and using
$r_1=r_2=\frac12$, $r_3\to0$, and $y=-2p\tau$,
we obtain
\begin{align}
B_{h\zeta\zeta}^{\lambda,(0)}
&=4\mathcal E_\lambda\frac{F_{S*}k^2}{H_*^2}
P_h^{(0)}(q)[P_\zeta^{(0)}(k)]^2
\operatorname{Re}\!\left[i\mathcal I(p,\tau_o)\right]
\bigg|_{\tau_o\to0^-}.
\label{eq:sg_leading_inin}
\end{align}
where $\tau_o<0$ is a finite observation endpoint. With the
Bunch--Davies prescription at early times, the integral is
\begin{equation}
        \mathcal I(p,\tau_o)
\equiv\int_{-\infty(1-i0)}^{\tau_o}
\frac{d\tau}{\tau^2}(1-ip\tau)^2\mathrm e^{2ip\tau}
 = -\frac1{\tau_o}-\frac{3ip}{2}+O(p^2\tau_o) \;\;
 \text{as}  \;\;\tau_o \rightarrow 0^{-}.
\label{eq:ss_vertex2_integral}
\end{equation}
The first term is real and does not contribute to
$\operatorname{Re}(i\mathcal I)$. Substitution into
Eq.\eqref{eq:sg_leading_inin} gives
\begin{align*}
B_{h\zeta\zeta}^{\lambda,(0)}
&=\frac32P_h^{(0)}(q)P_\zeta^{(0)}(k)
\mathcal E_\lambda.
\end{align*}
The leading consistency relation is therefore reproduced with the
same Fourier and helicity normalization as the full calculation.

We now retain the first-order corrections. All rates below are evaluated
at the common pivot, and their stars are suppressed. We now consider the
$q=1$ temporal function in Eq.\eqref{eq:app_zzh_T12} for  the tensor-soft
configuration. With the variables of
Eqs.\eqref{eq:modes_frequency_variables} and \eqref{eq:app_zzh_y},
\begin{equation}
\mathcal T_{1,\mathrm{soft}}^{(\zeta\zeta h)}(y)
\equiv
\left.\mathcal T_1^{(\zeta\zeta h)}(y)
\right|_{(r_1,r_2,r_3)=(1/2,1/2,0)}
=
\frac{(1+iy/2)^2}{y^2}.
\label{eq:sg_soft_temporal_function}
\end{equation}
The symbols $F$ and $\ell$ below denote the corresponding soft limits
of $F_1$ and $\ell_1$ in Eq.\eqref{eq:temporal_moments}. The common
factor $\mathrm e^{-iy}$ is included in the projection of
Eq.\eqref{eq:app_zzh_regulated_projection}. The limiting fractions are
used only in finite hard-mode integrals. For the external and soft bulk
corrections, $r_3$ is kept positive so that the two endpoints and the
soft power-spectrum normalization are combined before the soft limit is
taken.

The running insertion follows from the bare coefficient
$c_1(\tau)=F_S(\tau)$, whose logarithmic rate is
$\sigma_1^{(c)}=f_S$. Together with the scale-factor correction, this
produces
\begin{align}
\label{eq:sg_vertex_insertion}
        a^2(\tau)F_S(\tau)
&=a_0^2(\tau)F_{S*}
\left[1+2\epsilon-(2\epsilon+f_S)\ln y
+\mathcal O(\varepsilon^2)\right],\\
\Delta_{\mathrm{vert}} &=\mathscr R\left[\left(2 \epsilon-\left(2 \epsilon+f_S\right) \ln y\right) \mathcal{T}\right]
=2\epsilon F-(2\epsilon+f_S)\ell,
    \end{align}
where $\Delta_{\mathrm{vert}}$ is a correction from the time dependence of $a^2 F_S$, and notation and application of $\mathscr R$ is given by \eqref{eq:app_zzh_regulated_projection}. 
The constant correction to $a^2$ is important here. Retaining only its
logarithmic part would remove the first term of
Eq.\eqref{eq:sg_vertex_insertion}.

It remains to organize the corrections to the propagators. The scalar and tensor boundary-to-bulk propagators have the same
first-order structure \eqref{eq:complete_propagator_correction}. 
For a differentiated field in the interaction, $\partial_\tau$
acts only on the mode evaluated at the interaction time $\tau$.
Each corrected propagator already includes the corrections to both
the late-time external mode and the mode at the interaction time,
so no additional external-mode factor is added.

Also the soft tensor simplifies before integration. The small-argument
behavior of the auxiliary function is
$\mathfrak h^*(x)=-i(\gamma_2+\ln x)+O(x^2\ln x)$, see \eqref{eq:modes_hankel_order_function}.
Using the tensor expression in Eq.\eqref{eq:app_zzh_delta_P}
with $x_3=r_3y$, we find
\begin{align*}
\delta P_T(r_3y)
=-(\epsilon+s_T)+\delta_T\ln y
-\delta_T[\gamma_2+\ln(r_3y)]+o(1)=E_T(q)+o(1).
\end{align*}
Here $E_T(q)$ denotes $E_{T,3}$ at small positive $q$. The time-dependent
logarithms cancel, as they must for the constant long mode. Its bulk
insertion is consequently $\Delta_{\mathrm{soft}}^{(P)}=E_T(q)F$.
Together with the external tensor endpoint, this gives $2E_T(q)$,
precisely the relative correction to the soft tensor two-point
function. In the same limit the two scalar external factors coincide;
we denote either one by $E_S(k)$, so that
$E_{\mathrm{ext}}=2E_S(k)+E_T(q)$.

The only first-order mode corrections that remain to be evaluated are
those on the two hard scalar lines of the $q=1$ interaction. According
to Eq.\eqref{eq:app_zzh_T12}, both scalar lines are undifferentiated.
Therefore, for either hard scalar line, the scalar case of
Eq.\eqref{eq:app_zzh_DeltaP_general} gives
\begin{equation}
\Delta_{\mathrm{hard}}^{(P)}
=
-\epsilon F
+s\left[
-F+X_{\mathrm{hard}}^{(P)}
-Y_{\mathrm{hard}}^{(P)}
\right]
+\delta_S\left[
\ell+\mathfrak H_{\mathrm{hard}}^{(P)}
\right].
\label{eq:sg_hard_insertion_structure}
\end{equation}
Here
\begin{align*}
&F\equiv\left.F_1\right|_{\mathrm{soft}},
\qquad
\ell\equiv\left.\ell_1\right|_{\mathrm{soft}},
\\
&X_{\mathrm{hard}}^{(P)}
\equiv
\left.X_{11}^{(P)}\right|_{\mathrm{soft}}
=
\left.X_{12}^{(P)}\right|_{\mathrm{soft}},
\qquad
Y_{\mathrm{hard}}^{(P)}
\equiv
\left.Y_{11}^{(P)}\right|_{\mathrm{soft}}
=
\left.Y_{12}^{(P)}\right|_{\mathrm{soft}}.
\end{align*}
The leading and logarithmic projections $F_q$ and $\ell_q$ are defined
in Eq.\eqref{eq:temporal_moments}. The functions
$X_{qi}^{(P)}$ and $Y_{qi}^{(P)}$ are defined in
Eq.\eqref{eq:app_zzh_XY_P_definition}. They arise from the terms
$+s x_i^2$ and $-s x_i^2\ln y$, respectively, in the scalar-mode
correction of Eq.\eqref{eq:app_zzh_delta_P_scalar}. The function
$\mathfrak H_{qi}^{(P)}$, defined in
Eq.\eqref{eq:app_zzh_HP_definition}, contains the contribution
generated by the shift of the Hankel index.

For the tensor-soft configuration,
\begin{equation}
r_1=r_2=\frac12,
\qquad
r_3\to0,
\qquad
S_2=\frac14,
\qquad
S_3=0.
\end{equation}
Equation\eqref{eq:app_zzh_Fell1} therefore gives
\begin{equation}
F=\frac34,
\qquad
\ell=1-\frac34\gamma_E.
\end{equation}
For the first interaction, the general hard-line expressions are
Eqs.\eqref{eq:app_zzh_XP1i} and
\eqref{eq:app_zzh_YP1i}. Evaluating them for either hard scalar line
gives
\begin{align}
X_{\mathrm{hard}}^{(P)}
&=
\left.
r_i^2
\left[
1+r_j+r_3+2r_jr_3
\right]
\right|_{r_i=r_j=1/2,\,r_3=0}
=
\frac38,
\nonumber\\
Y_{\mathrm{hard}}^{(P)}
&=
\left.
\left\{
-\gamma_E X_{1i}^{(P)}
+r_i^2
\left[
r_j+r_3+3r_jr_3
\right]
\right\}
\right|_{r_i=r_j=1/2,\,r_3=0}
=
\frac18-\frac38\gamma_E.
\label{eq:sg_polynomial_integrals}
\end{align}
The master integral in
Eq.\eqref{eq:regulated_moment_master} is the tool used to evaluate
these projections; the quantities above are not independent
corrections. They are the polynomial terms that will be substituted
into Eq.\eqref{eq:sg_hard_insertion_structure}. The only remaining
term in that equation is
$\mathfrak H_{\mathrm{hard}}^{(P)}$, which is evaluated next.

The remaining Hankel-order contribution is evaluated with
\begin{align}
\mathfrak h^*(x)
&=\left[2i-\frac{\pi}{2}(1+ix)\right]\mathrm e^{-ix}
+i(1-ix)\mathrm e^{ix}E_1(2ix).
\label{eq:nlf_hankel_E1}
\end{align}
Here $E_1(z)\equiv\int_z^\infty dt\,\mathrm e^{-t}/t$ is the exponential
integral in the principal-value convention
\cite{NISTDLMF} (Eq.6.2.1 therein); it is unrelated to the external-mode
corrections $E_S$ and $E_T$. Using the common prescription in
Eq.\eqref{eq:app_zzh_regulated_projection}, one finally obtains
\begin{equation}
\mathfrak H_{\mathrm{hard}}^{(P)}
=
\operatorname{Re}\operatorname{FP}_{\alpha=0}
\int_0^\infty dy\,
y^{\alpha-2}(1+iy/2)\mathrm e^{-iy/2}\mathfrak h^*(y/2)
=\frac14.
\label{eq:sg_hankel_value}
\end{equation}
The pole of the intermediate gamma function cancels in the complete
combination before the finite part is extracted.

We can now combine the hard-mode terms. The undifferentiated-leg
insertion of Eq.\eqref{eq:app_zzh_DeltaP_general} gives
\begin{align}
\Delta_{\mathrm{hard}}^{(P)}
&=-(\epsilon+s)F+sX_{\mathrm{hard}}^{(P)}
+\delta_S\ell-sY_{\mathrm{hard}}^{(P)}
+\delta_S\mathfrak H_{\mathrm{hard}}^{(P)}
\nonumber\\
&=-\frac34\epsilon
+\left(-\frac12+\frac38\gamma_E\right)s
+\left(\frac54-\frac34\gamma_E\right)\delta_S.
\label{eq:sg_hard_insertion}
\end{align}
Adding the two hard scalar insertions to the coefficient and
scale-factor contribution gives
\begin{align*}
\Delta_{\mathrm{vert}}+2\Delta_{\mathrm{hard}}^{(P)}
&=-2\epsilon-f_S-s+\frac52\delta_S
+\frac{3\gamma_E}{4}
\left(2\epsilon+f_S+s-2\delta_S\right)=\frac12\delta_S,
\end{align*}
where $\Delta_{\mathrm{vert}}$ is given under \eqref{eq:sg_vertex_insertion}.
The cancellation of the Euler--Mascheroni constant follows from the
same index shift $\delta_S=\epsilon+f_S/2+s/2$ that enters the
quadratic spectrum. In particular, omitting the Hankel-order term would
spoil this equality. Neither the corrected vertex nor an individual
mode insertion has the required universal form on its own.

Finally, the first-vertex assembly in
Eq.\eqref{eq:app_zzh_delta_J_general} has
$J_*^{(1)}=-\Lambda_1F$. At leading soft order its relative correction
is therefore
\begin{align*}
\frac{\delta J_1}{J_*^{(1)}}
&=E_{\mathrm{ext}}
+\frac{\Delta_{\mathrm{vert}}
+2\Delta_{\mathrm{hard}}^{(P)}
+\Delta_{\mathrm{soft}}^{(P)}}{F}
=2E_S(k)+2E_T(q)+\frac23\delta_S.
\end{align*}
Restoring the common leading normalization, the leading soft part of
the bispectrum is thus
\begin{align*}
B_{h\zeta\zeta}^{\lambda}\big|_{\mathrm{leading\ soft}}
&=\frac32P_h^{(0)}(q)P_\zeta^{(0)}(k)
\mathcal E_\lambda
\left[1+2E_T(q)+2E_S(k)+\frac23\delta_S
+\mathcal O(\varepsilon^2)\right].
\end{align*}
The external corrections are real after phase cancellation, so the
corrected product of spectra is
\begin{align*}
P_h(q)P_\zeta(k)
&=P_h^{(0)}(q)P_\zeta^{(0)}(k)
\left[1+2E_T(q)+2E_S(k)+\mathcal O(\varepsilon^2)\right].
\end{align*}
Dividing the two expressions and expanding the ratio, rather than
normalizing the bispectrum by leading spectra alone, gives
\begin{align}
\mathcal R_{h\zeta\zeta}(k)
&=\frac32+\delta_{S*}+\mathcal O(\varepsilon^2)
=\frac32+\epsilon_*+\frac12f_{S*}+\frac12s_*
+\mathcal O(\varepsilon^2).
\label{eq:sg_final_result}
\end{align}
This agrees with the independent prediction
\eqref{eq:sg_symmetry_prediction}. All dependence on the soft tensor
normalization, including its first-order $f_T$, $s_T$, and
$\ln r_3$ terms, cancels against the soft tensor spectrum. The scalar
normalization corrections cancel similarly, leaving only the scalar
spectral slope.

The result also has a useful interpretation in terms of the Hankel
index. For constant slow-variation rates on the growing-mode solution,
$P_\zeta(k)\propto k^{-2\nu_S}$, with
$\nu_S=1/2+(1+f_S/2-s/2)/(1-\epsilon-s)$
\cite{Kobayashi:2011nu}. Equation\eqref{eq:sg_tilt_identity} then gives
$\mathcal R_{h\zeta\zeta}=\nu_S$, whose first-order expansion is
Eq.\eqref{eq:sg_final_result}. In canonical inflation,
$F_S=\epsilon M_{\mathrm{pl}}^2$ and $c_S=1$, this reduces to
$\mathcal R_{h\zeta\zeta}=3/2+\epsilon+\eta_H/2+\mathcal O(\varepsilon^2)$,
where $\eta_H=d\ln\epsilon/dN$. Exact scale invariance instead requires
$2\epsilon+f_S+s=0$ at first order, and returns the value $3/2$
without requiring the individual rates to vanish.

The soft limit must be interpreted within the perturbative expansion.
At finite squeezing, the unresummed mode formulas require the
logarithmic terms, in particular $|\delta_T\ln r_3|$, to remain
controlled. The leading soft coefficient is extracted only after the
complete first-order numerator and denominator have been combined;
it is not obtained by setting $r_3=0$ inside individual logarithmic
insertions. The calculation verifies the known tensor-soft consistency
relation \cite{Maldacena:2002vr,Hinterbichler:2013dpa}, rather than
establishing a new soft theorem. Its value here is that the explicitly
evaluated NLO integrals reproduce an answer fixed independently by the
two-point function. This provides a nontrivial check of the normalized
mixed bispectrum, while leaving its momentum dependence away from the
squeezed limit unconstrained by this particular identity.


\subsection{Soft scalar in \texorpdfstring{$\zeta hh$}{zeta hh}}

The soft-scalar limit provides an independent check of the mixed bispectrum
$\langle \zeta h h\rangle$ beyond leading order. In contrast to the
soft-graviton limit, where only a single cubic vertex survives at leading
order in the soft momentum, here several interactions can in principle
contribute at the same order, while the required consistency relation
emerges only after their sum is taken. Therefore, the
scalar-soft limit provides an especially sensitive check of the
relative normalization of different cubic operators and their first-order
corrections.

As in the previous analysis, we work at tree level in a stable
expanding attractor regime, assume the Bunch--Davies initial state
and regular background coefficients. The long-wavelength scalar
mode is treated as a constant adiabatic mode, while the hard
tensor modes are evaluated after their decaying solutions have become negligible.
Field redefinitions and boundary terms are understood under the same assumptions
as in the derivation of the reduced cubic action; in the
attractor regime\footnote{The attractor nature of the contracting solution considered here was established in Ref. \cite{Ageeva2026att}.}  under consideration the corresponding late-time corrections vanish.

First, we use the dimensional spectra defined in \eqref{eq:sg_spectra}. Next, we take the scalar momentum to be soft while maintaining exact momentum
conservation,
\begin{equation*}
    		\mathbf k_1 = \mathbf q,\quad
		\mathbf k_2 = \mathbf k-\frac{\mathbf q}{2},\quad
		\mathbf k_3 = -\mathbf k-\frac{\mathbf q}{2}.
\end{equation*}
We define
\begin{align*}
	B_{\zeta hh}^{\lambda\lambda'}(\mathbf q,\mathbf k)
	\equiv
	\left\langle
	\zeta(\mathbf q)
	h^{(\lambda)}
	\left(\mathbf k-\frac{\mathbf q}{2}\right)
	h^{(\lambda')}
	\left(-\mathbf k-\frac{\mathbf q}{2}\right)
	\right\rangle'.
\end{align*}
In addition to $q/k\ll1$, we require
\begin{align*}
	\frac{c_{S*}q}{c_{T*}k}\ll1,
\end{align*}
so that the long-wavelength scalar mode is already outside its
propagation horizon when the hard tensor modes interact.
We again use the normalized coefficient
\begin{align}
	\frac{
		B_{\zeta hh}^{\lambda\lambda'}(\mathbf q,\mathbf k)
	}{
		P_\zeta(q)P_h(k)
	}
	=
	\delta_{\lambda\lambda'}
	\mathcal R_{\zeta hh}(k)+o(1),
	\qquad
	\frac{q}{k}\to0.
	\label{eq:ss_R_definition}
\end{align}
The leading soft structure is isotropic; therefore, unlike the tensor-soft
limit \eqref{Eps_large}, there is no angular factor of the type $\mathcal E_\lambda$.

The constant long-wavelength scalar mode $\zeta_L$ in the parametrization $g_{ij}=a^2\mathrm e^{2\zeta}(\mathrm e^h)_{ij}$
is locally removed by an isotropic dilation
\begin{align*}
	\widetilde x^i=\mathrm e^{\zeta_L}x^i,
	\qquad
	\widetilde k_i=\mathrm e^{-\zeta_L}k_i.
\end{align*}
In contrast to the tensor shear, the Jacobian of this transformation is not
equal to unity. Therefore, for the dimensionful tensor spectrum we have
\begin{align*}
	P_h(\mathbf k\mid\zeta_L)
	=
	\mathrm e^{-3\zeta_L}
	P_h(\mathrm e^{-\zeta_L}k).
\end{align*}
To first order in the long mode
\begin{align*}
	\delta_{\zeta_L}P_h(k)
	=
	-\zeta_L
	\left(
	3+\frac{d}{d\ln k}
	\right)
	P_h(k).
\end{align*}
Contracting the long mode with the external scalar gives the scalar-soft
consistency relation
\begin{align*}
	\lim_{q/k\to0}
	\frac{
		B_{\zeta hh}^{\lambda\lambda'}(\mathbf q,\mathbf k)
	}{
		P_\zeta(q)
	}
	=
	-\delta_{\lambda\lambda'}
	\left(
	3+\frac{d}{d\ln k}
	\right)
	P_h(k),
\end{align*}
or, equivalently,
\begin{equation}
\mathcal R_{\zeta hh}(k)
=
-\frac{d\ln\mathcal P_{h,\mathrm{one\,helicity}}(k)}
{d\ln k}
=
-n_T(k).
\label{eq:ss_soft_theorem_prediction}
\end{equation}
In particular, for an exactly scale-invariant tensor spectrum the leading
scalar-soft coefficient vanishes.

The reduced cubic action for one scalar and two tensor modes
contains seven operators. In particular, the first two vertices have the form
\begin{align}
	S_{\zeta hh}^{(3)}
	\supset
	\int d\tau\,d^3x\,
	a^2
	\left[
	b_1\zeta h'_{ij}h'_{ij}
	+
	b_2\zeta\partial_kh_{ij}\partial_kh_{ij}
	\right].
\label{eq:ss_soft_action}
\end{align}
The remaining five operators contain $\Psi=\partial^{-2}\zeta'$,
$\zeta'$, $\partial^2\zeta$, or $\partial_i\partial_j\zeta$.
For the constant growing soft scalar mode
\begin{equation*}
    		\frac{\zeta_q(\tau)}{\zeta_q(0)}
		=
		1+\mathcal O(c_{S*}^2q^2\tau^2),\qquad
		\frac{\zeta_q'(\tau)}{\zeta_q(0)}
		=
		\mathcal O(c_{S*}^2q^2\tau).
\end{equation*}
Consequently,
\begin{align*}
	\partial_i\Psi_q=\mathcal O(q\tau),
	\qquad
	\partial_i\partial_j\Psi_q=\mathcal O(q^2\tau),
	\qquad
	\partial^2\zeta_q=\mathcal O(q^2).
\end{align*}
Therefore, the first and second vertices contribute at order $O(q^0)$, while the third
is suppressed as $O(q)$ and the vertices $b_4,\ldots,b_7$ -- at least as
$O(q^2)$. The same power suppression holds for the first-order corrections
to the growing soft mode. Thus, the leading scalar-soft coefficient is obtained from the two
interactions in Eq.\eqref{eq:ss_soft_action}; the other five interactions
are suppressed by at least one power of the soft momentum.

We set
$
p=c_{T*}k$ for the simplicity.
At leading order we again have
\begin{equation*}
    \begin{split}
        		G_\zeta^{(0)}(q;0,\tau)
		&=
		P_\zeta^{(0)}(q)
		+\mathcal O(q^2\tau^2),
		\\
		G_h^{(0)}(k;0,\tau)
		&=
		P_h^{(0)}(k)
		(1-ip\tau)\mathrm \mathrm e^{ip\tau},
		\\
		\partial_\tau G_h^{(0)}(k;0,\tau)
		&=
		P_h^{(0)}(k)
		p^2\tau \mathrm e^{ip\tau},
    \end{split}
\end{equation*}
where $P_h^{(0)}(k)
	=
	\frac{2H_*^2}
	{F_{T*}c_{T*}k^3}$.
For the first vertex in \eqref{eq:ss_soft_action}, the time dependence reduces to
\begin{align*}
	a_0^2
	\partial_\tau G_h^{(0)}
	\partial_\tau G_h^{(0)}
	\propto \mathrm e^{2ip\tau}.
\end{align*}
Taking into account the two tensor Wick contractions, we obtain
\begin{align*}
	B_1^{\lambda\lambda',(0)}
	=
	4b_{1*}^{\mathrm{lead}}
	\delta_{\lambda\lambda'}
	\frac{p^4}{H_*^2}
	P_\zeta^{(0)}(q)
	[P_h^{(0)}(k)]^2
	\operatorname{Re}
	\left[
	i\int_{-\infty(1-i0)}^0d\tau\,
	\mathrm e^{2ip\tau}
	\right],
\end{align*}
what equals to
\begin{align*}
	B_1^{\lambda\lambda',(0)}
	=
	\frac{4b_{1*}^{\mathrm{lead}}}{G_{T*}}
	\delta_{\lambda\lambda'}
	P_\zeta^{(0)}(q)
	P_h^{(0)}(k),
\end{align*}
and so this contribution to \eqref{eq:ss_R_definition} is
\begin{align*}
	\mathcal R_1^{(0)}
	=
	\frac{4b_1^{\mathrm{lead}}}{G_T}.
\end{align*}

For the second interaction in \eqref{eq:ss_soft_action}, the temporal integral is
Eq.\eqref{eq:ss_vertex2_integral}, now with $p=c_{T*}k$. Its projected
late-time value is $3p/2$.
Consequently,
\begin{align*}
	B_2^{\lambda\lambda',(0)}
	=
	\frac{12b_{2*}^{\mathrm{lead}}}{F_{T*}}
	\delta_{\lambda\lambda'}
	P_\zeta^{(0)}(q)
	P_h^{(0)}(k),
\end{align*}
and
\begin{align*}
	\mathcal R_2^{(0)}
	=
	\frac{12b_2^{\mathrm{lead}}}{F_T}.
\end{align*}

Using the exact expression for $F_S$ in
Eq.\eqref{eq:ss_FS_identity}, its leading slow-variation form is
\begin{align*}
F_S
=
\frac{HG_T^2}{\Theta}
-
F_T
+
\mathcal O(\varepsilon).
\end{align*}
Therefore,
\begin{equation}\label{eq:ss_b_leading}
    		b_1^{\mathrm{lead}}
		=
		-\frac{3G_TF_S}{8F_T},\qquad
		b_2^{\mathrm{lead}}
		=
		\frac{F_S}{8}.
\end{equation}
Hence
\begin{equation}
    		\mathcal R_1^{(0)}
	=
		-\frac32\frac{F_S}{F_T},
		\qquad
		\mathcal R_2^{(0)}
		=
		+\frac32\frac{F_S}{F_T}.
\label{eq:ss_leading_vertex_responses}
\end{equation}
and, consequently,
\begin{align}
		\mathcal R_{\zeta hh}^{(0)}
		=
		\mathcal R_1^{(0)}
		+
		\mathcal R_2^{(0)}
		=
		0.
	\label{eq:ss_leading_cancellation}
\end{align}
Thus, the zeroth-order limit of the scalar-soft consistency relation
derived in Ref. \cite{Maldacena:2002vr} is reproduced within the
general leading-order mixed bispectrum of Ref. \cite{Gao:2012ib}
through the exact cancellation between the time-derivative and
spatial-gradient interactions.

We now evaluate the first-order correction to the scalar-soft response
defined in Eq.\eqref{eq:ss_R_definition}. All slow-variation
quantities in this calculation are evaluated at the common pivot, and
their stars are suppressed. The labels $1$ and $2$ below refer to the
two interactions proportional to $b_1$ and $b_2$, while $q$ continues
to denote the soft scalar momentum.
We use the variables of
Eqs.\eqref{eq:modes_common_pivot},
\eqref{eq:modes_frequency_variables}, and
\eqref{eq:app_zzh_y}, with
\begin{equation}
(X_1,X_2,X_3)=(S,T,T).
\end{equation}
In the scalar-soft limit,
\begin{equation}
K_*\longrightarrow 2c_{T*}k,
\qquad
(r_1,r_2,r_3)
\longrightarrow
\left(0,\frac12,\frac12\right).
\label{eq:ss_soft_frequency_fractions}
\end{equation}
The limit $r_1\to0$ is taken only after the complete first-order
contribution has been assembled.
The organization of the calculation follows directly from
Eq.\eqref{eq:app_zhh_delta_I_explicit}. According to
Eq.\eqref{eq:app_zhh_line_assignments}, interaction $1$ contains an
undifferentiated scalar line and two differentiated tensor lines,
whereas interaction $2$ contains three undifferentiated lines:
\begin{equation}
(d_{11},d_{12},d_{13})=(P,D,D),
\qquad
(d_{21},d_{22},d_{23})=(P,P,P),
\qquad
p_1=p_2=2.
\label{eq:ss_two_interaction_assignments}
\end{equation}
Using
$I_*^{(a)}=\eta_a\Lambda_aF_a$ from
Eq.\eqref{eq:app_zhh_I_eta_Lambda}, the relative corrections to the
two temporal functions are therefore
\begin{align}
\frac{\delta I_1}{I_*^{(1)}}
&=
E_{\mathrm{ext}}^{(\zeta hh)}
+
\frac{
\Delta_1^{[b+a]}
+\Delta_{11}^{(P)}
+\Delta_{12}^{(D)}
+\Delta_{13}^{(D)}
}{F_1},
\nonumber\\
\frac{\delta I_2}{I_*^{(2)}}
&=
E_{\mathrm{ext}}^{(\zeta hh)}
+
\frac{
\Delta_2^{[b+a]}
+\Delta_{21}^{(P)}
+\Delta_{22}^{(P)}
+\Delta_{23}^{(P)}
}{F_2}.
\label{eq:ss_relative_I_corrections}
\end{align}
Equation\eqref{eq:ss_relative_I_corrections} specifies where every
quantity evaluated below will be substituted.

The temporal functions of the two surviving interactions are given in
Eq.\eqref{eq:app_zhh_T12}. Using
$P_i=1+ir_i y$ and $D_i=r_i y$ from
Eqs.\eqref{eq:app_zzh_Pi} and \eqref{eq:app_zzh_Di}, their
scalar-soft limits are
\begin{align}
\left.
\mathcal T_1^{(\zeta hh)}
\right|_{\mathrm{soft}}
&=
\left.
\frac{P_1D_2D_3}{y^2}
\right|_{\mathrm{soft}}
=
\frac14,
\nonumber\\
\left.
\mathcal T_2^{(\zeta hh)}
\right|_{\mathrm{soft}}
&=
\left.
\frac{P_1P_2P_3}{y^2}
\right|_{\mathrm{soft}}
=
\frac{(1+iy/2)^2}{y^2}
=
\mathcal T_{1,\mathrm{soft}}^{(\zeta\zeta h)}.
\label{eq:ss_soft_temporal_functions}
\end{align}
The corresponding leading and logarithmic moments follow from
Eqs.\eqref{eq:app_zhh_Fell1} and
\eqref{eq:app_zhh_Fell2}:
\begin{equation}
F_1=\frac14,
\qquad
\ell_1=-\frac{\gamma_E}{4},
\qquad
F_2=\frac34,
\qquad
\ell_2=1-\frac{3\gamma_E}{4}.
\label{eq:ss_soft_moments}
\end{equation}

For both interactions the soft scalar line is undifferentiated.
Using the scalar correction in
Eq.\eqref{eq:app_zzh_delta_P_scalar}, together with the
small-argument behavior following from
Eq.\eqref{eq:nlf_hankel_E1}, we find at small positive $r_1$
\begin{align}
\delta P_S(r_1y)
&=
-(\epsilon+s)
+\delta_S\ln y
-\delta_S\left(\gamma_2+\ln(r_1y)\right)
+o(1)
\nonumber\\
&=
-(\epsilon+s)
-\delta_S\left(\gamma_2+\ln r_1\right)
+o(1)
=
E_S(q)+o(1),
\label{eq:ss_soft_scalar_mode}
\end{align}
where the last equality uses Eq.\eqref{eq:gamma_2}.
The scalar case of Eq.\eqref{eq:app_zzh_DeltaP_general} consequently
gives
\begin{equation}
\Delta_{11}^{(P)}=E_S(q)F_1,
\qquad
\Delta_{21}^{(P)}=E_S(q)F_2.
\label{eq:ss_soft_scalar_insertions}
\end{equation}
The external-mode contribution in
Eq.\eqref{eq:app_zhh_Eext} becomes
\begin{equation}
E_{\mathrm{ext}}^{(\zeta hh)}
=
E_S(q)+2E_T(k).
\end{equation}
After adding the correction to the soft scalar mode at the interaction
time, both interactions contain the same relative normalization factor,
\begin{equation}
E_{\mathrm{norm}}
\equiv
2E_S(q)+2E_T(k).
\label{eq:ss_common_normalization}
\end{equation}
Equation\eqref{eq:ss_relative_I_corrections} can therefore be written
as
\begin{align}
\frac{\delta I_1}{I_*^{(1)}}
&=
E_{\mathrm{norm}}+\kappa_1,
&
\kappa_1
&\equiv
\frac{
\Delta_1^{[b+a]}
+2\Delta_{\mathrm{hard},1}^{(D)}
}{F_1},
\nonumber\\
\frac{\delta I_2}{I_*^{(2)}}
&=
E_{\mathrm{norm}}+\kappa_2,
&
\kappa_2
&\equiv
\frac{
\Delta_2^{[b+a]}
+2\Delta_{\mathrm{hard},2}^{(P)}
}{F_2}.
\label{eq:ss_kappa_definition}
\end{align}
Here the factor of two counts the two identical hard tensor lines.

The common term does not contribute to the total response, because
Eqs.\eqref{eq:ss_leading_vertex_responses} and
\eqref{eq:ss_leading_cancellation} imply
\begin{equation}
E_{\mathrm{norm}}
\left(
\mathcal R_1^{(0)}+\mathcal R_2^{(0)}
\right)
=0.
\label{eq:ss_common_normalization_cancellation}
\end{equation}
Thus the potentially large terms proportional to
$\delta_S\ln r_1$ cancel before the scalar-soft limit is taken.
It remains to evaluate $\kappa_1$ and $\kappa_2$ from \eqref{eq:ss_kappa_definition}.

For interaction $1$, both hard tensor lines are differentiated.
The correspondence
\begin{equation}
\mathcal T_1^{(\zeta hh)}
=
\mathcal T_3^{(\zeta\zeta h)}
\end{equation}
is part of Eq.\eqref{eq:app_zhh_temporal_correspondence}. Therefore,
the required polynomial responses follow from
Eq.\eqref{eq:app_zzh_D_responses_q3}. For either hard tensor line,
\begin{equation}
X_{\mathrm{hard},1}^{(D)}
=-\frac18,
\qquad
Y_{\mathrm{hard},1}^{(D)}
=1-\frac{\gamma_E}{2},
\qquad
Z_{\mathrm{hard},1}^{(D)}
=\frac{1-\gamma_E}{8}.
\label{eq:ss_D_polynomial_responses}
\end{equation}
The tensor case of Eq.\eqref{eq:app_zzh_DeltaD_general} gives
\begin{align}
\Delta_{\mathrm{hard},1}^{(D)}
={}&
-\epsilon F_1
+s_T
\left[
X_{\mathrm{hard},1}^{(D)}
-2\ell_1
+Z_{\mathrm{hard},1}^{(D)}
\right]
\nonumber\\
&+
\delta_T
\left[
Y_{\mathrm{hard},1}^{(D)}
+\ell_1
+\mathfrak H_{\mathrm{hard},1}^{(D)}
\right].
\label{eq:ss_D_insertion_structure}
\end{align}
This equation shows explicitly where the quantities in
Eq.\eqref{eq:ss_D_polynomial_responses} are used.
The last square bracket in
Eq.\eqref{eq:ss_D_insertion_structure} must be evaluated as one
complete expression. The definition of
$\mathfrak H^{(D)}$ is Eq.\eqref{eq:app_zzh_HD_definition_1}.
Differentiating the representation in
Eq.\eqref{eq:nlf_hankel_E1} gives, for $x=y/2$,
\begin{equation}
\left(\frac1x+i\right)
+x\ln y
-i\mathrm e^{ix}\mathfrak h^{*\prime}(x)
=
x\left[
\ln y+\frac{i\pi}{2}
+\mathrm e^{iy}E_1(iy)
\right].
\label{eq:ss_D_regular_combination}
\end{equation}
The apparent $1/x$ terms cancel before the projection is performed.
Applying the common projection of
Eq.\eqref{eq:app_zzh_regulated_projection}, with the elementary
moments from Eq.\eqref{eq:regulated_moment_master}, gives
\begin{equation}
Y_{\mathrm{hard},1}^{(D)}
+\ell_1
+\mathfrak H_{\mathrm{hard},1}^{(D)}
=
\frac{1-\gamma_E}{4}.
\label{eq:ss_D_complete_response}
\end{equation}
The coefficient multiplying \(s_T\) is
\begin{equation}
X_{\mathrm{hard},1}^{(D)}
-2\ell_1
+Z_{\mathrm{hard},1}^{(D)}
=
\frac{3\gamma_E}{8}.
\end{equation}
Substitution into
Eq.\eqref{eq:ss_D_insertion_structure} therefore yields the
correction associated with either hard tensor line,
\begin{equation}
\Delta_{\mathrm{hard},1}^{(D)}
=
-\frac14\epsilon
+\frac{3\gamma_E}{8}s_T
+\frac{1-\gamma_E}{4}\delta_T.
\label{eq:ss_hard_D_insertion}
\end{equation}

For interaction $2$, both hard tensor lines are undifferentiated.
Equation\eqref{eq:app_zhh_temporal_correspondence} gives
\begin{equation}
\mathcal T_2^{(\zeta hh)}
=
\mathcal T_1^{(\zeta\zeta h)}.
\end{equation}
The required polynomial responses are therefore
Eqs.\eqref{eq:app_zzh_XP1i} and
\eqref{eq:app_zzh_YP1i}, evaluated at
$(r_1,r_2,r_3)=(0,1/2,1/2)$ with either hard tensor line selected.
Together with the Hankel-order result
Eq.\eqref{eq:sg_hankel_value}, they give
\begin{equation}
X_{\mathrm{hard},2}^{(P)}
=\frac38,
\qquad
Y_{\mathrm{hard},2}^{(P)}
=\frac18-\frac{3\gamma_E}{8},
\qquad
\mathfrak H_{\mathrm{hard},2}^{(P)}
=\frac14.
\label{eq:ss_P_responses}
\end{equation}
The tensor case of Eq.\eqref{eq:app_zzh_DeltaP_general} now reads
\begin{align}
\Delta_{\mathrm{hard},2}^{(P)}
={}&
-\epsilon F_2
+s_T
\left[
-F_2
+X_{\mathrm{hard},2}^{(P)}
-Y_{\mathrm{hard},2}^{(P)}
\right]
\nonumber\\
&+
\delta_T
\left[
\ell_2
+\mathfrak H_{\mathrm{hard},2}^{(P)}
\right].
\label{eq:ss_P_insertion_structure}
\end{align}
Using Eqs.\eqref{eq:ss_soft_moments} and
\eqref{eq:ss_P_responses}, we obtain
\begin{equation}
\Delta_{\mathrm{hard},2}^{(P)}
=
-\frac34\epsilon
+
\left(
-\frac12+\frac{3\gamma_E}{8}
\right)s_T
+
\left(
\frac54-\frac{3\gamma_E}{4}
\right)\delta_T.
\label{eq:ss_hard_P_insertion}
\end{equation}
Equivalently, this is the tensor specialization of
Eq.\eqref{eq:sg_hard_insertion}.

The correction generated by the time dependence of \(b_a\) and by
the two powers of the scale factor is
Eq.\eqref{eq:app_zhh_Delta_ba}:
\begin{equation}
\Delta_a^{[b+a]}
=
2\epsilon F_a
-
\left(
\sigma_a^{(b)}+2\epsilon
\right)\ell_a,
\qquad
a=1,2.
\end{equation}
Using Eq.\eqref{eq:ss_soft_moments}, this gives
\begin{align}
\Delta_1^{[b+a]}
&=
\frac{\epsilon}{2}
+
\frac{\gamma_E}{4}
\left(
\sigma_1^{(b)}+2\epsilon
\right),
\nonumber\\
\Delta_2^{[b+a]}
&=
\frac{3\epsilon}{2}
-
\left(
\sigma_2^{(b)}+2\epsilon
\right)
\left(
1-\frac{3\gamma_E}{4}
\right).
\label{eq:ss_ba_corrections}
\end{align}
The leading coefficients in Eq.\eqref{eq:ss_b_leading} imply
\begin{equation}
\sigma_1^{(b)}
=
g_T+f_S-f_T,
\qquad
\sigma_2^{(b)}
=
f_S.
\end{equation}
Using
$g_T=f_T-2s_T$ from Eq.\eqref{eq:tensor_slow_rates}, we obtain
\begin{equation}
\sigma_1^{(b)}
=
f_S-2s_T,
\qquad
\sigma_2^{(b)}
=
f_S.
\label{eq:ss_sigma_b}
\end{equation}
Substituting Eqs.\eqref{eq:ss_hard_D_insertion},
\eqref{eq:ss_hard_P_insertion}, and
\eqref{eq:ss_ba_corrections} into
Eq.\eqref{eq:ss_kappa_definition}, and using
$\delta_T=\epsilon+f_T/2+s_T/2$ from
Eq.\eqref{eq:tensor_hankel_index}, gives
\begin{align}
\kappa_1
&=
2\epsilon
+(1-\gamma_E)f_T
+(1+2\gamma_E)s_T
+\gamma_E\sigma_1^{(b)},
\nonumber\\
\kappa_2
&=
\frac23\epsilon
+
\left(
\frac53-\gamma_E
\right)f_T
+
\frac13s_T
+
\left(
\gamma_E-\frac43
\right)\sigma_2^{(b)}.
\label{eq:ss_kappa_explicit}
\end{align}
After substituting Eq.\eqref{eq:ss_sigma_b}, all dependence on
\(\gamma_E\) cancels:
\begin{equation}
\kappa_1-\kappa_2
=
\frac23
\left(
2\epsilon-f_T+s_T+2f_S
\right).
\label{eq:ss_kappa_difference}
\end{equation}
The contribution of the two corrected temporal integrals to the
normalized scalar-soft response is
\begin{equation}
\Delta\mathcal R_{\mathrm{time}}
=
\mathcal R_1^{(0)}\kappa_1
+
\mathcal R_2^{(0)}\kappa_2.
\end{equation}
Since
$\mathcal R_2^{(0)}=-\mathcal R_1^{(0)}$ by
Eq.\eqref{eq:ss_leading_vertex_responses}, only the difference in
Eq.\eqref{eq:ss_kappa_difference} remains:
\begin{equation}
\Delta\mathcal R_{\mathrm{time}}
=
-\frac{F_S}{F_T}
\left(
2\epsilon-f_T+s_T+2f_S
\right).
\label{eq:ss_time_correction}
\end{equation}
This is not yet the complete answer. The result still contains
\(F_S/F_T\) and \(f_S\), because the first-order corrections to the
pivot values of \(b_1\) and \(b_2\) have not yet been added.

The exact coefficients are given in
Eq.\eqref{eq:z_hh_b_coefficients}. The factors multiplying them in
the two leading soft contributions show that their pivot-value
correction enters through
\begin{equation}
\Delta\mathcal R_{\mathrm{pivot}}
\equiv
\left[
\frac{4b_1}{G_T}
+
\frac{12b_2}{F_T}
\right]_{\mathcal O(\varepsilon)}.
\label{eq:ss_pivot_correction_definition}
\end{equation}
The zeroth-order part of this combination vanishes by
Eq.\eqref{eq:ss_leading_cancellation}.
To evaluate the first-order remainder, define
\begin{equation}
U\equiv\frac{G_T^2}{\Theta},
\qquad
u\equiv\frac{\dot U}{HU}.
\label{eq:ss_U_definition}
\end{equation}
The exact expressions for \(b_1\) and \(b_2\), together with the
background identity in Eq.\eqref{eq:ss_FS_identity}, give
\begin{equation}
\frac{4b_1}{G_T}
+
\frac{12b_2}{F_T}
=
\frac{HU}{2F_T}
\left(
4u-g_T-f_T
\right).
\label{eq:ss_exact_coefficient_sum}
\end{equation}
The same background identity can be written as
\begin{equation}
F_S+F_T
=
HU+\dot U
=
HU(1+u),
\end{equation}
and hence
\begin{equation}
1+\frac{F_S}{F_T}
=
\frac{HU}{F_T}(1+u).
\label{eq:ss_U_background_relation}
\end{equation}
Moreover,
\begin{equation}
\frac{d}{dN}
\ln\left(\frac{HU}{F_T}\right)
=
-\epsilon+u-f_T.
\end{equation}
At zeroth order,
$HU/F_T=1+F_S/F_T$. Using the definitions of \(f_S\) and \(f_T\)
in Eqs.\eqref{eq:slowroll_parameters_general} and
\eqref{eq:tensor_slow_rates}, we therefore obtain
\begin{equation}
u
=
\epsilon+f_T
+
\frac{F_S/F_T}{1+F_S/F_T}
\left(
f_S-f_T
\right)
+
\mathcal O(\varepsilon^2).
\label{eq:ss_u_solution}
\end{equation}
The bracket in Eq.\eqref{eq:ss_exact_coefficient_sum} is already
first order, so its prefactor is needed only at zeroth order:
\begin{equation}
\frac{HU}{F_T}
=
1+\frac{F_S}{F_T}
+\mathcal O(\varepsilon).
\end{equation}
Substituting Eq.\eqref{eq:ss_u_solution} and
$g_T=f_T-2s_T$ into
Eq.\eqref{eq:ss_exact_coefficient_sum}, we find
\begin{equation}
\Delta\mathcal R_{\mathrm{pivot}}
=
2\epsilon+f_T+s_T
+
\frac{F_S}{F_T}
\left(
2\epsilon-f_T+s_T+2f_S
\right).
\label{eq:ss_pivot_correction}
\end{equation}

The model-dependent term in
Eq.\eqref{eq:ss_pivot_correction} cancels exactly against
Eq.\eqref{eq:ss_time_correction}:
\begin{equation}
\Delta\mathcal R_{\mathrm{pivot}}
+
\Delta\mathcal R_{\mathrm{time}}
=
2\epsilon+f_T+s_T.
\end{equation}
Since the leading response vanishes by
Eq.\eqref{eq:ss_leading_cancellation}, the complete scalar-soft
coefficient through first order is
\begin{equation}
\mathcal R_{\zeta hh}(k)
=
2\epsilon_*+f_{T*}+s_{T*}
+
\mathcal O(\varepsilon^2).
\label{eq:ss_final_response}
\end{equation}
Equivalently,
\begin{align}
B_{\zeta hh}^{\lambda\lambda'}(\bm q,\bm k)
\big|_{\mathrm{leading\ soft}}
={}&
\delta_{\lambda\lambda'}
P_\zeta(q)P_h(k)
\left[
2\epsilon_*+f_{T*}+s_{T*}
+\mathcal O(\varepsilon^2)
\right]
+o(1).
\label{eq:ss_final_bispectrum}
\end{align}
The tensor spectrum in Eq.\eqref{eq:modes_corrected_spectra} has
\begin{equation}
n_T
=
-2\epsilon-f_T-s_T
+
\mathcal O(\varepsilon^2).
\end{equation}
Consequently,
\begin{equation}
\mathcal R_{\zeta hh}(k)
=
-n_T(k)
+
\mathcal O(\varepsilon^2),
\end{equation}
in agreement with the independent prediction
Eq.\eqref{eq:ss_soft_theorem_prediction}.

The mechanism of this cancellation differs substantially from the tensor-soft case.
There, the leading bispectrum is already nonzero and the single surviving vertex
reproduces the universal coefficient, after which the NLO correction leaves
the scalar spectral tilt. In the scalar-soft case, the two vertices $b_1$ and $b_2$
give nonzero but opposite contributions at zeroth order. At the next
order, neither the time integrals nor the corrections to the cubic
coefficients are universal separately. Only after they are combined do
the dependences on $F_S/F_T$, $f_S$, $\gamma_E$, and the normalization
of the soft scalar mode cancel, leaving only the tensor
spectral tilt.

As in the tensor-soft analysis, the soft limit must be understood within
the perturbative expansion. At finite squeezing, the unresummed
mode functions require the logarithmic terms to remain controlled, in particular
$|\delta_S\ln r_1|$. Therefore, one must not set $r_1=0$ inside
individual external or bulk insertions. One must first combine the complete
numerator to first order and only then extract the leading soft
coefficient. Since the normalized scalar-soft bispectrum itself starts at
$\mathcal O(\varepsilon)$, the first-order correction to the product
$P_\zeta(q)P_h(k)$ changes the final result only
at $\mathcal O(\varepsilon^2)$.

\section{Nonlinearity parameters and predictions for the contracting Universe}
\label{sec:nonlinearity_nlo}

The three amplitudes entering this section are
Eqs.\eqref{eq:zeta_zeta_h_final_nlo_amplitude},
\eqref{eq:z_hh_final_nlo_amplitude}, and
\eqref{eq:hhh_final_nlo_amplitude}. We first convert them to the
normalization of Eq.\eqref{eq:fnl_definition} and then specialize the
result to the nonsingular model.

\subsection{From the bispectra to \texorpdfstring{$f_{\rm NL}$}{fNL}}

The amplitudes above are written with leading pivot spectra, whereas
Eq.\eqref{eq:fnl_definition} uses corrected spectra at $k_0$. We
therefore separate the numerator from the denominator correction:
\begin{subequations}
\label{eq:nlf_reference_conversion}
\begin{align}
f_{\mathrm{NL,ref}}^A(k_i)
&=\frac{10}{3}\,
\frac{\mathcal A_{A,0}(k_i)+\mathcal A_{A,1}(k_i)}
{\sum_i k_i^3},
\\
\mathcal P_A(k_0)
&=\mathcal P_{A*}^{(0)}
\left[1+2E_{X_A}(k_0)\right]
+\mathcal O(\varepsilon^2),
\nonumber\\
f_{\mathrm{NL}}^A
&=f_{\mathrm{NL,ref}}^A
-4E_{X_A}(k_0)f_{\mathrm{NL},0}^A
+\mathcal O(\varepsilon^2).
\label{eq:nlf_denominator_conversion}
\end{align}
\end{subequations}
The subscript ``ref'' denotes normalization by the leading reference
spectrum $\mathcal P_{A*}^{(0)}$ used in the reduced amplitudes
Eqs.\eqref{eq:zeta_zeta_h_final_nlo_amplitude},
\eqref{eq:z_hh_final_nlo_amplitude}, and
\eqref{eq:hhh_final_nlo_amplitude}. It does not refer to a separate
momentum configuration and should not be confused with the reference
wavenumber $k_0$. Thus, $f_{\mathrm{NL,ref}}^A$ is an intermediate
quantity whose numerator contains the complete first-order amplitude,
while its normalization has not yet been converted to the corrected
spectrum $\mathcal P_A(k_0)$. The last line of
Eq.\eqref{eq:nlf_reference_conversion} performs precisely this
conversion.
Here $X_A=S$ for $A=\zeta\zeta h,\zeta hh$ and $X_A=T$ for
$A=hhh$. The functions $E_S$ and $E_T$ are given in
Eqs.\eqref{eq:gamma_2} and \eqref{eq:tensor_external_mode}.

For the mixed sectors, the temporal functions are assembled in
Eqs.\eqref{eq:app_zzh_delta_J_general} and
\eqref{eq:app_zhh_delta_I_explicit}. Their leading parts are
$J_*^{(q)}=\eta_q\Lambda_qF_q$ and
$I_*^{(q)}=\eta_q\Lambda_qF_q$, as shown in
Eqs.\eqref{eq:app_zzh_J_eta_Lambda} and
\eqref{eq:app_zhh_I_eta_Lambda}. The corrections to an
undifferentiated or differentiated line are
Eqs.\eqref{eq:app_zzh_DeltaP_general} and
\eqref{eq:app_zzh_DeltaD_general}. We collect the corresponding
coefficients as
\begin{subequations}
\label{eq:nlf_UV_dictionary}
\begin{align}
U_{qi}&=-F_q+X_{qi}^{(P)}-Y_{qi}^{(P)},&
V_{qi}&=\ell_q+\mathfrak H_{qi}^{(P)},&
d_{qi}&=P,
\\
U_{qi}&=X_{qi}^{(D)}-2\ell_q+Z_{qi}^{(D)},&
V_{qi}&=\ell_q+Y_{qi}^{(D)}+\mathfrak H_{qi}^{(D)},&
d_{qi}&=D.
\end{align}
\end{subequations}
Thus a scalar line contributes
$-\epsilon F_q+sU_{qi}+\delta_SV_{qi}$, while a tensor line
contributes
$-\epsilon F_q+s_TU_{qi}+\delta_TV_{qi}$. The two mixed-sector sums
are
\begin{align}
\Sigma_{q,\mathrm{mode}}^{(\zeta\zeta h)}
&=s(U_{q1}+U_{q2})+\delta_S(V_{q1}+V_{q2})
+s_TU_{q3}+\delta_TV_{q3},
\nonumber\\
\Sigma_{q,\mathrm{mode}}^{(\zeta hh)}
&=sU_{q1}+\delta_SV_{q1}
+s_T(U_{q2}+U_{q3})+\delta_T(V_{q2}+V_{q3}).
\label{eq:nlf_species_mode_sum}
\end{align}
The common frequencies and external-mode combinations are
\begin{subequations}
\begin{align}
K_*&=c_S(k_1+k_2)+c_Tk_3,&
\overline E_{\zeta\zeta h}
&=E_S(k_1)+E_S(k_2)+E_T(k_3)-4E_S(k_0),
\label{eq:nlf_ssh_assignment}\\
K_*&=c_Sk_1+c_T(k_2+k_3),&
\overline E_{\zeta hh}
&=E_S(k_1)+E_T(k_2)+E_T(k_3)-4E_S(k_0).
\label{eq:nlf_shh_assignment}
\end{align}
\end{subequations}

Combining the pivot correction, the logarithmic evolution of the bare
coefficient, the scale factor, the three line corrections, and
Eq.\eqref{eq:nlf_denominator_conversion} gives
\begin{align}
\widehat g_q
={}&g_{q*}^{\mathrm{lead}}+\delta g_{q*}
-\frac{\dot g_{q*}^{\mathrm{lead}}}{H_*}\frac{\ell_q}{F_q}
\nonumber\\
&+g_{q*}^{\mathrm{lead}}
\left[
\overline E_A+(p_q-3)\epsilon
-p_q\epsilon\frac{\ell_q}{F_q}
+\frac{\Sigma_{q,\mathrm{mode}}^{(A)}}{F_q}
\right],
\qquad g_q=c_q\ \text{or}\ b_q.
\label{eq:nlf_effective_bare_coefficients}
\end{align}
The first three terms follow from
Eq.\eqref{eq:cubic_coefficient_additive_convention}; the terms
containing $p_q$ follow from
Eqs.\eqref{eq:app_zzh_Delta_ca} and
\eqref{eq:app_zhh_Delta_ba}; the remaining terms follow from
Eqs.\eqref{eq:app_zzh_delta_J_general},
\eqref{eq:app_zhh_delta_I_explicit}, and
\eqref{eq:nlf_denominator_conversion}. All functions in
Eq.\eqref{eq:nlf_effective_bare_coefficients} are evaluated
explicitly in Appendix\ref{app:nonlinearity_evaluation}.

Substituting Eq.\eqref{eq:nlf_effective_bare_coefficients} into the
three amplitudes and then using Eq.\eqref{eq:fnl_definition} gives
the arbitrary-triangle results
\eqref{eq:fnl_zzh_general}, \eqref{eq:fnl_zhh_general}, and
\eqref{eq:fnl_hhh_general}.

\subsection{Equilateral configuration for \texorpdfstring{$f_{\text{NL}}$}{fNL} in general Horndeski theory}

Equations\eqref{eq:fnl_zzh_general},
\eqref{eq:fnl_zhh_general}, and
\eqref{eq:fnl_hhh_general} give the complete nonlinearity parameters
for arbitrary momentum triangles. We now specialize these expressions
to equilateral momenta as a compact benchmark rather than as an
assumption about the full momentum dependence of the bispectra.

A local-type characterization is not unique for correlators containing
tensor modes. In a mixed correlator one must specify which species
carries the soft momentum, and the result retains a nontrivial
dependence on the tensor helicities and polarization contractions.
Moreover, the tensor-soft and scalar-soft configurations have already
been studied separately in Sec.\ref{sec:Maldacena}. For a regular
single-clock attractor, their leading behavior is constrained by the
corresponding consistency relations and is determined by the scalar or
tensor spectral tilt
\cite{Maldacena:2002vr}. A squeezed amplitude therefore mainly tests
the soft relations and does not, by itself, display all independent
combinations of cubic interactions entering the complete bispectrum.

The equilateral configuration provides complementary information.
When $k_1=k_2=k_3$, no momentum is treated as a long-wavelength
background for the other two, and the contributions of the different
cubic interactions can be compared without taking a soft limit. This
is particularly useful in general Horndeski theory, where several
interactions containing spatial and time derivatives contribute to the
mixed and tensor bispectra. The equilateral specialization also reduces
the momentum and polarization contractions to closed algebraic
expressions, making the first-order corrections and their helicity
dependence transparent. It furthermore permits a direct comparison
with the known leading-order Horndeski results of
Refs.\cite{Gao:2011vs,Gao:2012ib}.

We emphasize that the quantities derived below are representative
equilateral amplitudes, not substitutes for the arbitrary-triangle
results. Different interactions need not have identical momentum
dependence or attain their maximal value exactly at equilateral
momenta. The complete shapes remain those in
Eqs.\eqref{eq:fnl_zzh_general},
\eqref{eq:fnl_zhh_general}, and
\eqref{eq:fnl_hhh_general}, while the finite-squeezing behavior of the
$h\zeta\zeta$ sector is examined separately below.

For two scalars and one graviton,
\begin{subequations}
\label{eq:nlf_ssh_equilateral_data}
\begin{align}
\Omega_c&=2c_S+c_T,
\label{eq:nlf_ssh_equilateral_fractions}\\
f_{\mathrm{NL}}^{(\zeta\zeta h),\mathrm{eq}}
&=\frac{5}{6F_Tc_T}\sum_{q=1}^{6}W_q^{(c)}\widehat c_q
+\mathcal O(\varepsilon^2),
\label{eq:nlf_ssh_equilateral}
\end{align}
\end{subequations}
and we also are going to use $(r_1,r_2,r_3)
=\left(\frac{c_S}{\Omega_c},\frac{c_S}{\Omega_c},
\frac{c_T}{\Omega_c}\right)$.
This follows from Eq.\eqref{eq:fnl_zzh_general}, the temporal
assignments in Eq.\eqref{eq:nlf_ssh_temporal_assignment}, and the
polarization contraction in Appendix\ref{app:nlf_polarizations}.
The weights are
\begin{subequations}
\label{eq:nlf_ssh_weights}
\begin{align}
W_1^{(c)}
&=\frac{6c_S^3+6c_S^2c_T+4c_Sc_T^2+c_T^3}{\Omega_c^2},
&
W_2^{(c)}
&=-\frac{Hc_T^2(10c_S^2+6c_Sc_T+c_T^2)}{\Omega_c^3},
\\
W_3^{(c)}
&=-\frac{c_S^2c_T^2(3c_S+c_T)}{\Omega_c^2},
&
W_4^{(c)}
&=-\frac{Hc_S^2(6c_S^2+9c_Sc_T+2c_T^2)}{\Omega_c^3},
\\
W_5^{(c)}
&=-\frac{2H^2(10c_S^3+20c_S^2c_T+8c_Sc_T^2+c_T^3)}
{\Omega_c^4},
&
W_6^{(c)}
&=-\frac{2c_S^4(c_S+c_T)}{\Omega_c^2}.
\end{align}
\end{subequations}
For one scalar and two gravitons, let
\begin{subequations}
\label{eq:nlf_shh_equilateral_data}
\begin{align}
\chi_a&=\frac{\lambda_a}{2}=\pm1,&
\sigma&=\chi_2\chi_3,&
v_\sigma&=\frac{(1+2\sigma)^2}{16},&
w_\sigma&=\frac{3(1+2\sigma)}{32},
\label{eq:nlf_shh_equilateral_fractions}
\end{align}
\end{subequations}
with $\Omega_b=c_S+2c_T,$ and where now $(r_1,r_2,r_3)
=\left(\frac{c_S}{\Omega_b},\frac{c_T}{\Omega_b},
\frac{c_T}{\Omega_b}\right)$.
Using Eq.\eqref{eq:fnl_zhh_general}, the assignments in
Eq.\eqref{eq:nlf_shh_temporal_assignment}, and the contractions in
Appendix\ref{app:nlf_polarizations}, we obtain
\begin{align}
f_{\mathrm{NL}}^{(\zeta hh),\mathrm{eq},\sigma}
&=\frac{160F_Sc_S}{9F_T^2c_T^2}
\left[
v_\sigma\sum_{q=1}^{5}W_q^{(b)}\widehat b_q
+w_\sigma\sum_{q=6}^{7}W_q^{(b)}\widehat b_q
\right]
+\mathcal O(\varepsilon^2).
\label{eq:nlf_shh_equilateral}
\end{align}
The weights are
\begin{subequations}
\label{eq:nlf_shh_weights}
\begin{align}
W_1^{(b)}
&=\frac{2c_T^4(c_S+c_T)}{\Omega_b^2},
&
W_2^{(b)}
&=\frac{c_S^3+4c_S^2c_T+6c_Sc_T^2+6c_T^3}
{2\Omega_b^2},
\\
W_3^{(b)}
&=-\frac{c_S^2c_T^2(c_S+3c_T)}{2\Omega_b^2},
&
W_4^{(b)}=W_6^{(b)}
&=\frac{2Hc_S^2c_T^4}{\Omega_b^3},
\\
W_5^{(b)}=W_7^{(b)}
&=\frac{4H^2c_T^4(2c_S+c_T)}{\Omega_b^4}.
\end{align}
\end{subequations}
The equality of the temporal functions for $q=4,6$ and for $q=5,7$
does not identify the logarithmic evolution of the corresponding bare
coefficients.

For three gravitons, the arbitrary-triangle correction is
Eq.\eqref{eq:nlf_tensor_general_rho}. At equilateral momenta,
Appendix\ref{app:nlf_tensor_reduction} gives
\begin{subequations}
\begin{align}
L_{\rm eq}&=\ln\frac32, \quad
\sigma_{\mu_5}
=\frac{\dot\mu_{5*}^{\mathrm{lead}}}
{H_*\mu_{5*}^{\mathrm{lead}}},
\nonumber\\
\rho_{\mathrm{GR}}^{\mathrm{eq}}
&=\frac{64L_{\rm eq}-8}{17}\,\delta_T,
\label{eq:nlf_tensor_GR_rho}\\
\rho_{\mathrm{new}}^{\mathrm{eq}}
&=\left(\gamma_E-\frac52\right)\epsilon
+\left(3\gamma_E-\frac{11}{2}\right)s_T
+\left(80L_{\rm eq}-2\gamma_E-29\right)\delta_T
+\left(\gamma_E-\frac32\right)\sigma_{\mu_5}.
\label{eq:nlf_tensor_new_rho}
\end{align}
\end{subequations}
Substitution into Eq.\eqref{eq:fnl_hhh_general} gives
\begin{subequations}
\begin{align}
f_{\mathrm{NL}}^{(hhh),+++,\mathrm{eq}}
&=\frac{255}{1024}\left(1+\rho_{\mathrm{GR}}^{\mathrm{eq}}\right)
-\frac{5H}{1152G_T}
\left[
\mu_{5*}^{\mathrm{lead}}+\delta\mu_{5*}
+\mu_{5*}^{\mathrm{lead}}\rho_{\mathrm{new}}^{\mathrm{eq}}
\right]
+\mathcal O(\varepsilon^2),
\label{eq:nlf_tensor_all_equal}\\
f_{\mathrm{NL}}^{(hhh),++-,\mathrm{eq}}
&=\frac{85}{27648}\left(1+\rho_{\mathrm{GR}}^{\mathrm{eq}}\right)
-\frac{5H}{10368G_T}
\left[
\mu_{5*}^{\mathrm{lead}}+\delta\mu_{5*}
+\mu_{5*}^{\mathrm{lead}}\rho_{\mathrm{new}}^{\mathrm{eq}}
\right]
+\mathcal O(\varepsilon^2).
\label{eq:nlf_tensor_one_opposite}
\end{align}
\end{subequations}

\subsection{Application to the nonsingular contracting model}
\label{sec:contracting_uni}

We now apply the general nonlinearity parameters derived in the preceding
subsections to the nonsingular cosmological model introduced in
Eqs.\eqref{eq:j_frame}--\eqref{1.1.1}. We first summarize the model
parameters that enter the three-point functions and then perform the
specialization in the $\zeta\zeta h$, $\zeta hh$, and $hhh$ sectors. All
quantities below are evaluated at the common physical pivot, and stars are
suppressed when no ambiguity can arise.

At early times, the cosmological evolution in the Jordan frame is described
by the power-law contracting solution
\begin{align}
 a_J(t_J)&=d(-t_J)^\chi,
 &H_J&=\frac{\chi}{t_J},
 &t_J&<0,
 &0&<\chi<1.
\label{eq:bounce_contracting_background}
\end{align}
The parameter $\chi$ controls the contraction rate, whereas $d$ fixes only
the normalization of the scale factor. After the conformal transformation
in Eq.\eqref{eq:conformal_factor}, the model is described in the Einstein
frame by the G-inflation action in Eq.\eqref{1.1.1}. The three-point
functions are evaluated in this Einstein-frame description; the
Jordan-frame solution fixes the physical interpretation and the values of
the model parameters.

The original parametrization contains
$\{\mu,\chi,\bar c_2,d_2,\bar c_3,d_3,\hat g\}$, see \eqref{1.1.1}. The parameter $\mu$ controls the
time dependence of the effective gravitational coupling and enters the
conformal factor relating the two frames. Together with $\chi$, it fixes the
scalar and tensor tilts \cite{Ageeva2023},
\begin{align}
 n_s-1=n_T
 &=2\frac{1-\mu}{1-\chi}.
\label{eq:bounce_tilts}
\end{align}
Thus the observed red scalar tilt requires $\mu>1$ \cite{Ageeva2023}. The corresponding
Einstein-frame background is power-law inflation,
\begin{align}
 a_E(t_E)\propto t_E^{(\mu-\chi)/(\mu-1)},\qquad
 \epsilon=\frac{\mu-1}{\mu-\chi}.
\label{eq:bounce_epsilon}
\end{align}
The limit $\mu\simeq1$ therefore corresponds to a nearly de Sitter
Einstein-frame solution.

The constants $d_2$ and $\bar c_3$ enter the background and perturbation
equations through
\begin{align}
 \kappa&\equiv \bar c_3(1+2\mu)-d_2.
\end{align}
We use the parameter choice
\begin{align}
 d_3&=-2,
\label{eq:d3_choice}
\end{align}
which gives a small scalar propagation speed and hence a suppressed
tensor-to-scalar ratio. With the background lapse set to $N=1$, the
background equations give \cite{Ageeva2023}
\begin{align}
 \bar c_2
 =-3(\chi^2-2\chi),\qquad
 \kappa
 =2+4\chi-4\mu(\chi-1).
\label{eq:c2_background}
\end{align}
After these relations are imposed, the spectral shape and the dimensionless
three-point functions may be described by $\mu$ and $\chi$. The positive
constant $\hat g$ fixes the overall normalization of the primordial spectra
through Eq.\eqref{eq:conformal_factor}; it can be fixed by the observed
scalar amplitude and does not provide an additional parameter in the
nonlinearity parameters.

For this choice of parameters, the propagation speeds satisfy
\begin{align}
 c_T^2&=1,
 &c_S^2&=\frac23(\mu-1)(1-\chi).
\end{align}
The tensor modes therefore propagate at unit speed, whereas the scalar
propagation speed can be small in the phenomenologically relevant region.
This property suppresses the tensor-to-scalar ratio and makes the mixed
three-point functions sensitive to the remaining model parameter $\chi$.

It is also convenient to replace $\mu$ by the scalar tilt. We define
\begin{align}
 \Delta&\equiv1-n_s.
\label{eq:bounce_Delta}
\end{align}
Using Eqs.\eqref{eq:bounce_tilts} and \eqref{eq:bounce_epsilon}, we obtain
\begin{align}
 \mu
 =1+\frac{\Delta}{2}(1-\chi),\qquad
 \epsilon
 =\frac{\Delta}{2+\Delta}.
\label{eq:bounce_tilt_parameters}
\end{align}
For $d_3=-2$ and $M_{\rm pl}=1$, substitution of the functions in
Eq.\eqref{1.1.1} into the scalar and tensor quadratic actions \eqref{eq:horndeski_quadratic_coefficients}, \eqref{Gamma_big} gives
\begin{subequations}
\label{eq:bounce_background_data}
\begin{align}
 &F_T=G_T=\Gamma=1, \quad c_T=1, \quad
 \mu_5=0,
\label{eq:bounce_tensor_background}\\
 &\Theta=(1-\epsilon)H,
\label{eq:bounce_Gamma_Theta}\\
 &F_S=\Delta,\quad
 G_S=\frac{3}{(1-\chi)^2},
\label{eq:bounce_scalar_background}\\
 &c_S\equiv c
 =(1-\chi)\sqrt{\frac{\Delta}{3}}.
\label{eq:bounce_cs}
\end{align}
\end{subequations}
All coefficients entering the quadratic actions are constant on this
power-law solution. Therefore \eqref{eq:slowroll_parameters_general} reads
\begin{align}
 f_S&=g_S=s=f_T=g_T=s_T=0,
 &\delta_S&=\delta_T=\epsilon.
\label{eq:bounce_rates}
\end{align}
Thus, after fixing the scalar tilt, all first-order mode corrections are
proportional to the single parameter $\epsilon$, while the remaining
dependence on $\chi$ enters through $c_S=c$.

The specialization is performed in the same order in every sector. We first
insert the model background into the exact cubic coefficients. We then
evaluate the equilateral frequency fractions using
Eqs.\eqref{eq:nlf_ssh_equilateral_fractions} or
\eqref{eq:nlf_shh_equilateral_fractions}. The leading temporal functions
$F_q$ and their logarithmic moments $\ell_q$ are taken from
Eq.\eqref{eq:nlf_Fell_explicit}. The first-order corrections on the three
mode functions are combined according to Eq.\eqref{eq:nlf_UV_definition}.
Finally, all terms are assembled into the corrected coefficients
$\widehat c_q$ or $\widehat b_q$ through
Eq.\eqref{eq:nlf_effective_bare_coefficients} and substituted into the
corresponding equilateral formula for $f_{\rm NL}$. 

\subsubsection{Two scalars and one graviton}

We begin with the $\zeta\zeta h$ sector. Substitution of the model
background into the six coefficients of
Eq.\eqref{eq:lagrangian_of_two_grav_and_one_scal} gives
\begin{align}
 c_1&=\Delta,
 &Hc_2&=\frac{\Delta(2+\Delta)}{4},
 &c_3&=-\frac{\Delta^2}{2c^2},
 \nonumber\\
 c_4&=c_5=0,
 &c_6&=\frac{\Delta^2}{4c^4}.
\label{eq:bounce_cq}
\end{align}
Therefore, only the interactions $q=1,2,3,6$ contribute.

For $k_1=k_2=k_3=k$, $c_T=1$, and $c_S=c$, the frequency fractions in
Eq.\eqref{eq:nlf_ssh_equilateral_fractions} become
\begin{align}
 (r_1,r_2,r_3)
 &=\left(
 \frac{c}{1+2c},
 \frac{c}{1+2c},
 \frac{1}{1+2c}
 \right).
\label{eq:bounce_zzh_fractions}
\end{align}
The general equilateral formula is Eq.\eqref{eq:nlf_ssh_equilateral}. At
leading order, inserting Eq.\eqref{eq:bounce_cq} into the weights of
Eq.\eqref{eq:nlf_ssh_weights} gives the four weighted contributions
$T_q=W_q^{(c)}c_q$,
\begin{align}
 T_1&=
 \frac{\Delta(6c^3+6c^2+4c+1)}{(1+2c)^2},
 \nonumber\\
 T_2&=-
 \frac{\Delta(2+\Delta)(10c^2+6c+1)}{4(1+2c)^3},
 \nonumber\\
 T_3&=
 \frac{\Delta^2(3c+1)}{2(1+2c)^2},
 &T_6&=-
 \frac{\Delta^2(c+1)}{2(1+2c)^2}.
\label{eq:bounce_Tq}
\end{align}
The one-index quantities $T_q$ are weighted interaction contributions and
must not be confused with the temporal functions $\mathcal T_q(y)$ used in
the time integrals.

Substituting Eq.\eqref{eq:bounce_Tq} into
Eq.\eqref{eq:nlf_ssh_equilateral} gives
\begin{align}
 f_{{\rm NL},0}^{(\zeta\zeta h),{\rm eq}}
 ={}&
 \frac{5\Delta}{24(1+2c)^3}
 \left[
 48c^4+72c^3+36c^2+12c+2
 -\Delta(2c^2+2c+1)
 \right].
\label{eq:bounce_fNL_zzh_lo}
\end{align}
The terms $T_1$ and $T_2$ start at order $\Delta$, whereas $T_3$ and
$T_6$ start at order $\Delta^2$.

We next evaluate the first-order factor multiplying each $T_q$. The
external-mode terms in Eq.\eqref{eq:nlf_ssh_assignment}, together with
the subtraction from the corrected scalar spectrum in
Eq.\eqref{eq:nlf_denominator_conversion}, give
\begin{subequations}
\label{eq:bounce_Bc_data}
\begin{align}
 \mathcal B_c
 &=\gamma_2+1+\ln\left(\frac{c^2}{1+2c}\right),
\label{eq:bounce_Bc}\\
 \overline E_{\zeta\zeta h}
 &=\epsilon\mathcal B_c.
\end{align}
\end{subequations}
Here $\gamma_2$ is defined in Eq.\eqref{eq:gamma_2}. Thus the logarithm
in $\mathcal B_c$ is the combined external-mode and denominator correction
evaluated at the mixed acoustic pivot.

The time dependence of the surviving bare coefficients must be treated
separately from their values at the pivot. From
Eq.\eqref{eq:bounce_cq}, $c_1$, $c_3$, and $c_6$ are constant, whereas
$c_2\propto H^{-1}$. Their logarithmic rates are
\begin{align}
 \sigma_1^{(c)}
 &=\sigma_3^{(c)}
 =\sigma_6^{(c)}=0,
 &\sigma_2^{(c)}&=\epsilon.
\label{eq:bounce_c_rates}
\end{align}

Because $s=s_T=0$ and $\delta_S=\delta_T=\epsilon$, only the terms
multiplying the Hankel-index shifts survive in the mode-function part of
Eq.\eqref{eq:nlf_effective_bare_coefficients}. For an undifferentiated
line, Eq.\eqref{eq:nlf_UV_definition} gives
$V_{qi}=\ell_q+\mathfrak H_{qi}^{(P)}$; for a differentiated line it gives
$V_{qi}=\ell_q+Y_{qi}^{(D)}+\mathfrak H_{qi}^{(D)}$. Substitution of the
line assignments from Appendix\ref{app:nonlinearity_evaluation} into
Eq.\eqref{eq:nlf_effective_bare_coefficients} gives
\begin{subequations}
\label{eq:bounce_Rc}
\begin{align}
 \mathcal R_1^{(c)}
 &=\mathcal B_c-1
 +\frac{
 \ell_1
 +2\mathfrak H_{11}^{(P)}
 +\mathfrak H_{13}^{(P)}
 }{F_1},
\label{eq:bounce_Rc1}\\
 \mathcal R_2^{(c)}
 &=\mathcal B_c-2
 +\frac{
 \ell_2
 +2\mathfrak H_{21}^{(P)}
 +Y_{23}^{(D)}
 +\mathfrak H_{23}^{(D)}
 }{F_2},
\label{eq:bounce_Rc2}\\
 \mathcal R_3^{(c)}
 &=\mathcal B_c-1
 +\frac{
 \ell_3
 +\mathfrak H_{31}^{(P)}
 +Y_{32}^{(D)}
 +\mathfrak H_{32}^{(D)}
 +Y_{33}^{(D)}
 +\mathfrak H_{33}^{(D)}
 }{F_3},
\label{eq:bounce_Rc3}\\
 \mathcal R_6^{(c)}
 &=\mathcal B_c-1
 +\frac{
 \ell_6
 +Y_{61}^{(D)}
 +\mathfrak H_{61}^{(D)}
 +Y_{62}^{(D)}
 +\mathfrak H_{62}^{(D)}
 +\mathfrak H_{63}^{(P)}
 }{F_6}.
\label{eq:bounce_Rc6}
\end{align}
\end{subequations}
Every quantity on the right-hand sides is evaluated at
Eq.\eqref{eq:bounce_zzh_fractions}. The functions $F_q$ and $\ell_q$ are
the corresponding entries of Eq.\eqref{eq:nlf_Fell_explicit}. The
quantities $Y_{qi}^{(D)}$ and $\mathfrak H_{qi}^{(P,D)}$ are the pieces of
the differentiated and undifferentiated mode corrections combined in
Eq.\eqref{eq:nlf_UV_definition}; their analytic evaluation is given in
Appendices\ref{app:zeta_zeta_h_slow_variation} and
\ref{app:nonlinearity_evaluation}. The factors $2\mathfrak H_{11}^{(P)}$
and $2\mathfrak H_{21}^{(P)}$ arise because the two scalar lines are
identical. The constants $-1$ and $-2$ follow from the scale-factor powers
and, for $q=2$, from the nonzero rate
$\sigma_2^{(c)}=\epsilon$.

The complete equilateral result is therefore
\begin{align}
 f_{\rm NL}^{(\zeta\zeta h),{\rm eq}}
 &=\frac56
 \sum_{q=1,2,3,6}
 T_q\left[1+\epsilon\mathcal R_q^{(c)}\right]
 +\mathcal O(\epsilon^2).
\label{eq:bounce_fNL_zzh_nlo}
\end{align}
This is obtained by inserting
$W_q^{(c)}\widehat c_q=T_q[1+\epsilon\mathcal R_q^{(c)}]$ into
Eq.\eqref{eq:nlf_ssh_equilateral}. Setting the four factors
$\mathcal R_q^{(c)}$ to zero reproduces
Eq.\eqref{eq:bounce_fNL_zzh_lo}.

\subsubsection{One scalar and two gravitons}

We now turn to the $\zeta hh$ sector. Substitution of the model background
into the seven coefficients of
Eq.\eqref{eq:general_z_hh_cubic_action_cosmic} gives
\begin{align}
 b_1&=-\frac{\Delta}{8},
 &b_2&=\frac{\Delta}{8},
 &b_3&=-\frac{\Delta}{4c^2},
 \nonumber\\
 b_4&=b_5=b_6=b_7=0.
\label{eq:bounce_bq}
\end{align}
Only three of the seven general Horndeski interactions remain. In
particular, the interactions proportional to the second equilateral
polarization structure in Eq.\eqref{eq:nlf_shh_equilateral} are absent.

For equilateral momenta, the frequency fractions in
Eq.\eqref{eq:nlf_shh_equilateral_fractions} become
\begin{align}
 (r_1,r_2,r_3)
 &=\left(
 \frac{c}{c+2},
 \frac{1}{c+2},
 \frac{1}{c+2}
 \right).
\label{eq:bounce_zhh_fractions}
\end{align}
The common helicity factor is
\begin{align}
 v_\sigma
 &=\frac{(1+2\sigma)^2}{16},
 &v_+&=\frac9{16},
 &v_-&=\frac1{16}.
\label{eq:bounce_vsigma}
\end{align}
Inserting Eq.\eqref{eq:bounce_bq} into the weights of
Eq.\eqref{eq:nlf_shh_weights} gives the three combinations
$U_q=W_q^{(b)}b_q$,
\begin{align}
 U_1&=-\frac{\Delta(c+1)}{4(c+2)^2},
 \nonumber\\
 U_2&=\frac{\Delta(c^3+4c^2+6c+6)}{16(c+2)^2},
 \nonumber\\
 U_3&=\frac{\Delta(c+3)}{8(c+2)^2}.
\label{eq:bounce_Uq}
\end{align}
These one-index quantities $U_q$ are weighted interaction contributions and
are distinct from the lower-index functions $U_{qi}$ in
Eq.\eqref{eq:nlf_UV_definition}.

Substitution of Eq.\eqref{eq:bounce_Uq} into
Eq.\eqref{eq:nlf_shh_equilateral} gives
\begin{align}
 f_{{\rm NL},0}^{(\zeta hh),{\rm eq},\sigma}
 &=\frac{10}{9}v_\sigma\Delta^2c
 \frac{c^3+4c^2+4c+8}{(c+2)^2}.
\label{eq:bounce_fNL_zhh_lo}
\end{align}
For the two independent helicity choices,
\begin{subequations}
\label{eq:bounce_fNL_zhh_helicity_lo}
\begin{align}
 f_{{\rm NL},0}^{(\zeta hh),{\rm eq},+}
 &=\frac58\,\Delta^2c\,
 \frac{c^3+4c^2+4c+8}{(c+2)^2},
\label{eq:bounce_fNL_zhh_plus_lo}\\
 f_{{\rm NL},0}^{(\zeta hh),{\rm eq},-}
 &=\frac5{72}\,\Delta^2c\,
 \frac{c^3+4c^2+4c+8}{(c+2)^2}.
\label{eq:bounce_fNL_zhh_minus_lo}
\end{align}
\end{subequations}
The common factor $\Delta^2c$ makes this sector parametrically smaller than
$\zeta\zeta h$ when $c$ is small.

The external-mode and denominator terms follow from
Eqs.\eqref{eq:nlf_shh_assignment} and
\eqref{eq:nlf_denominator_conversion}. With the model values in
Eq.\eqref{eq:bounce_rates} and the fractions in
Eq.\eqref{eq:bounce_zhh_fractions}, they give
\begin{subequations}
\label{eq:bounce_Bb_data}
\begin{align}
 \mathcal B_b
 &=\gamma_2+1+\ln\left(\frac{c^3}{c+2}\right),
\label{eq:bounce_Bb}\\
 \overline E_{\zeta hh}
 &=\epsilon\mathcal B_b.
\end{align}
\end{subequations}
All three surviving coefficients are constant on the background,
\begin{align}
 \sigma_1^{(b)}
 &=\sigma_2^{(b)}
 =\sigma_3^{(b)}=0.
\label{eq:bounce_b_rates}
\end{align}

Using the line assignments of Appendix\ref{app:nonlinearity_evaluation}
and the definitions in Eq.\eqref{eq:nlf_UV_definition},
Eq.\eqref{eq:nlf_effective_bare_coefficients} gives
\begin{subequations}
\label{eq:bounce_Rb}
\begin{align}
 \mathcal R_1^{(b)}
 &=\mathcal B_b-1
 +\frac{
 \ell_1
 +\mathfrak H_{11}^{(P)}
 +Y_{12}^{(D)}
 +\mathfrak H_{12}^{(D)}
 +Y_{13}^{(D)}
 +\mathfrak H_{13}^{(D)}
 }{F_1},
\label{eq:bounce_Rb1}\\
 \mathcal R_2^{(b)}
 &=\mathcal B_b-1
 +\frac{
 \ell_2
 +\mathfrak H_{21}^{(P)}
 +\mathfrak H_{22}^{(P)}
 +\mathfrak H_{23}^{(P)}
 }{F_2},
\label{eq:bounce_Rb2}\\
 \mathcal R_3^{(b)}
 &=\mathcal B_b-1
 +\frac{
 \ell_3
 +Y_{31}^{(D)}
 +\mathfrak H_{31}^{(D)}
 +Y_{32}^{(D)}
 +\mathfrak H_{32}^{(D)}
 +\mathfrak H_{33}^{(P)}
 }{F_3}.
\label{eq:bounce_Rb3}
\end{align}
\end{subequations}
All functions are evaluated at Eq.\eqref{eq:bounce_zhh_fractions}. The
functions $F_q$ and $\ell_q$ follow from
Eq.\eqref{eq:nlf_Fell_explicit}, while the remaining terms are the
explicit pieces of Eq.\eqref{eq:nlf_UV_definition} evaluated in the
appendices.

The complete equilateral result is
\begin{align}
 f_{\rm NL}^{(\zeta hh),{\rm eq},\sigma}
 &=\frac{160\Delta c}{9}v_\sigma
 \sum_{q=1}^{3}
 U_q\left[1+\epsilon\mathcal R_q^{(b)}\right]
 +\mathcal O(\epsilon^2).
\label{eq:bounce_fNL_zhh_nlo}
\end{align}
This is obtained by inserting
$W_q^{(b)}\widehat b_q=U_q[1+\epsilon\mathcal R_q^{(b)}]$ into
Eq.\eqref{eq:nlf_shh_equilateral}. Because all three surviving
interactions are proportional to the same polarization contraction, the
helicity dependence factorizes at both orders:
\begin{align}
 f_{\rm NL}^{(\zeta hh),{\rm eq},+}
 &=9f_{\rm NL}^{(\zeta hh),{\rm eq},-}.
\label{eq:bounce_helicity_ratio}
\end{align}
At leading order, the same factor-of-nine relation follows from the
general helicity contractions derived in
Ref.\cite{Gao:2012ib}. In the present model only the interactions
proportional to $b_1$, $b_2$, and $b_3$ survive, and all of them carry
the same helicity factor, with $v_+=9/16$ and $v_-=1/16$ in the
equilateral configuration. Equation\eqref{eq:bounce_helicity_ratio}
shows that this leading-order helicity ratio is preserved by the
first-order slow-variation corrections.

\subsubsection{Three gravitons}

The three-graviton sector simplifies for a different reason. The
Einstein-frame model in Eq.\eqref{1.1.1} has $G_5=0$, and therefore
\begin{align}
 \mu_5&=\dot\phi XG_{5X}=0.
\end{align}
The additional Horndeski interaction in
Eq.\eqref{eq:hhh_cubic_lagrangian_cosmic} is absent, and the tensor
bispectrum is generated entirely by the universal GR interaction.

The general equilateral correction is Eq.\eqref{eq:nlf_tensor_GR_rho}.
Using $f_T=s_T=0$ and $\delta_T=\epsilon$ from
Eq.\eqref{eq:bounce_rates}, we obtain
\begin{align}
 \rho_{\rm GR}^{\rm eq}
 &=\frac{64\ln(3/2)-8}{17}\epsilon
 =\frac{64\ln(3/2)-8}{17}\frac{\Delta}{2+\Delta}.
\label{eq:bounce_rho_GR}
\end{align}
Substitution into Eqs.\eqref{eq:nlf_tensor_all_equal} and
\eqref{eq:nlf_tensor_one_opposite} gives
\begin{subequations}
\label{eq:bounce_fNL_hhh}
\begin{align}
 f_{\rm NL}^{(hhh),+++,{\rm eq}}
 &=\frac{255}{1024}
 \left[
 1+\frac{64\ln(3/2)-8}{17}\frac{\Delta}{2+\Delta}
 \right]
 +\mathcal O(\epsilon^2),
\label{eq:bounce_fNL_hhh_ppp}\\
 f_{\rm NL}^{(hhh),++-,{\rm eq}}
 &=\frac{85}{27648}
 \left[
 1+\frac{64\ln(3/2)-8}{17}\frac{\Delta}{2+\Delta}
 \right]
 +\mathcal O(\epsilon^2).
\label{eq:bounce_fNL_hhh_ppm}
\end{align}
\end{subequations}
Parity and permutation symmetry determine the remaining equilateral
helicity configurations.

The parameter dependence of the three sectors is therefore
\begin{align}
 f_{\rm NL}^{(\zeta\zeta h),{\rm eq}}
 &=f_{\rm NL}^{(\zeta\zeta h),{\rm eq}}(n_s,\chi),
 &
 f_{\rm NL}^{(\zeta hh),{\rm eq},\sigma}
 &=f_{\rm NL}^{(\zeta hh),{\rm eq},\sigma}(n_s,\chi),
 \nonumber\\
 f_{\rm NL}^{(hhh),{\rm eq}}
 &=f_{\rm NL}^{(hhh),{\rm eq}}(n_s).
\label{eq:bounce_parameter_dependence}
\end{align}
The mixed sectors retain information about $\chi$ through
Eq.\eqref{eq:bounce_cs}, whereas the purely tensor sector is fixed once
$n_s$ has been specified. For the central value $n_s=0.9649$,
\begin{subequations}
\label{eq:bounce_central_values}
\begin{align}
 \Delta&=0.0351,
 &\epsilon&=\frac{0.0351}{2.0351}
 \simeq1.72\times10^{-2},
\label{eq:bounce_central_background}\\
 f_{\rm NL}^{(hhh),+++,{\rm eq}}
 &\simeq2.54\times10^{-1},
 &f_{\rm NL}^{(hhh),++-,{\rm eq}}
 &\simeq3.13\times10^{-3}.
\label{eq:bounce_tensor_numbers}
\end{align}
\end{subequations}
Thus the mixed bispectra can provide constraints on $\chi$ that are
independent of the overall spectrum normalization fixed by $\hat g$, while
the tensor bispectrum constitutes a definite prediction of the model at
fixed scalar tilt.

Direct observational constraints on tensor and mixed tensor--scalar
non-Gaussianity have recently been obtained from the Planck PR4
temperature and polarization bispectra in
Ref.\cite{Philcox:2024wqx}. Of particular relevance to the present
model is the parity-even squeezed tensor--scalar--scalar template considered
there, which contains one long-wavelength tensor mode and two
short-wavelength scalar modes and is therefore most closely related to our
$h\zeta\zeta$ bispectrum. In the template normalization of
Ref.\cite{Philcox:2024wqx}, the combined Planck $T+E+B$ analysis
gives
\begin{align}
f_{\rm NL}^{tss,\mathrm{sq}}
&=11\pm11
\qquad (\textsc{sevem}),
&
f_{\rm NL}^{tss,\mathrm{sq}}
&=13\pm10
\qquad (\textsc{smica}),
\end{align}
with no statistically significant evidence for mixed tensor--scalar
non-Gaussianity. Here \textsc{sevem} and \textsc{smica} denote two independent
component-separation pipelines applied to the same Planck PR4
multifrequency data. The former removes foregrounds using internally
constructed templates in map space, whereas the latter combines the
frequency maps by matching their covariance structure in harmonic
space. Quoting both results tests the sensitivity of the inferred
bispectrum amplitude to residual foregrounds and instrumental
systematics.

These bounds cannot be converted directly into a constraint on $\chi$ by
substituting Eqs.\eqref{eq:bounce_fNL_zzh_lo} or
\eqref{eq:bounce_fNL_zzh_nlo}. The observational analysis of
Ref.\cite{Philcox:2024wqx} assumes a particular squeezed template
and amplitude convention, whereas Eq.\eqref{eq:fnl_zzh_general} gives the
complete momentum and helicity dependence predicted by the present model,
including the first-order slow-variation corrections. A model-specific
constraint requires projecting this primordial bispectrum onto the CMB
$T$-, $E$-, and $B$-mode bispectra, using the scalar and tensor transfer
functions, and determining its correlation and relative normalization with
the template employed in Ref. \cite{Philcox:2024wqx}.

If this correlation is sufficiently close to unity, the published Planck
likelihood can be recast in terms of the model parameters by using
Eq.\eqref{eq:bounce_cs} to convert the resulting constraint on $c_S$ into
a constraint on $\chi$. Otherwise, the exact bispectrum in
Eq.\eqref{eq:fnl_zzh_general} must be implemented as a separate
observational template. Performing this CMB projection and deriving the
corresponding joint constraints on $(n_s,\chi)$ are left for future work.

The tensor--tensor--scalar template constrained in
Ref.\cite{Philcox:2024wqx} is parity odd and is motivated by a
Chern--Simons interaction. It therefore does not directly constrain the
parity-even $\zeta hh$ bispectrum predicted by
Eq.\eqref{eq:fnl_zhh_general}; a dedicated parity-even template is also
required for that sector.

\subsection{Finite squeezing and the acoustic crossover}

The tensor-soft relation determines the normalized $h\zeta\zeta$
bispectrum in the strict soft limit. In the present mixed sector, however,
the relevant hierarchy is not only $q/k\ll1$ but
$c_Tq/(c_Sk)\ll1$. This distinction is important when $c_S$ is small: a
triangle can be strongly squeezed in momentum space while the tensor and
scalar acoustic frequencies are still comparable. We therefore evaluate
the full first-order result at finite squeezing and determine how it
approaches the independently known soft value.

We consider a one-parameter set of momentum triangles with two equal scalar
momenta,
\begin{subequations}
\label{eq:finite_squeezing_geometry}
\begin{align}
 \bm k_1+\bm k_2+\bm k_3&=0,
 &k_1&=k_2=k,
 &k_3&=q=\xi k,
 &k_0&=k,
\label{eq:finite_squeezing_momenta}
\end{align}
\end{subequations}
where $ 0<\xi<2.$
The parameter $\xi=q/k$ changes the spatial triangle continuously. The
limit $\xi\ll1$ is squeezed, $\xi=1$ is equilateral, and $\xi\to2$
approaches a flattened triangle. In what follows, the emphasis is on the
squeezed region.

For $c_T=1$ and $c_S=c$, the common frequency and the fractions defined in
Eqs.\eqref{eq:modes_common_pivot} and
\eqref{eq:modes_frequency_variables} are
\begin{subequations}
\label{eq:finite_squeezing_fractions}
\begin{align}
 &K_*=(2c+\xi)k,
\label{eq:finite_squeezing_K}\\
 &r_1=r_2=\frac{c}{2c+\xi},\quad
 r_3=\frac{\xi}{2c+\xi}.
\label{eq:finite_squeezing_r}
\end{align}
\end{subequations}
It is useful to compare the tensor and scalar acoustic frequencies directly:
\begin{subequations}
\label{eq:finite_squeezing_acoustic_ratio}
\begin{align}
 &u\equiv\frac{c_Tq}{c_Sk}=\frac{\xi}{c},
\label{eq:finite_squeezing_u}\\
 &r_1=r_2=\frac{1}{2+u}, \quad
 r_3=\frac{u}{2+u}.
\end{align}
\end{subequations}
Thus $u\ll1$ is the tensor-soft regime, while $u\simeq1$ marks the
acoustic crossover at which $c_Tq\simeq c_Sk$. In terms of the ordinary
momentum ratio, the crossover occurs at
\begin{align}
 \frac{q}{k}&=c_S.
\label{eq:finite_squeezing_transition}
\end{align}
Consequently, a smaller scalar propagation speed moves the crossover to a
more strongly squeezed spatial triangle.

To compare the leading and first-order calculations with the same
normalization convention, we define
\begin{subequations}
\label{eq:finite_squeezing_coefficient}
\begin{align}
 \frac{
 B_{h\zeta\zeta}^{\lambda,(0)}(\bm q,\bm k)
 }{
 P_h^{(0)}(q)P_\zeta^{(0)}(k)
 }
 &=\mathcal E_\lambda\,
 \mathcal R_{h\zeta\zeta}^{\rm LO}(\xi,c),
\label{eq:finite_squeezing_LO}\\
 \frac{
 B_{h\zeta\zeta}^{\lambda,(0)}(\bm q,\bm k)
 +B_{h\zeta\zeta}^{\lambda,(1)}(\bm q,\bm k)
 }{
 P_h(q)P_\zeta(k)
 }
 &=\mathcal E_\lambda\,
 \mathcal R_{h\zeta\zeta}^{\rm NLO}(\xi,c),
\label{eq:finite_squeezing_NLO}\\
 \Delta\mathcal R_{h\zeta\zeta}(\xi,c)
 &\equiv
 \mathcal R_{h\zeta\zeta}^{\rm NLO}(\xi,c)
 -\mathcal R_{h\zeta\zeta}^{\rm LO}(\xi,c).
\label{eq:finite_squeezing_delta}
\end{align}
\end{subequations}
The leading bispectrum is divided by the leading spectra, whereas the
first-order bispectrum is divided by the spectra evaluated to the same
slow-variation order. This implements the denominator conversion in
Eq.\eqref{eq:nlf_denominator_conversion} and prevents the external-mode
corrections from being counted twice.

The coefficient in Eq.\eqref{eq:finite_squeezing_coefficient} is related
to the momentum-dependent nonlinearity parameter by
\begin{align}
 \mathcal E_\lambda\,
 \mathcal R_{h\zeta\zeta}(\xi,c)
 &=\frac65(2+\xi^3)
 \frac{\mathcal P_\zeta(k)}
 {\mathcal P_{h,\mathrm{one\ helicity}}(\xi k)}
 f_{\rm NL}^{(\zeta\zeta h),\lambda}(k,k,\xi k;k).
\label{eq:finite_squeezing_fnl_conversion}
\end{align}
This relation follows from the definition of $f_{\rm NL}$ given at the
beginning of this section and from the spectra in Eq.\eqref{eq:sg_spectra}.
In particular, $\Delta\mathcal R_{h\zeta\zeta}$ is not obtained by simply
subtracting two values of $f_{\rm NL}$; the corresponding power-spectrum
normalization has to be used at each order.

We now state explicitly how the finite-squeezing result is evaluated. The
model coefficients in Eq.\eqref{eq:bounce_cq} imply that only
$q=1,2,3,6$ survive. For every value of $\xi$, the fractions in
Eq.\eqref{eq:finite_squeezing_fractions} are substituted into the
functions $F_q$ and $\ell_q$ of Eq.\eqref{eq:nlf_Fell_explicit} and into
the mode-function corrections assembled in Eq.\eqref{eq:nlf_UV_definition}.
The temporal form and the scale-factor power of every interaction are those
listed in Appendix\ref{app:nonlinearity_evaluation}. These quantities,
together with the external-mode combination in
Eq.\eqref{eq:nlf_ssh_assignment} and the rates in
Eq.\eqref{eq:bounce_c_rates}, are inserted into
Eq.\eqref{eq:nlf_effective_bare_coefficients} to obtain the four
$\widehat c_q$.

Substitution into the arbitrary-triangle $\zeta\zeta h$ formula derived
above gives, for the present momentum configuration,
\begin{align}
 f_{\rm NL}^{(\zeta\zeta h),\lambda}(k,k,\xi k;k)
 ={}&
 \frac{10H^2}{3k^3(2+\xi^3)}
 \sum_{q\in\{1,2,3,6\}}
 \Big[
 V_\lambda^{(q)}\eta_q\Lambda_qF_q\widehat c_q
 +(k_1\leftrightarrow k_2)
 \Big].
\label{eq:finite_squeezing_fnl}
\end{align}
Equation\eqref{eq:finite_squeezing_fnl_conversion} then gives
$\mathcal R_{h\zeta\zeta}^{\rm NLO}$. The leading quantity is obtained by
using the leading coefficients and leading spectra in the same sequence.

The strict soft value is fixed independently by the tensor-soft calculation.
Equation\eqref{eq:sg_final_result} gives
$\mathcal R_{h\zeta\zeta}=3/2+\delta_S+\mathcal O(\varepsilon^2)$, while
the leading term is $3/2$. Since $\delta_S=\epsilon$ in
Eq.\eqref{eq:bounce_rates}, the finite-squeezing correction must satisfy
\begin{align}
 \lim_{u\to0}\Delta\mathcal R_{h\zeta\zeta}
 &=\epsilon
 =\frac{\Delta}{2+\Delta}.
\label{eq:finite_squeezing_soft_limit}
\end{align}
The finite-$u$ calculation therefore determines how the complete
momentum-dependent bispectrum approaches this independently derived value
and identifies the range in which the strict soft approximation is
accurate.

Finally, we again emphasize that the fixed-order expressions are used only while the logarithmic
terms generated by the transfer to the common pivot remain perturbative.
With $r_3=u/(2+u)$, the required condition is
\begin{align}
 \left|\epsilon\ln r_3\right|&\ll1.
\label{eq:finite_squeezing_control}
\end{align}

To display separately the physical acoustic dependence and its translation
into the ordinary momentum ratio, we present the same finite-squeezing
correction in two complementary forms. At fixed $n_s$, after the model
coefficients have been substituted, the normalized quantity
$\Delta\mathcal R_{h\zeta\zeta}/\epsilon$ depends on the triangle through
$u=q/(c_Sk)$. Plotting it against $u$ therefore removes the horizontal shift
caused by changing $c_S$ and provides a direct test of the soft limit. We
then plot the unnormalized correction against $q/k$ for several values of
$c_S$ in order to show where the same acoustic transition occurs in the
usual momentum variable.

For the numerical curves we use $n_s=0.9649$. The colored bands are obtained
by varying $n_s$ within $0.9649\pm0.0042$ while keeping $c_S$ fixed. In the
numerical implementation of Eq.\eqref{eq:finite_squeezing_control}, only
points satisfying $|\epsilon\ln r_3|<0.12$ are retained.

\begin{figure}[t]
\centering
\includegraphics[width=\textwidth]{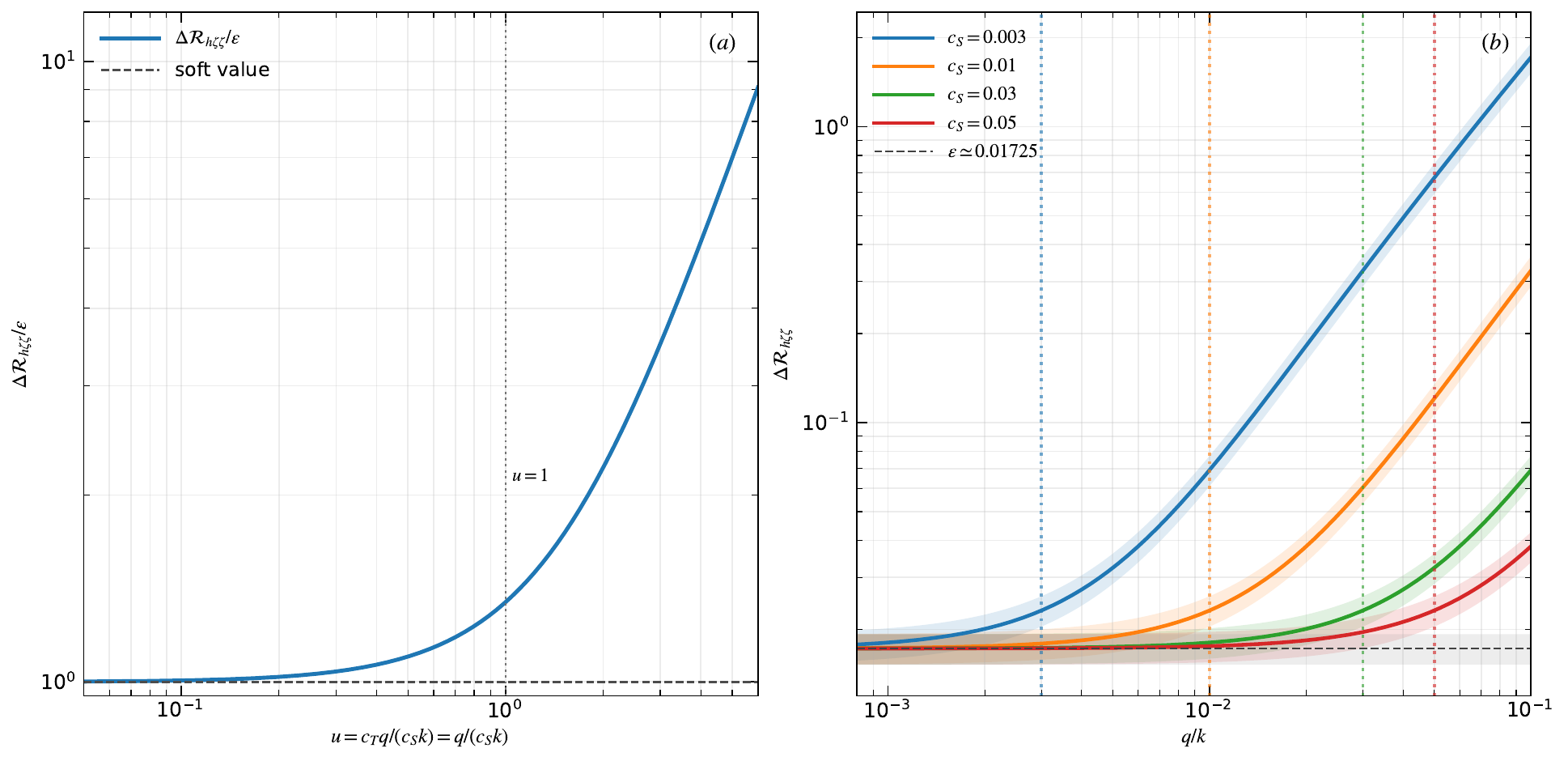}
\caption{
Two complementary representations of the first-order finite-squeezing
correction. Panel(a) shows
$\Delta\mathcal R_{h\zeta\zeta}/\epsilon$ as a function of the acoustic
frequency ratio $u=q/(c_Sk)$. The horizontal dashed line is the tensor-soft
value $1$, and the vertical dotted line marks the acoustic crossover $u=1$.
At fixed $n_s$, the curves obtained for different $c_S$ coincide in these
variables. Panel(b) shows the same correction as a function of $q/k$ for
$c_S=0.003$, $0.01$, $0.03$, and $0.05$. The colored vertical dotted lines
mark $q/k=c_S$, while the horizontal dashed line gives the central soft
value $\epsilon\simeq0.01725$ and the grey band shows its variation over the
quoted range of $n_s$. The colored bands in both panels are obtained by
varying $n_s$ at fixed $c_S$.
}
\label{fig:finite_squeezing_two_panel}
\end{figure}

Panel(a) shows that the complete result approaches the independently known
soft value for $u\ll1$ and begins to depart from it when the tensor and
scalar acoustic frequencies become comparable. Panel(b) shows that a
smaller $c_S$ moves this same transition to a smaller value of $q/k$.
Thus the separation of the colored curves in the right panel is primarily
a horizontal displacement of one acoustic dependence, rather than an
independent enhancement generated by changing $c_S$.



\section{Outlook}
\label{sec:conclusion}

We have obtained analytic primordial bispectra for the
$\zeta\zeta h$, $\zeta hh$, and $hhh$ sectors through first order in
slow variation, retaining their complete momentum and helicity
dependence. All temporal integrals have been evaluated, and the
corrections associated with the cubic coefficients, the scale factor,
and the scalar and tensor mode functions have been separated
explicitly. The agreement with the two mixed Maldacena consistency
relations provides an independent check of these results. The
finite-squeezing analysis further shows that, when $c_S\neq c_T$, a
geometrically squeezed triangle need not be acoustically soft: the
full momentum dependence becomes important near
$c_Tq/(c_Sk)\simeq1$.

For the nonsingular model considered here, the mixed bispectra retain
information about the contraction parameter $\chi$ through
Eq.\eqref{eq:bounce_cs}. The results obtained in this work are,
however, primordial predictions rather than direct observational
constraints. Deriving a bound on $\chi$ requires projecting the
momentum- and helicity-dependent primordial bispectra onto observable
CMB temperature and polarization bispectra.

The closest existing analysis is the Planck PR4 search of
Ref.\cite{Philcox:2024wqx}, which constrained eleven tensor and mixed
tensor--scalar templates using $T$-, $E$-, and $B$-mode bispectra and
found no statistically significant signal. Its parity-even squeezed
tensor--scalar--scalar template is most closely related to our
$h\zeta\zeta$ sector. In the normalization of
Ref.\cite{Philcox:2024wqx}, the combined $T+E+B$ constraints are
\begin{align}
f_{\rm NL}^{tss,\mathrm{sq}}
&=
11\pm11
\qquad
(\textsc{sevem}),
&
f_{\rm NL}^{tss,\mathrm{sq}}
&=
13\pm10
\qquad
(\textsc{smica}),
\end{align}
where \textsc{sevem} and \textsc{smica} denote two
component-separated versions of the same Planck PR4 data. These
results improve substantially on the earlier direct search using WMAP
temperature data \cite{Shiraishi:2017tss} and demonstrate the
importance of polarization information; see also
Ref.\cite{Shiraishi:2019tensorNG}.

These bounds cannot be converted into a constraint on $\chi$ by simply
substituting the squeezed or equilateral value of
$f_{\rm NL}^{(\zeta\zeta h)}$. Ref. \cite{Philcox:2024wqx} constrains
a particular phenomenological squeezed template, whereas
Eq.\eqref{eq:fnl_zzh_general} predicts the full momentum and helicity
dependence, including the first-order corrections and the acoustic
crossover. Agreement in the strict soft limit does not guarantee a
large correlation between the two CMB shapes after the radiation
transfer functions have been included.

A model-specific analysis therefore requires
\begin{align}
B_{h\zeta\zeta}^{\lambda}
(k_1,k_2,k_3;n_s,\chi)
&\longrightarrow
b_{\ell_1\ell_2\ell_3}^{XYZ}
(n_s,\chi)
\longrightarrow
\mathcal L_{\rm CMB}(n_s,\chi),
\qquad
X,Y,Z\in\{T,E,B\}.
\label{eq:outlook_cmb_pipeline}
\end{align}
The projected CMB shape should first be compared with the template of
Ref.\cite{Philcox:2024wqx} using the same covariance-weighted inner
product. If the correlation is sufficiently large, the published
likelihood may be approximately recast as a likelihood for $\chi$.
Otherwise, Eq.\eqref{eq:fnl_zzh_general} must be implemented as a
separate observational template. This comparison should be performed
for both the leading and first-order shapes, since the latter change
not only the amplitude but also the logarithmic and finite-squeezing
momentum dependence.

The $\zeta hh$ sector requires a dedicated analysis. The
tensor--tensor--scalar template constrained in
Ref.\cite{Philcox:2024wqx} is parity odd and is motivated by a
Chern--Simons interaction, whereas Eq.\eqref{eq:fnl_zhh_general} is
parity even. The corresponding published likelihood therefore cannot
be applied directly. A parity-even template retaining both tensor
helicities is needed; the factor-of-nine relation in
Eq.\eqref{eq:bounce_helicity_ratio} simplifies its model
specialization but does not replace the full CMB projection.

Before constructing such templates, the bispectra should be surveyed
over the complete momentum-triangle domain. The equilateral
configuration provides a useful benchmark and the consistency
relations determine the strict soft limits, but neither identifies the
dominant intermediate configurations, possible cancellations between
interactions, or sign changes of the complete result. For efficient
data analysis, the resulting shapes should then be represented in a
separable, modal, or optimized binned basis. Masks, residual
foregrounds, weak lensing, and inhomogeneous polarization noise should
be included consistently, as emphasized in
Ref.\cite{Philcox:2024wqx}.

Future polarization surveys are particularly relevant for these
signals \cite{Shiraishi:2019tensorNG,Philcox:2024wqx}. In a squeezed
$h\zeta\zeta$ configuration, the long tensor mode is probed by
large-scale polarization, while the two hard scalar modes retain
information at higher multipoles. Full-sky, low-noise polarization
from LiteBIRD \cite{LiteBIRD:2022overview} is therefore complementary
to the deeper and higher-resolution measurements of the Simons
Observatory \cite{SimonsObservatory:2025eLAT}. Their combination,
together with delensing and multifrequency foreground removal, offers
a natural route toward improved constraints on mixed tensor--scalar
non-Gaussianity.

Several theoretical extensions are also natural. For very small
$c_S$ or extremely squeezed configurations, the logarithmic
corrections associated with the common acoustic pivot may require
resummation or an exact power-law treatment. In the nonsingular model,
the primordial correlators should ultimately be evolved through the
transition and the subsequent expanding phase to determine whether
mode mixing modifies their momentum or helicity dependence. Finally,
the same first-order formalism can be applied to other Horndeski
backgrounds and to broader scalar--tensor theories. Connecting these
model-specific primordial predictions with polarization likelihoods is
the main next step toward using mixed and tensor bispectra as
quantitative probes of the early Universe.

\section*{Acknowledgments}

The authors are grateful to Sergei Mironov and Victoria Volkova for useful comments and careful reading of the manuscript. This work was supported by the Russian Science Foundation, Grant No. 26-72-00152.

\newpage

\begin{appendices}


\section{First-order tensor mode functions}
\label{app:tensor_mode_functions}

We apply the scalar slow-variation procedure of
Ref.\cite{DeFelice:2013ar} to
Eq.\eqref{eq:tensor_quadratic_action}, with the canonical
normalization of Ref. \cite{Kobayashi:2011nu} and the stable
expanding-attractor assumptions of the main text. Terms of second
slow-variation order are omitted, and sound-time derivatives are
written explicitly to distinguish them from the conformal-time
derivatives in the interaction vertices.

Starting from $z_T$ in Eq.\eqref{eq:tensor_canonical_variables},
the logarithmic evolution of the canonical normalization $z_T$
and of the tensor sound-horizon time are
\begin{align}
\frac{d\ln z_T}{dN}
&=1+\frac12f_T-\frac12s_T,
&\frac{dy_T}{dN}&=\frac{c_T}{aH},
\nonumber\\
y_T&=-\frac{c_T}{(1-\epsilon-s_T)aH}
 \left[1+\mathcal O(\varepsilon^2)\right].
\label{eq:app_tm_sound_time}
\end{align}
The additive constant is fixed by $y_T\to0^-$ as $c_T/(aH)\to0$.
The last relation is exact for constant rates; for slowly varying
rates, the local index defined by
$d\ln z_T/dy_T=-(\nu_T-1/2)/y_T$ satisfies
\begin{align}
\nu_T
&=\frac12+\frac{1+f_T/2-s_T/2}{1-\epsilon-s_T}
 +\mathcal O(\varepsilon^2)
=\frac32+\delta_T+\mathcal O(\varepsilon^2),
\nonumber\\
\frac1{z_T}\frac{d^2z_T}{dy_T^2}
&=\frac{\nu_T^2-\frac14}{y_T^2}
 -\frac1{y_T}\frac{d\nu_T}{dy_T}
=\frac{\nu_T^2-\frac14}{y_T^2}
 +\mathcal O\!\left(\frac{\varepsilon^2}{y_T^2}\right).
\label{eq:app_tm_pump_index}
\end{align}
For strictly constant rates, the rational expression for $\nu_T$
is exact and the index derivative vanishes. The conditions
$1-\epsilon-s_T>0$ and $3-\epsilon+g_T>0$ select a shrinking
sound horizon and a decaying second superhorizon solution,
respectively, as in Ref. \cite{Kobayashi:2011nu}.

The canonical mode therefore obeys a Hankel equation through the
retained order, with the positive-frequency condition
$v_k\to \mathrm e^{-iky_T}/\sqrt{2k}$ for $-ky_T\gg1$ fixing the solution:
\begin{align}
&\frac{d^2v_k}{dy_T^2}
+\left(k^2-\frac{\nu_T^2-\frac14}{y_T^2}\right)v_k
=0,
\nonumber\\
&v_k(y_T)
=\frac{\sqrt\pi}{2}
 \mathrm e^{\frac{i\pi}{2}(\nu_T+\frac12)}
 \sqrt{-y_T}\,H_{\nu_T}^{(1)}(-ky_T).
\label{eq:app_tm_canonical_solution}
\end{align}
To recover Eq.\eqref{eq:tensor_mode_slow_variation}, divide by
$z_T$, use $-ky_T=(1+\epsilon+s_T)y_h$ to first order, and rephase
the mode by $\mathrm e^{-i\pi s_{T*}/4}$. The opposite oscillator
rephasing leaves the field unchanged, while the remaining mode
phase is $\theta_{T*}$.

To obtain the corrections used in the conformal-time integrals,
expand the background functions about the physical pivot of
Eq.\eqref{eq:modes_common_pivot}. With $L=\ln(\tau/\tau_*)$,
\begin{align}
H(\tau)&=H_*\left[1+\epsilon_*L+\mathcal O(\varepsilon^2)\right],
&F_T(\tau)&=F_{T*}\left[1-f_{T*}L+\mathcal O(\varepsilon^2)\right],
\nonumber\\
c_T(\tau)&=c_{T*}\left[1-s_{T*}L+\mathcal O(\varepsilon^2)\right],
&a(\tau)&=-\frac{1}{H_*\tau}
 \left[1+\epsilon_*\{1-L\}+\mathcal O(\varepsilon^2)\right].
\label{eq:app_tm_background_expansion}
\end{align}
In particular, the constant $\epsilon_*$ term in $a(\tau)$ is
retained: for constant $\epsilon_*$, the physical crossing occurs at
$\tau(t_*)=-[(1-\epsilon_*)K_*]^{-1}$ rather than at the logarithmic
reference $\tau_*=-K_*^{-1}$. For $x=x_{hi}$, this expansion gives
\begin{align}
y_h&=x\left[1-\epsilon_*-s_{T*}L
 +\mathcal O(\varepsilon^2)\right],
\nonumber\\
-k_i y_T&=x+\Delta x+\mathcal O(\varepsilon^2x),
&\Delta x&=s_{T*}x(1-L).
\label{eq:app_tm_argument_expansion}
\end{align}
Thus the change of the Hankel argument is distinct from the
first-order shift of its index.

Writing $\mathcal U_\nu(x)\equiv x^{3/2}H_\nu^{(1)}(x)$, the
phase-stripped mode can be organized as
\begin{align}
\widetilde h_{k_i}(\tau)
&=-\sqrt{\frac\pi2}\,A_{T,i}
 \left(1-\epsilon_*-s_{T*}+\delta_{T*}L\right)
 \mathcal U_{3/2+\delta_{T*}}(x+\Delta x)
 +\mathcal O(\varepsilon^2A_{T,i}),
\label{eq:app_tm_compact_expansion}\\
\mathcal U_{3/2+\delta_{T*}}(x+\Delta x)
&=\mathcal U_{3/2}(x)
 +\delta_{T*}\left.\partial_\nu\mathcal U_\nu(x)\right|_{3/2}
 +\Delta x\,\partial_x\mathcal U_{3/2}(x)
 +\mathcal O(\varepsilon^2),
\nonumber\\
\mathcal U_{3/2}(x)
&=-i\sqrt{\frac2\pi}\,(1-ix)\mathrm e^{ix}.
\nonumber
\end{align}
The prefactor, the argument shift, and the order derivative give the
three distinct contributions to the correction; retaining each of
them and taking the complex conjugate yields
\begin{align}
\widetilde h_{k_i}^*(\tau)
&=-iA_{T,i}\mathrm e^{-ix}\left[P(x)+\delta P_T(x)\right],
&P(x)&=1+ix,
\nonumber\\
\partial_\tau\widetilde h_{k_i}^*(\tau)
&=iA_{T,i}c_{T*}k_i \mathrm e^{-ix}\left[D(x)+\delta D_T(x)\right],
&D(x)&=x,
\label{eq:app_tm_factored_modes}
\end{align}
where the rates in the following expressions are evaluated at $t_*$:
\begin{align}
\delta P_T(x)
={}&-(\epsilon+s_T)P(x)+s_Tx^2
 +\left[\delta_TP(x)-s_Tx^2\right]L
 -i\delta_T\mathrm e^{ix}\mathfrak h^*(x),
\label{eq:app_tm_delta_P}\\
\delta D_T(x)
={}&-\epsilon x-is_Tx^2+\delta_T(1/x+i)
 +\left[(\delta_T-2s_T)x+is_Tx^2\right]L
\nonumber\\
&\hspace{3cm}-i\delta_T\mathrm e^{ix}\mathfrak h^{*\prime}(x).
\label{eq:app_tm_delta_D}
\end{align}
The derivative follows from
$\delta D_T=(\partial_x-i)\delta P_T$, with
$\partial_xL=1/x$, and uses the Hankel-order function defined in
Eq.\eqref{eq:modes_hankel_order_function}. In particular,
$\mathfrak h^{*\prime}(x)=-i/x+\mathcal O(x\ln x)$ cancels the
explicit inverse-power term in Eq.\eqref{eq:app_tm_delta_D}, so
neither term is interpreted separately at the late-time endpoint.
These expressions reproduce the tensor replacements in
Eq.\eqref{eq:tensor_mode_insertions} with their stated canonical
normalization.

Finally, the small-argument behavior
$\mathfrak h^*(x)=-i(\gamma_2+\ln x)+\mathcal O(x^2\ln x)$ and
$x=r_i(-K_*\tau)$ imply
\begin{align}
\delta P_T(x)
&=-(\epsilon_*+s_{T*})
 +\delta_{T*}\left[L-\gamma_2-\ln x\right]+o(1)
\nonumber\\
&=-(\epsilon_*+s_{T*})
 -\delta_{T*}(\gamma_2+\ln r_i)+o(1)
=E_T(k_i)+o(1).
\label{eq:app_tm_external_limit}
\end{align}
The time-dependent logarithm cancels before the superhorizon
amplitude is extracted, leaving the pivot-transfer logarithm already
present in Eq.\eqref{eq:tensor_external_mode}. Consequently, $\Delta h_{k_i}(0)=iA_{T,i}E_T(k_i)$, and the complete
tensor propagator correction is the $X=T$ specialization of
Eq.\eqref{eq:complete_propagator_correction}. Differentiation acts only on the internal endpoint,
and no additional external-mode correction is added.


\section{First-order slow-variation corrections to the scalar--scalar--tensor bispectrum}
\label{app:zeta_zeta_h_slow_variation}
\label{app:zzh_hankel_integrals}

In this appendix we evaluate the six corrections $\delta\mathcal Q_q$
entering the amplitude in
Eq.\eqref{eq:zeta_zeta_h_final_nlo_amplitude}. The pivot corrections to
the reduced coefficients are already separated in that equation. Inside
the time integrals we therefore expand only the bare coefficients $c_q(t)$,
using Eq.\eqref{eq:cubic_coefficient_additive_convention}; the reduced
normalization is defined in Eq.\eqref{eq:reduced_c_coefficients} and is not
expanded a second time.

We first perform the calculation for the unsymmetrized ordering in which
the scalar perturbations carry momenta $\bm k_1$ and $\bm k_2$, while the
tensor perturbation carries momentum $\bm k_3$. The second scalar ordering
is restored at the end by the interchange $k_1\leftrightarrow k_2$.

We use the common variables of
Eqs.\eqref{eq:modes_common_pivot},
\eqref{eq:modes_frequency_variables}, and \eqref{eq:app_zzh_y},
with $(X_1,X_2,X_3)=(S,S,T)$. The integration measure and logarithm are
therefore those implied directly by these equations, and are not
redefined in this appendix. All slow-variation parameters in this appendix are evaluated at the common
pivot. In intermediate formulas we suppress the star in order to keep the
expressions readable. Thus, for example, $\epsilon$, $f_S$, and $s_T$
below stand for $\epsilon_*$, $f_{S*}$, and $s_{T*}$. The bare coefficients follow
Eq.\eqref{eq:cubic_coefficient_additive_convention}, with the rates
$\sigma_q^{(c)}$ defined by Eq.\eqref{eq:cubic_running_rate}. The pivot
correction $\delta c_{q*}$ is already included through
Eq.\eqref{eq:reduced_c_coefficients}; only the logarithmic time dependence
contributes to $\delta\mathcal Q_q$.
The scale-factor contribution follows from
Eq.\eqref{eq:z_hh_scale_factor_power_expansion}. For the six operators in
this sector,
\begin{equation}
(p_1,p_2,p_3,p_4,p_5,p_6)=(2,1,2,1,0,2).
\end{equation}

After the constant normalizations of the leading mode functions have been
factored out, an undifferentiated conjugate mode contributes the polynomial
\begin{align}
P_i(y)
&=
1+ir_i y,
\label{eq:app_zzh_Pi}
\end{align}
whereas a differentiated conjugate mode contributes
\begin{align}
D_i(y)
&=
r_i y.
\label{eq:app_zzh_Di}
\end{align}
The product of the three oscillatory factors is
\begin{align*}
\mathrm e^{-ix_1}\mathrm e^{-ix_2}\mathrm e^{-ix_3}
&=
\mathrm e^{-i(r_1+r_2+r_3)y}
=
\mathrm e^{-iy}.
\end{align*}
The six dimensionless leading temporal functions are therefore
\begin{align}
\mathcal T_1(y)
&=
\frac{P_1P_2P_3}{y^2},
&
\mathcal T_2(y)
&=
\frac{P_1P_2D_3}{y},
\label{eq:app_zzh_T12}
\\
\mathcal T_3(y)
&=
\frac{P_1D_2D_3}{y^2},
&
\mathcal T_4(y)
&=
\frac{P_1D_2P_3}{y},
\nonumber
\\
\mathcal T_5(y)
&=
P_1P_2P_3,
&
\mathcal T_6(y)
&=
\frac{D_1D_2P_3}{y^2}.
\nonumber
\end{align}
These expressions also make the derivative structure of every vertex
transparent. In the first and fifth vertices all three internal modes are
undifferentiated. In the second vertex the tensor mode is differentiated.
In the third vertex the scalar mode with momentum $k_2$ and the tensor mode
are differentiated. In the fourth vertex only the scalar mode with
momentum $k_2$ is differentiated. Finally, in the sixth vertex both scalar
modes are differentiated.

The common projection is defined with one early-time regulator and one
late-time regulator. For a generic function $F(y)$,
\begin{align}
\mathscr R_{\alpha,\varepsilon_{\mathrm{BD}}}[F]
&\equiv
\operatorname{Re}\left[
 i\int_0^\infty dy\,
 y^\alpha \mathrm e^{-(\varepsilon_{\mathrm{BD}}+i)y}F(y)
\right],
\label{eq:app_zzh_common_regulator}
\\
\mathscr R[F]
&\equiv
\lim_{\varepsilon_{\mathrm{BD}}\to0^+}
\operatorname{Re}\left\{
\operatorname{FP}_{\alpha=0}
\left[
 i\int_0^\infty dy\,
 y^\alpha \mathrm e^{-(\varepsilon_{\mathrm{BD}}+i)y}F(y)
\right]
\right\}.
\label{eq:app_zzh_regulated_projection}
\end{align}
The complex integral is evaluated first for
$\varepsilon_{\mathrm{BD}}>0$ and sufficiently large
$\operatorname{Re}\alpha$, then analytically continued to a neighborhood
of $\alpha=0$. All terms belonging to one complete mode insertion are
combined with the same regulators before the finite part and the real
projection are taken.
The leading functions and logarithmic moments are
\begin{equation}
F_q=\mathscr R[\mathcal T_q],
\qquad
\ell_q=\mathscr R[\mathcal T_q\ln y].
\label{eq:temporal_moments}
\end{equation}

\paragraph{Master integrals and leading functions.}

The required moments follow from the standard gamma-function integral
\cite[Eq.5.9.1]{NISTDLMF}:
\begin{equation}
\mathfrak I_n(\alpha,\varepsilon_{\mathrm{BD}})
\equiv
\int_0^\infty dy\,
y^{n+\alpha}\mathrm e^{-(\varepsilon_{\mathrm{BD}}+i)y}
=
\frac{\Gamma(n+\alpha+1)}
     {(\varepsilon_{\mathrm{BD}}+i)^{n+\alpha+1}}.
\label{eq:regulated_moment_master}
\end{equation}
Initially this formula holds in its convergence domain. Analytic
continuation and finite-part extraction then define
\begin{align}
I_n
&\equiv
\lim_{\varepsilon_{\mathrm{BD}}\to0^+}
\operatorname{FP}_{\alpha=0}
\mathfrak I_n(\alpha,\varepsilon_{\mathrm{BD}}),
\nonumber\\
I_n^{\ln}
&\equiv
\lim_{\varepsilon_{\mathrm{BD}}\to0^+}
\operatorname{FP}_{\alpha=0}
\partial_\alpha\mathfrak I_n(\alpha,\varepsilon_{\mathrm{BD}}).
\label{eq:app_zzh_In_definition}
\end{align}
For $n\geq0$,
\begin{equation}
I_n=\mathrm e^{-i\pi(n+1)/2}\Gamma(n+1),
\qquad
I_n^{\ln}=I_n\left[\psi(n+1)-\frac{i\pi}{2}\right].
\end{equation}
The negative-power values required here are
\begin{align}
I_{-2}&=-\frac\pi2+i(\gamma_E-1),
&I_{-1}&=-\gamma_E-\frac{i\pi}{2},
\label{eq:app_zzh_Inegative}\\
I_{-2}^{\ln}
&=\frac\pi2(\gamma_E-1)
-i\left(1-\gamma_E+\frac{\gamma_E^2}{2}-\frac{\pi^2}{24}\right),
&
I_{-1}^{\ln}
&=\frac{\gamma_E^2}{2}-\frac{\pi^2}{24}
+\frac{i\pi\gamma_E}{2}.
\nonumber
\end{align}
For nonnegative powers we use
\begin{align}
&I_0=-i,
\quad I_1=-1,
\quad I_2=2i,
\quad I_3=6,
\quad I_4=-24i,
\nonumber\\
&I_0^{\ln}=-\frac\pi2+i\gamma_E,
\quad I_1^{\ln}=\gamma_E-1+\frac{i\pi}{2},
\quad I_2^{\ln}=\pi+i(3-2\gamma_E),
\nonumber\\
&I_3^{\ln}=11-6\gamma_E-3i\pi,
\qquad
I_4^{\ln}=-12\pi+i(24\gamma_E-50).
\label{eq:app_zzh_I4_log}
\end{align}

It is useful to introduce
\begin{equation}\label{eq:S_2S_3}
S_2
=
r_1r_2+r_1r_3+r_2r_3,
\qquad
S_3
=
r_1r_2r_3.
\end{equation}
For the first temporal function,
Eqs.\eqref{eq:app_zzh_regulated_projection} and
\eqref{eq:regulated_moment_master} give
\begin{equation}
F_1=1-S_2-S_3,
\qquad
\ell_1=1-S_3-\gamma_EF_1.
\label{eq:app_zzh_Fell1}
\end{equation}
Applying the same procedure to the remaining temporal functions gives
\begin{equation*}
    \begin{split}
F_2
&=
r_3
\left[
1+r_1+r_2+2r_1r_2
\right],
\\
F_3
&=
r_2r_3(1+r_1),
\\
F_4
&=
r_2
\left[
1+r_1+r_3+2r_1r_3
\right],
\\
F_5
&=
2
\left[
1+S_2+3S_3
\right],
\\
F_6
&=
r_1r_2(1+r_3).
    \end{split}
\end{equation*}

The logarithmic moments are evaluated by repeating exactly the same
polynomial expansion and replacing every $I_n$ by $I_n^{\ln}$. The result
is
\begin{equation*}
    \begin{split}
\ell_2
&=
-\gamma_EF_2
+
r_3
\left[
r_1+r_2+3r_1r_2
\right],
\\
\ell_3
&=
-\gamma_EF_3+S_3,
\\
\ell_4
&=
-\gamma_EF_4
+
r_2
\left[
r_1+r_3+3r_1r_3
\right],
\\
\ell_5
&=
-\gamma_EF_5
+
1+3S_2+11S_3,
\\
\ell_6
&=
-\gamma_EF_6+S_3.
    \end{split}
\end{equation*}

The dimensional leading functions $J_*^{(q)}$ are recovered from the
dimensionless $F_q$ through
\begin{align}
J_*^{(q)}=
\eta_q\Lambda_qF_q,
\label{eq:app_zzh_J_eta_Lambda}
\end{align}
where
\begin{equation*}
\eta_1=
-1,
\qquad
\eta_2=
\eta_3
=
\eta_4
=
\eta_5
=
\eta_6
=
1,
\end{equation*}
and
\begin{equation*}
    \begin{split}
        \Lambda_1
=
\frac{K_*}{H_*^2},
\quad
\Lambda_2
=
\frac{\omega_3}{H_*},
\quad
\Lambda_3
=
\frac{K_*\omega_2\omega_3}{H_*^2},
\quad
\Lambda_4
=
\frac{\omega_2}{H_*},
\quad
\Lambda_5
=
\frac{1}{K_*},
\quad
\Lambda_6
=
\frac{K_*\omega_1\omega_2}{H_*^2}.
    \end{split}
\end{equation*}
As a direct check, the third function is
\begin{align*}
J_*^{(3)}=
\frac{K_*\omega_2\omega_3}{H_*^2}
r_2r_3(1+r_1)=
\frac{\omega_2^2\omega_3^2
(K_*+\omega_1)}
{H_*^2K_*^2}=
\frac{1}{H_*^2}
\frac{
c_{S*}^2c_{T*}^2
k_2^2k_3^2
(K_*+c_{S*}k_1)
}{
K_*^2
},
\end{align*}
which agrees with the leading function used in the main text.

Combining the coefficient evolution of
Eq.\eqref{eq:cubic_coefficient_additive_convention} with the scale-factor
power in Eq.\eqref{eq:z_hh_scale_factor_power_expansion} gives
\begin{equation}
\Delta_q^{[c+a]}
=
p_q\epsilon F_q-
(\sigma_q^{(c)}+p_q\epsilon)\ell_q.
\label{eq:app_zzh_Delta_ca}
\end{equation}
The pivot correction $\delta c_{q*}$ remains in the separate reduced
coefficient of Eq.\eqref{eq:reduced_c_coefficients}.

\paragraph{External and internal mode corrections.}

The scalar and tensor index shifts are defined in
Eqs.\eqref{eq:delta_s_definition} and
\eqref{eq:tensor_hankel_index}. The external-mode corrections are those of
Eqs.\eqref{eq:gamma_2} and \eqref{eq:tensor_external_mode}, evaluated with
the fractions of this sector. Their total contribution is
\begin{equation}
E_{\mathrm{ext}}^{(\zeta\zeta h)}
=E_S(k_1)+E_S(k_2)+E_T(k_3).
\label{eq:app_zzh_Eext}
\end{equation}
It contributes $E_{\mathrm{ext}}^{(\zeta\zeta h)}J_*^{(q)}$ to every
unsymmetrized temporal function.

We now organize the internal mode corrections. The functions
$\delta P_S$, $\delta D_S$, $\delta P_T$, and $\delta D_T$ are factored
versions of the mode corrections, not additional insertions. For the
phase-stripped mode on line $i$,
\begin{align}
\Delta m_i^*(\tau)
&=-iA_i\mathrm e^{-ix_i}\delta P_{X_i}(x_i),
\nonumber\\
\partial_\tau\Delta m_i^*(\tau)
&=iA_i\omega_i\mathrm e^{-ix_i}\delta D_{X_i}(x_i).
\label{eq:factored_mode_corrections}
\end{align}
These relations follow from
Eqs.\eqref{eq:scalar_mode_correction},
\eqref{eq:scalar_mode_derivative_correction}, and
\eqref{eq:tensor_mode_insertions}. Scalar lines use $s$ and $\delta_S$,
whereas tensor lines use $s_T$ and $\delta_T$.

For an undifferentiated mode, the two cases are
\begin{subequations}
\label{eq:app_zzh_delta_P}
\begin{align}
\delta P_S(x)
={}&-(\epsilon+s)(1+ix)+sx^2
+\left[\delta_S(1+ix)-sx^2\right]\ln y
-i\delta_S\mathrm e^{ix}\mathfrak h^*(x),
\label{eq:app_zzh_delta_P_scalar}\\
\delta P_T(x)
={}&-(\epsilon+s_T)(1+ix)+s_Tx^2
+\left[\delta_T(1+ix)-s_Tx^2\right]\ln y
-i\delta_T\mathrm e^{ix}\mathfrak h^*(x).
\label{eq:app_zzh_delta_P_tensor}
\end{align}
\end{subequations}
The first terms contain the non-logarithmic normalization and argument
corrections, the terms proportional to $\ln y$ contain the running, and
the final terms arise from the shift of the Hankel index away from $3/2$.

For a differentiated mode one obtains
\begin{subequations}
\label{eq:app_zzh_delta_D}
\begin{align}
\delta D_S(x)
={}&-\epsilon x-isx^2
+\delta_S\left(\frac{1}{x}+i\right)
+\left[(\delta_S-2s)x+isx^2\right]\ln y
-i\delta_S\mathrm e^{ix}\frac{d\mathfrak h^*(x)}{dx},
\label{eq:app_zzh_delta_D_scalar}\\
\delta D_T(x)
={}&-\epsilon x-is_Tx^2
+\delta_T\left(\frac{1}{x}+i\right)
+\left[(\delta_T-2s_T)x+is_Tx^2\right]\ln y
-i\delta_T\mathrm e^{ix}\frac{d\mathfrak h^*(x)}{dx}.
\label{eq:app_zzh_delta_D_tensor}
\end{align}
\end{subequations}
Although individual terms in Eqs.\eqref{eq:app_zzh_delta_D_scalar} and
\eqref{eq:app_zzh_delta_D_tensor} contain $1/x$, the complete in-in
integral is finite after the terms belonging to the same mode correction
are combined.

The auxiliary function is defined once in
Eq.\eqref{eq:modes_hankel_order_function}. A useful sine--cosine integral
representation is
\begin{equation}\label{eq:app_zzh_hfrak_explicit}
    \begin{split}
        \mathfrak h^*(x)
=&
2\sin x
+
(\sin x-x\cos x)\operatorname{Ci}(2x)
-
(\cos x+x\sin x)\operatorname{Si}(2x)
\\
+&
i\Big[
2\cos x
+\pi\sin x
-\pi x\cos x
\\-&
(\cos x+x\sin x)\operatorname{Ci}(2x)
-
(\sin x-x\cos x)\operatorname{Si}(2x)
\Big].
    \end{split}
\end{equation}
Differentiating with respect to the argument gives
\begin{equation*}
    \begin{split}
        \frac{d\mathfrak h^*(x)}{dx}
=&
\cos x
-\frac{\sin x}{x}
+
x\sin x\,\operatorname{Ci}(2x)-
x\cos x\,\operatorname{Si}(2x)
\\
+&
i\Bigg[
\pi x\sin x
-\sin x
-\frac{\cos x}{x}
-
x\cos x\,\operatorname{Ci}(2x)
-
x\sin x\,\operatorname{Si}(2x)
\Bigg].
    \end{split}
\end{equation*}

When the Hankel-order correction is inserted on an undifferentiated leg
$i$ of vertex $q$, the corresponding finite one-dimensional function is
\begin{align}
\mathfrak H^{(P)}_{qi}
=&
\operatorname{Re}
\operatorname{FP}
\int_0^\infty
dy\,
\mathrm e^{-i(1-r_i)y}
\frac{\mathcal T_q(y)}{P_i(y)}
\mathfrak h^*(r_i y).
\label{eq:app_zzh_HP_definition}
\end{align}
For a differentiated leg,
\begin{align}
\mathfrak H^{(D)}_{qi}
=&
\operatorname{Re}
\operatorname{FP}
\int_0^\infty
dy\,
\mathrm e^{-i(1-r_i)y}
\frac{\mathcal T_q(y)}{D_i(y)}
\left.
\frac{d\mathfrak h^*(x)}{dx}
\right|_{x=r_i y}.
\label{eq:app_zzh_HD_definition}
\end{align}
The factor $\mathrm e^{-i(1-r_i)y}$ follows directly from the product of the common
phase $\mathrm e^{-iy}$ and the factor $\mathrm e^{ir_i y}$ present in the Hankel-order
correction. The same $i0$ prescription as in
Eq.\eqref{eq:app_zzh_regulated_projection} is understood.

For example, the Hankel-order insertion on scalar leg $1$ of the first
vertex is
\begin{align*}
\mathfrak H^{(P)}_{11}
=&
\operatorname{Re}
\operatorname{FP}
\int_0^\infty
dy\,
\frac{P_2P_3}{y^2}
\mathrm e^{-i(1-r_1)y}
\mathfrak h^*(r_1y).
\end{align*}
The insertion on the differentiated tensor leg of the second vertex is
\begin{align*}
\mathfrak H^{(D)}_{23}
=&
\operatorname{Re}
\int_0^\infty
dy\,
\frac{P_1P_2}{y}
\mathrm e^{-i(1-r_3)y}
\mathfrak h^{*\prime}(r_3y),
\end{align*}
where the prime on $\mathfrak h^*$ denotes differentiation with respect
to its argument.

The polynomial terms in the correction to an undifferentiated line
are described by
\begin{equation}
\begin{aligned}
X^{(P)}_{qi}
&\equiv
\mathscr R
\left[
\frac{\mathcal T_q}{P_i}x_i^2
\right],
\\
Y^{(P)}_{qi}
&\equiv
\mathscr R
\left[
\frac{\mathcal T_q}{P_i}x_i^2\ln y
\right].
\end{aligned}
\label{eq:app_zzh_XY_P_definition}
\end{equation}
Substitution of Eq.\eqref{eq:app_zzh_delta_P} then gives the complete
contribution of one undifferentiated line:
\begin{align}
\Delta^{(P)}_{qi}
&=
\begin{cases}
-\epsilon F_q
+s\left[-F_q+X^{(P)}_{qi}-Y^{(P)}_{qi}\right]
+\delta_S\left[\ell_q+\mathfrak H^{(P)}_{qi}\right],
&\text{scalar line},\\[1mm]
-\epsilon F_q
+s_T\left[-F_q+X^{(P)}_{qi}-Y^{(P)}_{qi}\right]
+\delta_T\left[\ell_q+\mathfrak H^{(P)}_{qi}\right],
&\text{tensor line}.
\end{cases}
\label{eq:app_zzh_DeltaP_general}
\end{align}
In the $\zeta\zeta h$ sector, lines $1$ and $2$ use the scalar expression,
while line $3$ uses the tensor expression.

The polynomial responses are obtained directly by expanding
$\mathcal T_q/P_i$ and applying
Eq.\eqref{eq:regulated_moment_master}. The complete set of
undifferentiated responses is as follows.
For the first vertex, if $i$, $j$, and $k$ are all distinct,
\begin{align}
X^{(P)}_{1i}
&=
r_i^2
\left[
1+r_j+r_k+2r_jr_k
\right],
\label{eq:app_zzh_XP1i}
\\
Y^{(P)}_{1i}
&=
-\gamma_EX^{(P)}_{1i}
+
r_i^2
\left[
r_j+r_k+3r_jr_k
\right].
\label{eq:app_zzh_YP1i}
\end{align}
For the two undifferentiated scalar legs of the second vertex,
\begin{align}
X^{(P)}_{21}
&=
-2r_1^2r_3(1+3r_2),
\\
X^{(P)}_{22}
&=
-2r_2^2r_3(1+3r_1),
\nonumber
\\
Y^{(P)}_{21}
&=
-\gamma_EX^{(P)}_{21}
-r_1^2r_3(3+11r_2),
\label{eq:app_zzh_YP21}
\\
Y^{(P)}_{22}
&=
-\gamma_EX^{(P)}_{22}
-r_2^2r_3(3+11r_1).
\label{eq:app_zzh_YP22}
\end{align}
For the undifferentiated scalar leg of the third vertex,
\begin{equation*}
X^{(P)}_{31}
=
-2r_1^2r_2r_3,
\qquad
Y^{(P)}_{31}
=
-\gamma_EX^{(P)}_{31}
-3r_1^2r_2r_3.
\end{equation*}
For the undifferentiated scalar and tensor legs of the fourth vertex,
\begin{equation*}
    \begin{split}
        X^{(P)}_{41}
&=
-2r_1^2r_2(1+3r_3),
\\
X^{(P)}_{43}
&=
-2r_2r_3^2(1+3r_1),
\\
Y^{(P)}_{41}
&=
-\gamma_EX^{(P)}_{41}
-r_1^2r_2(3+11r_3),
\\
Y^{(P)}_{43}
&=
-\gamma_EX^{(P)}_{43}
-r_2r_3^2(3+11r_1).
    \end{split}
\end{equation*}
For the fifth vertex, all three legs are undifferentiated. If $i$, $j$,
and $k$ are all distinct,
\begin{align*}
X^{(P)}_{5i}
=&
-2r_i^2
\left[
1+3(r_j+r_k)+12r_jr_k
\right],
\\
Y^{(P)}_{5i}
=&
-\gamma_EX^{(P)}_{5i}
-
r_i^2
\left[
3+11(r_j+r_k)+50r_jr_k
\right].
\end{align*}
Finally, for the undifferentiated tensor leg of the sixth vertex,
\begin{align*}
X^{(P)}_{63}
=
-2r_1r_2r_3^2,
\qquad
Y^{(P)}_{63}
=
-\gamma_EX^{(P)}_{63}
-3r_1r_2r_3^2.
\end{align*}

For a differentiated internal mode, we define
\begin{equation}
\begin{aligned}
X^{(D)}_{qi}
&\equiv
\mathscr R
\left[
\frac{\mathcal T_q}{D_i}
(-ix_i^2)
\right],
\\
Y^{(D)}_{qi}
&\equiv
\mathscr R
\left[
\frac{\mathcal T_q}{D_i}
\left(
\frac{1}{x_i}+i
\right)
\right],
\\
Z^{(D)}_{qi}
&\equiv
\mathscr R
\left[
\frac{\mathcal T_q}{D_i}
ix_i^2\ln y
\right].
\end{aligned}
\label{eq:app_zzh_XYZ_D_definition}
\end{equation}
Using Eq.\eqref{eq:app_zzh_delta_D}, the complete differentiated-line
correction is
\begin{align}
\Delta^{(D)}_{qi}
&=
\begin{cases}
-\epsilon F_q
+s\left[X^{(D)}_{qi}-2\ell_q+Z^{(D)}_{qi}\right]
+\delta_S\left[Y^{(D)}_{qi}+\ell_q+\mathfrak H^{(D)}_{qi}\right],
&\text{scalar line},\\[1mm]
-\epsilon F_q
+s_T\left[X^{(D)}_{qi}-2\ell_q+Z^{(D)}_{qi}\right]
+\delta_T\left[Y^{(D)}_{qi}+\ell_q+\mathfrak H^{(D)}_{qi}\right],
&\text{tensor line}.
\end{cases}
\label{eq:app_zzh_DeltaD_general}
\end{align}

For the differentiated tensor line of the second interaction, direct
application of Eq.\eqref{eq:app_zzh_regulated_projection} gives
\begin{equation}
Y_{23}^{(D)}=\frac{F_1}{r_3}.
\end{equation}
This polynomial finite part is well defined at fixed $r_3>0$. Its factor
$1/r_3$ does not by itself determine the soft behavior of the complete
insertion: the corresponding $\mathfrak H_{23}^{(D)}$ term must be combined
with it according to Eq.\eqref{eq:app_zzh_DeltaD_general} before a soft
limit is taken.

The differentiated response functions needed for the second interaction are
\begin{align}
X^{(D)}_{23}
&=
-r_3^2
\left[
1+2(r_1+r_2)+6r_1r_2
\right],
\label{eq:app_zzh_XD23}
\\
Y^{(D)}_{23}
&=
\frac{F_1}{r_3},
\label{eq:app_zzh_YD23}
\\
Z^{(D)}_{23}
&=
\gamma_EX^{(D)}_{23}
+
r_3^2
\left[
1+3(r_1+r_2)+11r_1r_2
\right].
\label{eq:app_zzh_ZD23}
\end{align}
For the differentiated scalar and tensor legs of the third interaction,
\begin{equation}
\begin{aligned}
X^{(D)}_{32}
&=
-r_2^2r_3(1+2r_1),
&
X^{(D)}_{33}
&=
-r_2r_3^2(1+2r_1),
\\
Y^{(D)}_{32}
&=
\frac{r_3}{r_2}
\left[
1-r_1r_2-\gamma_Er_3
\right],
&
Y^{(D)}_{33}
&=
\frac{r_2}{r_3}
\left[
1-r_1r_3-\gamma_Er_2
\right],
\\
Z^{(D)}_{32}
&=
\gamma_EX^{(D)}_{32}
+
r_2^2r_3(1+3r_1),
&
Z^{(D)}_{33}
&=
\gamma_EX^{(D)}_{33}
+
r_2r_3^2(1+3r_1).
\end{aligned}
\label{eq:app_zzh_D_responses_q3}
\end{equation}
For the differentiated scalar leg of the fourth vertex,
\begin{align*}
X^{(D)}_{42}
&=
-r_2^2
\left[
1+2(r_1+r_3)+6r_1r_3
\right],
\\
Y^{(D)}_{42}
&=
\frac{F_1}{r_2},
\\
Z^{(D)}_{42}
&=
\gamma_EX^{(D)}_{42}
+
r_2^2
\left[
1+3(r_1+r_3)+11r_1r_3
\right].
\end{align*}
For the two differentiated scalar legs of the sixth vertex,
\begin{align*}
X^{(D)}_{61}
&=
-r_1^2r_2(1+2r_3),
\\
X^{(D)}_{62}
&=
-r_1r_2^2(1+2r_3),
\\
Y^{(D)}_{61}
&=
\frac{r_2}{r_1}
\left[
1-r_1r_3-\gamma_Er_2
\right],
\\
Y^{(D)}_{62}
&=
\frac{r_1}{r_2}
\left[
1-r_2r_3-\gamma_Er_1
\right],
\\
Z^{(D)}_{61}
&=
\gamma_EX^{(D)}_{61}
+
r_1^2r_2(1+3r_3),
\\
Z^{(D)}_{62}
&=
\gamma_EX^{(D)}_{62}
+
r_1r_2^2(1+3r_3).
\end{align*}

\paragraph{Assembly of the six corrected temporal functions.}

Let $d_{qi}=P$ or $D$ indicate whether line $i$ is undifferentiated or
differentiated. The complete unsymmetrized correction is
\begin{equation}
\delta J_q
=
E_{\mathrm{ext}}^{(\zeta\zeta h)}J_*^{(q)}
+
\eta_q\Lambda_q
\left[
p_q\epsilon F_q
-(\sigma_q^{(c)}+p_q\epsilon)\ell_q
+\sum_{i=1}^{3}\Delta_{qi}^{(d_{qi})}
\right].
\label{eq:app_zzh_delta_J_general}
\end{equation}
The required assignments are
\begin{equation}
\label{eq:app_zzh_line_assignments}
\begin{array}{c|c|c}
q&(d_{q1},d_{q2},d_{q3})&p_q\\ \hline
1&(P,P,P)&2\\
2&(P,P,D)&1\\
3&(P,D,D)&2\\
4&(P,D,P)&1\\
5&(P,P,P)&0\\
6&(D,D,P)&2
\end{array}
\end{equation}
The insertions $\Delta^{(P)}$ and $\Delta^{(D)}$ are given in
Eqs.\eqref{eq:app_zzh_DeltaP_general} and
\eqref{eq:app_zzh_DeltaD_general}. Only one internal mode is corrected in
each term; products of two first-order corrections are omitted.

The symmetrized shape correction is finally
\begin{align}
\delta\mathcal Q_q(k_1,k_2,k_3)
=&
V^{(q)}(k_1,k_2,k_3)
\delta J_q(k_1,k_2,k_3)+
V^{(q)}(k_2,k_1,k_3)
\delta J_q(k_2,k_1,k_3).
\label{eq:app_zzh_delta_Qq}
\end{align}
Under this interchange,
\begin{equation*}
r_1
\to
r_2,
\qquad
r_3
\to
r_3.
\end{equation*}
No additional numerical factor is to be inserted: the second scalar Wick
ordering is represented precisely by the explicit interchange in
Eq.\eqref{eq:app_zzh_delta_Qq}.

Substituting Eq.\eqref{eq:app_zzh_delta_Qq} into
Eq.\eqref{eq:zeta_zeta_h_final_nlo_amplitude} completes the corrected
amplitude. The reduced coefficients are those of
Eq.\eqref{eq:reduced_c_coefficients}.

Several checks are immediate. In the strict de Sitter limit,
\begin{align*}
\epsilon=
f_S
=
s
=
f_T
=
s_T
=
\sigma_q^{(c)}
=
0,
\end{align*}
one has
\begin{align*}
\delta_S
=
\delta_T
=
E_{\mathrm{ext}}
=
0,
\end{align*}
and all functions
$\Delta_q^{[c+a]}$, $\Delta^{(P)}_{qi}$, and
$\Delta^{(D)}_{qi}$ vanish. It follows that
\begin{equation*}
\delta J_q
=
0,
\qquad
\delta\mathcal Q_q
=
0,
\end{equation*}
as required.

The dimensionless quantities
\begin{align*}
F_q,
\quad
\ell_q,
\quad
X^{(P)}_{qi},
\quad
Y^{(P)}_{qi},
\quad
X^{(D)}_{qi},
\quad
Y^{(D)}_{qi},
\quad
Z^{(D)}_{qi},
\quad
\mathfrak H^{(P)}_{qi},
\quad
\mathfrak H^{(D)}_{qi}
\end{align*}
carry no dimensions. Consequently, $\delta J_q$ has the same dimension
as $J_*^{(q)}$ for every vertex, and
$V^{(q)}\delta J_q$ has the same dimension as the corresponding leading
shape.

The normalization in Eq.\eqref{eq:zeta_zeta_h_final_nlo_amplitude} assumes that
the reduced amplitude is defined using the leading scalar power spectrum,
with all first-order external-mode corrections included explicitly in
$E_{\mathrm{ext}}$. If instead the denominator is written in terms of a
corrected power spectrum,
\begin{align*}
\mathcal P_\zeta
=
\mathcal  P_\zeta^{(0)}
(1+\delta_{\mathcal P}),
\end{align*}
then
\begin{align*}
\frac{1}{\mathcal P_\zeta^2}
=
\frac{1}{(\mathcal P_\zeta^{(0)})^2}
\left[
1-2\delta_{\mathcal P}
\right]
+
\mathcal O(\varepsilon^2),
\end{align*}
and one must supplement the shape correction by
\begin{align}
\delta\mathcal Q_q^{[\mathrm{norm}]}
=
-2\delta_{\mathcal P}\,
\mathcal S_q^{(0)}.
\label{eq:app_zzh_normalization_correction}
\end{align}
This additional term must not be included when the leading power spectrum
is used in the definition of the reduced amplitude, since in that
convention the external-mode corrections have already been accounted for
once in Eq.\eqref{eq:app_zzh_Eext}.

For the Einstein-frame $G_2+G_3$ specialization, the coefficients $c_4$
and $c_5$ vanish. The final result then reduces to the four operators
$q=1,2,3,6$, with the corresponding expressions obtained directly from
Eq.\eqref{eq:zeta_zeta_h_final_nlo_amplitude}.


\paragraph{Common regularization and analytic evaluation of the Hankel-order integrals}

The functions $\mathfrak h^*(x)$ and
$d\mathfrak h^*(x)/dx$ obtained above enter the time integrals through
convolutions with the polynomial temporal functions $\mathcal T_q(y)$. Some separate
terms in these convolutions are singular at $y=0$, even though the
complete correction to a physical mode is regular. It is therefore
important to use one and the same regulator for all terms belonging to a
given mode insertion and to remove the regulator only after these terms
have been combined.

The common regulators and the physical projection were defined in
Eqs.\eqref{eq:app_zzh_common_regulator} and
\eqref{eq:app_zzh_regulated_projection}. In the present subsection they are
kept until all pieces of a complete insertion have been combined.

For an insertion on leg $i$, it is convenient to introduce
\begin{align*}
Q_i
&\equiv
\varepsilon_{\mathrm{BD}}+i(1-r_i),
&
z_i
&\equiv
\frac{Q_i}{r_i}.
\end{align*}
Since $0<r_i<1$ for a nondegenerate frequency momentum triangle, the
quantity $1-r_i$ is positive. Moreover,
$\operatorname{Re}Q_i=\varepsilon_{\mathrm{BD}}>0$, so all Laplace
transforms below are initially well defined.

The regulated Hankel-order insertion on an undifferentiated leg is
\begin{align}
\mathfrak H_{qi}^{(P)}
\left(
\alpha,\varepsilon_{\mathrm{BD}}
\right)
\equiv&
\operatorname{Re}
\int_0^\infty
dy\,
y^\alpha
\mathrm e^{-Q_i y}
\frac{\mathcal T_q(y)}{P_i(y)}
\mathfrak h^*(r_i y).
\label{eq:app_zzh_HP_definition_1}
\end{align}
For a differentiated leg, we define
\begin{align}
\mathfrak H_{qi}^{(D)}
\left(
\alpha,\varepsilon_{\mathrm{BD}}
\right)
\equiv&
\operatorname{Re}
\int_0^\infty
dy\,
y^\alpha
\mathrm e^{-Q_i y}
\frac{\mathcal T_q(y)}{D_i(y)}
\left.
\frac{d\mathfrak h^*(x)}{dx}
\right|_{x=r_i y}.
\label{eq:app_zzh_HD_definition_1}
\end{align}
The complex integrals inside the real-part operations in
Eqs.\eqref{eq:app_zzh_HP_definition_1} and
\eqref{eq:app_zzh_HD_definition_1} admit meromorphic continuation in
$\alpha$. Their finite parts and real projections are understood
according to the prescription above. Individual terms need not be
finite at $\alpha=0$.

\paragraph{Universal Laplace transforms.}

Every function in
Eqs.\eqref{eq:app_zzh_HP_definition_1} and
\eqref{eq:app_zzh_HD_definition_1} can be reduced to two universal
transforms. We define
\begin{align*}
\mathcal A_n^{(i)}
\left(
\alpha,\varepsilon_{\mathrm{BD}}
\right)
\equiv
\int_0^\infty
dy\,
y^{\alpha+n}
\mathrm e^{-Q_i y}
\mathfrak h^*(r_i y),
\quad
\mathcal B_n^{(i)}
\left(
\alpha,\varepsilon_{\mathrm{BD}}
\right)
\equiv
\int_0^\infty
dy\,
y^{\alpha+n}
\mathrm e^{-Q_i y}
\left.
\frac{d\mathfrak h^*(x)}{dx}
\right|_{x=r_i y}.
\end{align*}
The values of $n$ required for the scalar--scalar--tensor temporal functions lie
between $-2$ and $2$.

We use the standard Bessel--Laplace transform
\cite[Eq.10.22.49]{NISTDLMF},
\begin{align}
\mathcal J_{\mu,\nu}(z)
&\equiv
\int_0^\infty dx\,x^{\mu-1}\mathrm e^{-zx}J_\nu(x)
\nonumber\\
&=
\frac{\Gamma(\mu+\nu)}
{2^\nu\Gamma(\nu+1)z^{\mu+\nu}}
{}_2F_1\!\left(
\frac{\mu+\nu}{2},
\frac{\mu+\nu+1}{2};
\nu+1;
-\frac{1}{z^2}
\right),
\label{eq:app_zzh_J_transform}
\end{align}
initially for $\operatorname{Re}z>0$ and
$\operatorname{Re}(\mu+\nu)>0$.
The notation $\mathcal J_{\mu,\nu}$ denotes an auxiliary Laplace
transform and should not be confused with the temporal functions
$J_*^{(q)}$.

Using
\begin{align*}
H_\nu^{(2)}(x)
=i\,
\frac{J_{-\nu}(x)-\mathrm e^{i\pi\nu}J_\nu(x)}{\sin(\pi\nu)},
\end{align*}
we obtain
\begin{align}
\int_0^\infty dx\,x^{\mu-1}\mathrm e^{-zx}H_\nu^{(2)}(x)
=
i\,
\frac{
\mathcal J_{\mu,-\nu}(z)
-\mathrm e^{i\pi\nu}\mathcal J_{\mu,\nu}(z)
}{\sin(\pi\nu)}.
\label{eq:app_zzh_Hankel_Laplace}
\end{align}
Differentiation with respect to the order at $\nu=3/2$ gives
\begin{align}
&\left.
\frac{\partial}{\partial\nu}
\int_0^\infty dx\,x^{\mu-1}\mathrm e^{-zx}H_\nu^{(2)}(x)
\right|_{\nu=3/2}
\nonumber\\
&=
\left.
\frac{\partial\mathcal J_{\mu,\nu}(z)}{\partial\nu}
\right|_{\nu=3/2}
+i\left.
\frac{\partial\mathcal J_{\mu,\nu}(z)}{\partial\nu}
\right|_{\nu=-3/2}
+i\pi\mathcal J_{\mu,3/2}(z).
\label{eq:app_zzh_Hankel_order_transform}
\end{align}
The derivatives in Eq.\eqref{eq:app_zzh_Hankel_order_transform}
are obtained by differentiating Eq.\eqref{eq:app_zzh_J_transform}.
Their values are defined by analytic continuation from
$\varepsilon_{\mathrm{BD}}>0$, which also fixes the logarithms and
hypergeometric functions. In particular,
\begin{equation*}
w_i=-\frac{1}{z_i^2}
\longrightarrow
\left(\frac{r_i}{1-r_i}\right)^2+i0
\qquad
(\varepsilon_{\mathrm{BD}}\to0^+),
\end{equation*}
so for $r_i>1/2$ the continuation approaches the standard cut from
above. The complete combinations are formed before the damping is
removed.

As a check, for $0<r<1$ and $r\ne1/2$,
\begin{align*}
\mathcal A_0(r)
={}&
2+\frac{r}{1-2r}
+\frac{1-3r}{(1-2r)^2}\ln\!\left(\frac{1}{2r}\right)
+\frac{i\pi}{2}(1+r),
\end{align*}
and the apparent singularity is removable:
\begin{equation*}
\mathcal A_0(1/2)=\frac{11}{4}+\frac{3\pi i}{4}.
\end{equation*}

We now return to $\mathcal A_n^{(i)}$. Changing variables according to
\begin{align*}
x
=
r_i y,
\qquad
dy
=
\frac{dx}{r_i}
\end{align*}
gives
\begin{align*}
\mathcal A_n^{(i)}
=&
\sqrt{\frac{\pi}{2}}\,
r_i^{-\alpha-n-1}
\left.
\frac{\partial}{\partial\nu}
\int_0^\infty
dx\,
x^{\alpha+n+3/2}
\mathrm e^{-z_i x}
H_\nu^{(2)}(x)
\right|_{\nu=3/2}.
\end{align*}
Define
\begin{align*}
\mu_n
&=
\alpha+n+\frac52.
\end{align*}
Using Eq.\eqref{eq:app_zzh_Hankel_order_transform}, we obtain the closed
analytic result
\begin{equation}\label{eq:app_zzh_An_analytic}
    \begin{split}
        \mathcal A_n^{(i)}
\left(
\alpha,\varepsilon_{\mathrm{BD}}
\right)
=&
\sqrt{\frac{\pi}{2}}\,
r_i^{-\alpha-n-1}
\Bigg[
\left.
\frac{\partial\mathcal J_{\mu_n,\nu}(z_i)}
{\partial\nu}
\right|_{\nu=3/2}\\
+&
i
\left.
\frac{\partial\mathcal J_{\mu_n,\nu}(z_i)}
{\partial\nu}
\right|_{\nu=-3/2}
+
i\pi
\mathcal J_{\mu_n,3/2}(z_i)
\Bigg].
    \end{split}
\end{equation}
Thus every integral involving $\mathfrak h^*$ is expressed entirely in
terms of gamma functions, digamma functions, hypergeometric functions,
and hypergeometric parameter derivatives.

The derivative transform $\mathcal B_n^{(i)}$ follows directly by
integration by parts in the initial convergence domain:
\begin{align}
\mathcal B_n^{(i)}
\left(
\alpha,\varepsilon_{\mathrm{BD}}
\right)
=
\frac{Q_i}{r_i}
\mathcal A_n^{(i)}
\left(
\alpha,\varepsilon_{\mathrm{BD}}
\right)
-
\frac{\alpha+n}{r_i}
\mathcal A_{n-1}^{(i)}
\left(
\alpha,\varepsilon_{\mathrm{BD}}
\right).
\label{eq:app_zzh_Bn_analytic}
\end{align}
The endpoint terms vanish before analytic continuation because the
common lower-endpoint regulator and the positive early-time damping are
still present. Since both sides are meromorphic in $\alpha$, the same
identity holds after continuation to $\alpha=0$.

\paragraph{Reduction of all scalar--scalar--tensor Hankel-order integrals.}

We now express every function
$\mathfrak H_{qi}^{(P,D)}$ in terms of the analytic transforms
$\mathcal A_n^{(i)}$ and $\mathcal B_n^{(i)}$. To keep the following
formulas readable, the arguments
$(\alpha,\varepsilon_{\mathrm{BD}})$ of
$\mathcal A_n^{(i)}$, $\mathcal B_n^{(i)}$, and
$\mathfrak H_{qi}^{(P,D)}$ are temporarily suppressed.

For the first vertex, let $i$, $j$, and $k$ be pairwise distinct. Then
\begin{align}
\mathfrak H_{1i}^{(P)}
=&
\operatorname{Re}
\left[
\mathcal A_{-2}^{(i)}
+
i(1-r_i)\mathcal A_{-1}^{(i)}
-
r_jr_k\mathcal A_0^{(i)}
\right].
\label{eq:app_zzh_H1i_analytic}
\end{align}
For the second vertex,
\begin{align}
\mathfrak H_{21}^{(P)}
=&
r_3
\operatorname{Re}
\left[
\mathcal A_0^{(1)}
+
ir_2\mathcal A_1^{(1)}
\right],
\\
\mathfrak H_{22}^{(P)}
=&
r_3
\operatorname{Re}
\left[
\mathcal A_0^{(2)}
+
ir_1\mathcal A_1^{(2)}
\right],
\\
\mathfrak H_{23}^{(D)}
=&
\operatorname{Re}
\left[
\mathcal B_{-1}^{(3)}
+
i(r_1+r_2)\mathcal B_0^{(3)}
-
r_1r_2\mathcal B_1^{(3)}
\right].
\end{align}
For the third vertex,
\begin{align}
\mathfrak H_{31}^{(P)}
=&
r_2r_3
\operatorname{Re}
\mathcal A_0^{(1)},
\\
\mathfrak H_{32}^{(D)}
=&
r_3
\operatorname{Re}
\left[
\mathcal B_{-1}^{(2)}
+
ir_1\mathcal B_0^{(2)}
\right],
\\
\mathfrak H_{33}^{(D)}
=&
r_2
\operatorname{Re}
\left[
\mathcal B_{-1}^{(3)}
+
ir_1\mathcal B_0^{(3)}
\right].
\end{align}
For the fourth vertex,
\begin{align}
\mathfrak H_{41}^{(P)}
=&
r_2
\operatorname{Re}
\left[
\mathcal A_0^{(1)}
+
ir_3\mathcal A_1^{(1)}
\right],
\\
\mathfrak H_{42}^{(D)}
=&
\operatorname{Re}
\left[
\mathcal B_{-1}^{(2)}
+
i(r_1+r_3)\mathcal B_0^{(2)}
-
r_1r_3\mathcal B_1^{(2)}
\right],
\\
\mathfrak H_{43}^{(P)}
=&
r_2
\operatorname{Re}
\left[
\mathcal A_0^{(3)}
+
ir_1\mathcal A_1^{(3)}
\right].
\end{align}
For the fifth vertex, with $i$, $j$, and $k$ pairwise distinct,
\begin{align}
\mathfrak H_{5i}^{(P)}
=&
\operatorname{Re}
\left[
\mathcal A_0^{(i)}
+
i(1-r_i)\mathcal A_1^{(i)}
-
r_jr_k\mathcal A_2^{(i)}
\right].
\end{align}
For the sixth vertex,
\begin{align}
\mathfrak H_{61}^{(D)}
=&
r_2
\operatorname{Re}
\left[
\mathcal B_{-1}^{(1)}
+
ir_3\mathcal B_0^{(1)}
\right],
\\
\mathfrak H_{62}^{(D)}
=&
r_1
\operatorname{Re}
\left[
\mathcal B_{-1}^{(2)}
+
ir_3\mathcal B_0^{(2)}
\right],
\\
\mathfrak H_{63}^{(P)}
=&
r_1r_2
\operatorname{Re}
\mathcal A_0^{(3)}.
\label{eq:app_zzh_H63_analytic}
\end{align}
Equations\eqref{eq:app_zzh_An_analytic},
\eqref{eq:app_zzh_Bn_analytic}, and
\eqref{eq:app_zzh_H1i_analytic}--
\eqref{eq:app_zzh_H63_analytic} eliminate all remaining
one-dimensional conformal-time integrals from the
scalar--scalar--tensor calculation.

\paragraph{Laurent poles and their cancellation.}

We next determine which regulated intermediate quantities contain
poles at $\alpha=0$. The small-argument behavior obtained from
Eq.\eqref{eq:nlf_hankel_E1} is
\begin{align*}
\mathfrak h^*(x)
&=-i\left(\gamma_2+\ln x\right)
+\mathcal O(x^2\ln x),
&
\frac{d\mathfrak h^*(x)}{dx}
&=-\frac{i}{x}+\mathcal O(x\ln x).
\end{align*}
Since $x=r_i y$, define
\begin{align*}
C_i
\equiv
\gamma_2+\ln r_i.
\end{align*}
Then
\begin{align*}
\mathfrak h^*(r_i y)
=
-i
\left[
C_i+\ln y
\right]
+
\mathcal O(y^2\ln y).
\end{align*}

The elementary integrals
\begin{align*}
\int_0^1dy\,y^{\alpha-1}
=
\frac1\alpha,
\qquad
\int_0^1dy\,y^{\alpha-1}\ln y
=
-\frac1{\alpha^2}
\end{align*}
show that
\begin{align*}
\mathcal A_{-1}^{(i)}
=
\frac{i}{\alpha^2}
-
\frac{iC_i}{\alpha}
+
\mathcal O(1).
\end{align*}
For $\mathcal A_{-2}^{(i)}$, the pole at $\alpha=0$ comes from the
first-order term in $\mathrm e^{-Q_i y}
=
1-Q_i y+\mathcal O(y^2).$
Consequently,
\begin{align*}
\mathcal A_{-2}^{(i)}
=
-\frac{iQ_i}{\alpha^2}
+
\frac{iQ_iC_i}{\alpha}
+
\mathcal O(1).
\end{align*}
For $n\geq0$,
\begin{align*}
\mathcal A_n^{(i)}
&=
\mathcal O(1)
\qquad
\text{as }\alpha\to0.
\end{align*}
Using Eq.\eqref{eq:app_zzh_Bn_analytic}, we obtain
\begin{align*}
\mathcal B_{-1}^{(i)}
=
\frac{iQ_i}{r_i\alpha}
+
\mathcal O(1),
\quad
\mathcal B_0^{(i)}
=
-\frac{i}{r_i\alpha}
+
\mathcal O(1),
\quad
\mathcal B_1^{(i)}
=
\mathcal O(1).
\end{align*}
The double poles present separately in
$\mathcal A_{-1}^{(i)}$ and
$\mathcal A_{-2}^{(i)}$ have cancelled in
$\mathcal B_{-1}^{(i)}$.

Substitution into
Eqs.\eqref{eq:app_zzh_H1i_analytic}--
\eqref{eq:app_zzh_H63_analytic} shows that only four real
Hankel-order pieces have poles at $\alpha=0$:
\begin{align}
\mathfrak H_{32}^{(D)}
\left(
\alpha,\varepsilon_{\mathrm{BD}}
\right)
=&
-\frac{r_3^2}{r_2}
\frac1\alpha
+
\mathcal O(1),
\label{eq:app_zzh_H32_pole}
\\
\mathfrak H_{33}^{(D)}
\left(
\alpha,\varepsilon_{\mathrm{BD}}
\right)
=&
-\frac{r_2^2}{r_3}
\frac1\alpha
+
\mathcal O(1),
\label{eq:app_zzh_H33_pole}
\\
\mathfrak H_{61}^{(D)}
\left(
\alpha,\varepsilon_{\mathrm{BD}}
\right)
=&
-\frac{r_2^2}{r_1}
\frac1\alpha
+
\mathcal O(1),
\label{eq:app_zzh_H61_pole}
\\
\mathfrak H_{62}^{(D)}
\left(
\alpha,\varepsilon_{\mathrm{BD}}
\right)
=&
-\frac{r_1^2}{r_2}
\frac1\alpha
+
\mathcal O(1).
\label{eq:app_zzh_H62_pole}
\end{align}
All other real functions
$\mathfrak H_{qi}^{(P,D)}$ are regular at $\alpha=0$.

The poles in
Eqs.\eqref{eq:app_zzh_H32_pole}--
\eqref{eq:app_zzh_H62_pole} are cancelled by the explicit
$\delta_S(1/x_i+i)$ or $\delta_T(1/x_i+i)$ part of the corresponding
differentiated mode correction, according to whether that line is scalar
or tensor. When the elementary functions $Y_{qi}^{(D)}$ are evaluated
with the same regulator, their singular parts are
\begin{align*}
Y_{32}^{(D)}
\left(
\alpha,\varepsilon_{\mathrm{BD}}
\right)
=&
+\frac{r_3^2}{r_2}
\frac1\alpha
+
\mathcal O(1),
\\
Y_{33}^{(D)}
\left(
\alpha,\varepsilon_{\mathrm{BD}}
\right)
=&
+\frac{r_2^2}{r_3}
\frac1\alpha
+
\mathcal O(1),
\\
Y_{61}^{(D)}
\left(
\alpha,\varepsilon_{\mathrm{BD}}
\right)
=&
+\frac{r_2^2}{r_1}
\frac1\alpha
+
\mathcal O(1),
\\
Y_{62}^{(D)}
\left(
\alpha,\varepsilon_{\mathrm{BD}}
\right)
=&
+\frac{r_1^2}{r_2}
\frac1\alpha
+
\mathcal O(1).
\end{align*}
Therefore
\begin{align*}
\operatorname{Pole}_{\alpha=0}
\left[
Y_{32}^{(D)}
+
\mathfrak H_{32}^{(D)}
\right]
&=
0,
&
\operatorname{Pole}_{\alpha=0}
\left[
Y_{33}^{(D)}
+
\mathfrak H_{33}^{(D)}
\right]
&=
0,
\\
\operatorname{Pole}_{\alpha=0}
\left[
Y_{61}^{(D)}
+
\mathfrak H_{61}^{(D)}
\right]
&=
0,
&
\operatorname{Pole}_{\alpha=0}
\left[
Y_{62}^{(D)}
+
\mathfrak H_{62}^{(D)}
\right]
&=
0.
\end{align*}
It follows that every complete differentiated-mode insertion is finite.

A direct local check can be performed for the differentiated scalar leg
$2$ of the third vertex. In this case,
\begin{align*}
\frac{\mathcal T_3}{D_2}
&=
\frac{r_3P_1}{y}.
\end{align*}
At small $y$,
\begin{align*}
\mathrm e^{-i(1-r_2)y}P_1
&=
1-ir_3y+\mathcal O(y^2),
\\
\left.
\frac{d\mathfrak h^*(x)}{dx}
\right|_{x=r_2y}
&=
-\frac{i}{r_2y}
+
\mathcal O(y\ln y).
\end{align*}
The isolated Hankel-order integrand therefore behaves as
\begin{align}
\mathrm e^{-i(1-r_2)y}
\frac{\mathcal T_3}{D_2}
\left.
\frac{d\mathfrak h^*(x)}{dx}
\right|_{x=r_2y}
=
-\frac{ir_3}{r_2y^2}
-
\frac{r_3^2}{r_2y}
+
\mathcal O(\ln y).
\label{eq:app_zzh_q3_leg2_H_singular}
\end{align}
The explicit term proportional to $1/x_2+i$ contributes, including the
outer in-in factor $i$,
\begin{align}
i\mathrm e^{-iy}
\frac{\mathcal T_3}{D_2}
\left(
\frac1{x_2}+i
\right)
=
+\frac{ir_3}{r_2y^2}
+
\frac{r_3^2}{r_2y}
+
\mathcal O(1).
\label{eq:app_zzh_q3_leg2_Y_singular}
\end{align}
Adding Eqs.\eqref{eq:app_zzh_q3_leg2_H_singular} and
\eqref{eq:app_zzh_q3_leg2_Y_singular} cancels both the $y^{-2}$ and
$y^{-1}$ terms before integration. The remaining behavior is at worst
$\mathcal O(\ln y)$ and is integrable at $y=0$.

The complete undifferentiated and differentiated insertions are therefore
defined with the same scalar and tensor cases as in
Eqs.\eqref{eq:app_zzh_DeltaP_general} and
\eqref{eq:app_zzh_DeltaD_general}. All functions in the following two
expressions carry the arguments
$(\alpha,\varepsilon_{\mathrm{BD}})$:
\begin{align}
\mathcal W_{qi}^{(P)}
&=
\begin{cases}
-\epsilon F_q+s[-F_q+X_{qi}^{(P)}-Y_{qi}^{(P)}]
+\delta_S[\ell_q+\mathfrak H_{qi}^{(P)}],
&\text{scalar line},\\
-\epsilon F_q+s_T[-F_q+X_{qi}^{(P)}-Y_{qi}^{(P)}]
+\delta_T[\ell_q+\mathfrak H_{qi}^{(P)}],
&\text{tensor line},
\end{cases}
\nonumber\\
\Delta_{qi}^{(P)}
&=
\lim_{\varepsilon_{\mathrm{BD}}\to0^+}
\operatorname{FP}_{\alpha=0}
\mathcal W_{qi}^{(P)}(\alpha,\varepsilon_{\mathrm{BD}}),
\label{eq:app_zzh_DeltaP_common_regulator}\\
\mathcal W_{qi}^{(D)}
&=
\begin{cases}
-\epsilon F_q+s[X_{qi}^{(D)}-2\ell_q+Z_{qi}^{(D)}]
+\delta_S[Y_{qi}^{(D)}+\ell_q+\mathfrak H_{qi}^{(D)}],
&\text{scalar line},\\
-\epsilon F_q+s_T[X_{qi}^{(D)}-2\ell_q+Z_{qi}^{(D)}]
+\delta_T[Y_{qi}^{(D)}+\ell_q+\mathfrak H_{qi}^{(D)}],
&\text{tensor line},
\end{cases}
\nonumber\\
\Delta_{qi}^{(D)}
&=
\lim_{\varepsilon_{\mathrm{BD}}\to0^+}
\operatorname{FP}_{\alpha=0}
\mathcal W_{qi}^{(D)}(\alpha,\varepsilon_{\mathrm{BD}}).
\label{eq:app_zzh_DeltaD_common_regulator}
\end{align}
Here the regulated versions of $F_q$, $\ell_q$, $X_{qi}$,
$Y_{qi}$, $Z_{qi}$, and $\mathfrak H_{qi}$ are obtained from their earlier
definitions by replacing $\mathscr R$ with
$\mathscr R_{\alpha,\varepsilon_{\mathrm{BD}}}$. Since the finite-part
operation is linear when the same regulator is used throughout, the
finite formulas obtained earlier remain valid. The individual pieces,
however, are only bookkeeping devices; the physical quantities are the
complete sums in
Eqs.\eqref{eq:app_zzh_DeltaP_common_regulator} and
\eqref{eq:app_zzh_DeltaD_common_regulator}.

\paragraph{Checks of regulator independence.}

For every late-time singularity that occurs in a complete insertion,
analytic continuation and the specified common-cutoff subtraction give
the same finite constant:
\begin{equation}
\begin{array}{c|c|c}
f(y)
&
\displaystyle
\operatorname{FP}_{\alpha=0}
\int_0^1dy\,y^\alpha f(y)
&
\displaystyle
\operatorname{FP}_{\rho\to0^+}
\int_\rho^1dy\,f(y)
\\ \hline
y^{-2} & -1 & -1\\
y^{-2}\ln y & -1 & -1\\
y^{-1} & 0 & 0\\
y^{-1}\ln y & 0 & 0
\end{array}
\end{equation}
Here the cutoff finite part removes the divergent powers of $\rho$ and
$\ln\rho$ using the same subtraction for all terms of one insertion.

At the early-time endpoint, polynomial oscillatory integrals are
understood in the ordered Abel sense,
\begin{equation}
\lim_{\varepsilon_{\mathrm{BD}}\to0^+}
\lim_{R\to\infty}
\int_0^Rdy\,y^n
\mathrm e^{-(\varepsilon_{\mathrm{BD}}+ia)y}
=
\frac{\Gamma(n+1)}{(ia)^{n+1}},
\qquad a>0,\quad n\geq0.
\end{equation}
At fixed $\varepsilon_{\mathrm{BD}}>0$ the upper-endpoint terms vanish;
the damping is removed only afterward. Thus omission of oscillatory
upper-endpoint terms denotes this ordered limit, not an ordinary limit
of the undamped integral. The same convention is used in integrations
by parts involving the Hankel-order functions. Degenerate frequency
limits are taken only after the regulated analytic expression has been
obtained.

The same transforms
$\mathcal A_n^{(i)}$ and
$\mathcal B_n^{(i)}$ also evaluate the Hankel-order terms in the
scalar--tensor--tensor section. For any of the seven temporal functions in that
section, the ratios $\mathcal T_q/P_i$ and $\mathcal T_q/D_i$ are finite
polynomials in $y$ and $y^{-1}$. Expanding such a ratio as
\begin{align*}
\frac{\mathcal T_q(y)}{P_i(y)}
=
\sum_n a_n y^n,
\qquad
\frac{\mathcal T_q(y)}{D_i(y)}
=
\sum_n b_n y^n
\end{align*}
immediately gives
\begin{align}
\mathfrak H_{qi}^{(P)}
=
\operatorname{Re}
\sum_n
a_n\mathcal A_n^{(i)},
\qquad
\mathfrak H_{qi}^{(D)}
=
\operatorname{Re}
\sum_n
b_n\mathcal B_n^{(i)}.
\label{eq:app_zzh_hankel_general_reduction}
\end{align}
No additional special functions or time integrations are required for
that sector.


\section{First-order slow-variation corrections
to the scalar--tensor--tensor bispectrum}
\label{app:zeta_h_h_slow_variation}

In this appendix, we evaluate the seven corrected shape functions
$\delta\mathcal Q_{q,\lambda_2\lambda_3}^{(\zeta hh)}$ entering the
complete amplitude in Eq.\eqref{eq:z_hh_final_nlo_amplitude}. The
pivot-value corrections to the coefficients are already separated in
that equation and are not written a second time.

We first work with a fixed ordering in which the scalar perturbation
carries momentum $\bm k_1$, while the tensor perturbations carry momenta
$\bm k_2$ and $\bm k_3$. The second tensor ordering will be restored
at the end by the interchange $(k_2,\lambda_2)\to(k_3,\lambda_3)$.

\paragraph{Common frequency variables and leading temporal functions.}

We use the variables in Eqs.\eqref{eq:modes_common_pivot},
\eqref{eq:modes_frequency_variables}, and \eqref{eq:app_zzh_y}, with
$(X_1,X_2,X_3)=(S,T,T)$. In this assignment,
Eq.\eqref{eq:modes_common_pivot} gives
$K_*=c_{S*}k_1+c_{T*}(k_2+k_3)$. The mode polynomials $P_i$ and $D_i$
are those of Eqs.\eqref{eq:app_zzh_Pi} and \eqref{eq:app_zzh_Di}; the
common oscillatory factor is $\mathrm e^{-iy}$, as in the preceding appendix.

The seven cubic operators lead to the following dimensionless temporal
functions:
\begin{align}
\mathcal T_1(y)
&=
\frac{P_1D_2D_3}{y^2},
&
\mathcal T_2(y)
&=
\frac{P_1P_2P_3}{y^2},
\label{eq:app_zhh_T12}
\\
\mathcal T_3(y)
&=
\frac{D_1D_2P_3}{y^2},
&
\mathcal T_4(y)
&=
\frac{D_1D_2D_3}{y},
\label{eq:app_zhh_T34}
\\
\mathcal T_5(y)
&=
P_1D_2D_3,
&
\mathcal T_6(y)
&=
\frac{D_1D_2D_3}{y},
\nonumber
\\
\mathcal T_7(y)
&=
P_1D_2D_3.
\nonumber
\end{align}

For the third vertex, the expression in
\eqref{eq:app_zhh_T34} corresponds to the ordering in which the tensor
mode carrying momentum $k_2$ is differentiated. The second ordering is
generated by the interchange $(k_2,\lambda_2)\to(k_3,\lambda_3)$.

Notice that the first three temporal functions already appeared in the
calculation of the $\zeta\zeta h$ bispectrum:
\begin{equation}
\mathcal T_1^{(\zeta hh)}
=
\mathcal T_3^{(\zeta\zeta h)},
\qquad
\mathcal T_2^{(\zeta hh)}
=
\mathcal T_1^{(\zeta\zeta h)},
\qquad
\mathcal T_3^{(\zeta hh)}
=
\mathcal T_6^{(\zeta\zeta h)}.
\label{eq:app_zhh_temporal_correspondence}
\end{equation}
Therefore, the corresponding polynomial response functions for the first
three vertices can be taken directly from
Appendix\ref{app:zeta_zeta_h_slow_variation}, with the vertex labels
mapped according to \eqref{eq:app_zhh_temporal_correspondence}.

The real projection and the moments $F_q,\ell_q$ are those defined in
Eqs.\eqref{eq:app_zzh_regulated_projection} and
\eqref{eq:temporal_moments}. All required polynomial integrals follow
from Eqs.\eqref{eq:app_zzh_In_definition}--\eqref{eq:app_zzh_I4_log}.
Using \eqref{eq:S_2S_3} we obtain
\begin{align}
F_1
&=
r_2r_3(1+r_1),
&
\ell_1
&=
-\gamma_EF_1+S_3,
\label{eq:app_zhh_Fell1}
\\
F_2
&=
1-S_2-S_3,
&
\ell_2
&=
1-S_3-\gamma_EF_2,
\label{eq:app_zhh_Fell2}
\\
F_3
&=
r_1r_2(1+r_3),
&
\ell_3
&=
-\gamma_EF_3+S_3,
\nonumber
\\
F_4
&=
-2S_3,
&
\ell_4
&=
(2\gamma_E-3)S_3,
\nonumber
\\
F_5
&=
-2r_2r_3(1+3r_1),
&
\ell_5
&=
-\gamma_EF_5
-r_2r_3(3+11r_1),
\nonumber
\\
F_6
&=
F_4,
&
\ell_6
&=
\ell_4,
\nonumber
\\
F_7
&=
F_5,
&
\ell_7
&=
\ell_5.
\nonumber
\end{align}

The dimensionful leading functions used in the main text can conveniently
be written as
\begin{align}
I_*^{(q)}
&=
\eta_q\Lambda_qF_q,
\label{eq:app_zhh_I_eta_Lambda}
\end{align}
where
\begin{equation}
    (\eta_1,\eta_2,\eta_3,\eta_4,\eta_5,\eta_6,\eta_7)=(1,-1,1,-1,-1,-1,-1)\label{eq:app_zhh_eta}
\end{equation}
and
\begin{align*}
\Lambda_1
&=
\frac{K_*\omega_2\omega_3}{H_*^2},
&
\Lambda_2
&=
\frac{K_*}{H_*^2},
\\
\Lambda_3
&=
\frac{K_*\omega_1\omega_2}{H_*^2},
&
\Lambda_4
&=
\frac{\omega_1\omega_2\omega_3}{H_*},
\\
\Lambda_5
&=
\frac{\omega_2\omega_3}{K_*},
&
\Lambda_6
&=
\Lambda_4,
&
\Lambda_7
&=
\Lambda_5.
\end{align*}

Substitution of $r_i=\omega_i/K_*$ into
Eq.\eqref{eq:app_zhh_I_eta_Lambda} reproduces all seven leading
functions in Eq.\eqref{eq:z_hh_I1}; no separate example is needed.

\paragraph{Corrections from the coefficients and the scale factor.}

The bare coefficients follow the additive expansion in
Eq.\eqref{eq:cubic_coefficient_additive_convention}, with $g_q=b_q$
and logarithmic rate $\sigma_q^{(b)}$. Their pivot-value corrections
enter through the reduced coefficients in
Eq.\eqref{eq:reduced_b_coefficients}; only the logarithmic variation of
$b_q(\tau)$ enters the corrected temporal function.

The powers of the scale factor for the seven operators are
\begin{align*}
(p_1,p_2,p_3,p_4,p_5,p_6,p_7)
&=
(2,2,2,1,0,1,0).
\end{align*}
Using the general expansion
\eqref{eq:z_hh_scale_factor_power_expansion}, we obtain the combined correction from
the time dependence of the coefficient and the scale factor:
\begin{align}
\Delta_q^{[b+a]}
&=
p_q\epsilon F_q
-
\left(
\sigma_q^{(b)}+p_q\epsilon
\right)\ell_q.
\label{eq:app_zhh_Delta_ba}
\end{align}

In particular, for the fifth and seventh vertices there is no separate
scale-factor correction
$
\Delta_5^{[a]}
=
\Delta_7^{[a]}
=
0.
$

\paragraph{Boundary and bulk mode corrections.}

The index shifts are defined in Eqs.\eqref{eq:delta_s_definition}
and \eqref{eq:tensor_hankel_index}. Momentum-independent normalization
phases cancel within each propagator and are not counted as separate
first-order corrections.

The boundary corrections follow from Eqs.\eqref{eq:gamma_2} and
\eqref{eq:tensor_external_mode}, evaluated with the fractions of the
present sector. Their common sum is
\begin{equation}
E_{\mathrm{ext}}^{(\zeta hh)}
=E_S(k_1)+E_T(k_2)+E_T(k_3).
\label{eq:app_zhh_Eext}
\end{equation}
Its contribution to the unsymmetrized temporal function is
$E_{\mathrm{ext}}^{(\zeta hh)}I_*^{(q)}$.

No new mode-function expansions are required for the bulk-mode
corrections. The universal corrections $\delta P_X$ and $\delta D_X$
were already obtained in \eqref{eq:app_zzh_delta_P} and
\eqref{eq:app_zzh_delta_D}, while the response functions associated
with the shift of the Hankel index,
$\mathfrak H^{(P)}$ and $\mathfrak H^{(D)}$, were defined in
\eqref{eq:app_zzh_HP_definition_1} and
\eqref{eq:app_zzh_HD_definition_1}. Therefore, the general formulas for
the insertion of the correction on an undifferentiated and differentiated
leg remain those of \eqref{eq:app_zzh_DeltaP_general} and
\eqref{eq:app_zzh_DeltaD_general}, respectively.

Only the field content changes: line $1$ is scalar and therefore uses
the scalar cases in Eqs.\eqref{eq:app_zzh_DeltaP_general} and
\eqref{eq:app_zzh_DeltaD_general}, while lines $2$ and $3$ are tensor and
use the tensor cases.

For $q=1,2,3$, the polynomial and Hankel-order responses follow from
the scalar--scalar--tensor results through
Eq.\eqref{eq:app_zhh_temporal_correspondence}, with this scalar/tensor
assignment. 

The fourth and sixth vertices have the same temporal function,
\begin{align*}
\mathcal T_4
=
\mathcal T_6
=
\frac{D_1D_2D_3}{y}.
\end{align*}
All three legs are differentiated in this case. For pairwise distinct
$i,j,k$, the additional response functions are
\begin{align}
X^{(D)}_{4i}
&=
X^{(D)}_{6i}
=
6r_i^2r_jr_k,
\nonumber
\\
Y^{(D)}_{4i}
&=
Y^{(D)}_{6i}
=
\frac{r_jr_k}{r_i}(1+r_i),
\nonumber
\\
Z^{(D)}_{4i}
&=
Z^{(D)}_{6i}
=
(6\gamma_E-11)r_i^2r_jr_k.
\label{eq:app_zhh_DDD_responses}
\end{align}

The fifth and seventh interactions also have the same temporal function,
\begin{align*}
\mathcal T_5
=
\mathcal T_7
=
P_1D_2D_3.
\end{align*}
The scalar leg is undifferentiated, while both tensor legs are
differentiated. For the scalar leg, we obtain
\begin{align*}
X^{(P)}_{51}
=
X^{(P)}_{71}
=
24r_1^2r_2r_3,
\quad
Y^{(P)}_{51}
=
Y^{(P)}_{71}
=
-\gamma_E X^{(P)}_{51}
+
50r_1^2r_2r_3.
\end{align*}

For the differentiated tensor leg $2$, we have
\begin{align*}
X^{(D)}_{52}
&=
X^{(D)}_{72}
=
6r_2^2r_3(1+4r_1),
\\
Y^{(D)}_{52}
&=
Y^{(D)}_{72}
=
\frac{r_3}{r_2}
\left[
1+r_1+r_2+2r_1r_2
\right],
\\
Z^{(D)}_{52}
&=
Z^{(D)}_{72}
=
\gamma_E X^{(D)}_{52}
-
r_2^2r_3(11+50r_1).
\end{align*}

Similarly, for leg $3$,
\begin{align*}
X^{(D)}_{53}
&=
X^{(D)}_{73}
=
6r_2r_3^2(1+4r_1),
\\
Y^{(D)}_{53}
&=
Y^{(D)}_{73}
=
\frac{r_2}{r_3}
\left[
1+r_1+r_3+2r_1r_3
\right],
\\
Z^{(D)}_{53}
&=
Z^{(D)}_{73}
=
\gamma_E X^{(D)}_{53}
-
r_2r_3^2(11+50r_1).
\end{align*}

\paragraph{Explicit reduction of the Hankel-order responses.}

The Hankel-order responses are obtained using
Eq.\eqref{eq:app_zzh_hankel_general_reduction}.
For $q=1,2,3$, they follow the temporal-function correspondence in
Eq.\eqref{eq:app_zhh_temporal_correspondence}, evaluated with the
frequency fractions and field-species assignments of the present
sector.

For the remaining vertices, we give the reductions explicitly.
The common regulator arguments
$(\alpha,\varepsilon_{\mathrm{BD}})$ are suppressed below.
Analytic continuation, finite-part extraction, and real projection
are understood according to the common-regulator prescription.

For $q=4,6$, all three legs are differentiated. Since
\begin{equation*}
\frac{\mathcal T_4}{D_i}
=
\frac{\mathcal T_6}{D_i}
=
\frac{D_jD_k}{y}
=
r_jr_k y,
\qquad
\{i,j,k\}=\{1,2,3\},
\end{equation*}
we obtain
\begin{equation}
\mathfrak H_{4i}^{(D)}
=
\mathfrak H_{6i}^{(D)}
=
r_jr_k\operatorname{Re}\mathcal B_1^{(i)}.
\label{eq:app_zhh_HDDD}
\end{equation}
Here $i,j,k$ are pairwise distinct, and no summation is implied.

For $q=5,7$, the scalar leg is undifferentiated and both tensor
legs are differentiated. The required ratios are
\begin{align*}
\frac{\mathcal T_5}{P_1}
=
\frac{\mathcal T_7}{P_1}
&=
r_2r_3 y^2,
\\
\frac{\mathcal T_5}{D_2}
=
\frac{\mathcal T_7}{D_2}
&=
r_3 y+i r_1r_3 y^2,
\\
\frac{\mathcal T_5}{D_3}
=
\frac{\mathcal T_7}{D_3}
&=
r_2 y+i r_1r_2 y^2.
\end{align*}
Consequently,
\begin{align*}
\mathfrak H_{51}^{(P)}
=
\mathfrak H_{71}^{(P)}
&=
r_2r_3\operatorname{Re}\mathcal A_2^{(1)},
\\
\mathfrak H_{52}^{(D)}
=
\mathfrak H_{72}^{(D)}
&=
r_3\operatorname{Re}
\left[
\mathcal B_1^{(2)}
+i r_1\mathcal B_2^{(2)}
\right],
\\
\mathfrak H_{53}^{(D)}
=
\mathfrak H_{73}^{(D)}
&=
r_2\operatorname{Re}
\left[
\mathcal B_1^{(3)}
+i r_1\mathcal B_2^{(3)}
\right].
\end{align*}

The transforms needed here are already covered by
Eqs.\eqref{eq:app_zzh_An_analytic} and
\eqref{eq:app_zzh_Bn_analytic}; the required $\mathcal B_1^{(i)}$ and
$\mathcal B_2^{(i)}$ follow by setting $n=1,2$ in the latter identity.
Thus no additional independent time integral or special function is
required.

These formulas specify the Hankel-order response functions already
entering $\Delta_{qi}^{(P)}$ and $\Delta_{qi}^{(D)}$ through
Eqs.\eqref{eq:app_zzh_DeltaP_general} and
\eqref{eq:app_zzh_DeltaD_general}; they do not constitute additional
terms beyond those insertions.

\paragraph{Assembly of the seven corrected temporal functions.}

Let $d_{qi}=P$ or $D$ denote whether the mode on leg $i$ is
undifferentiated or differentiated. The assignments are
\begin{equation}
\label{eq:app_zhh_line_assignments}
\begin{array}{c|c|c}
q&(d_{q1},d_{q2},d_{q3})&p_q\\ \hline
1&(P,D,D)&2\\
2&(P,P,P)&2\\
3&(D,D,P)&2\\
4&(D,D,D)&1\\
5&(P,D,D)&0\\
6&(D,D,D)&1\\
7&(P,D,D)&0
\end{array}
\end{equation}
With one corrected mode at a time, the unsymmetrized result is
\begin{align}
\delta I_q
={}&
E_{\mathrm{ext}}^{(\zeta hh)}I_*^{(q)}
+
\eta_q\Lambda_q
\left[
p_q\epsilon F_q
-
\left(\sigma_q^{(b)}+p_q\epsilon\right)\ell_q
+
\sum_{i=1}^{3}\Delta_{qi}^{(d_{qi})}
\right],
\qquad q=1,\ldots,7.
\label{eq:app_zhh_delta_I_explicit}
\end{align}
The insertions $\Delta^{(P)}$ and $\Delta^{(D)}$ are those of
Eqs.\eqref{eq:app_zzh_DeltaP_general} and
\eqref{eq:app_zzh_DeltaD_general}; products of two first-order
corrections are omitted.

The symmetrized correction to the shape function is
\begin{align}
\delta\mathcal Q_{q,\lambda_2\lambda_3}^{(\zeta hh)}
=&
V_{\lambda_2\lambda_3}^{(q)}
(k_1,k_2,k_3)
\,
\delta I_q(k_1,k_2,k_3)
+
V_{\lambda_3\lambda_2}^{(q)}
(k_1,k_3,k_2)
\,
\delta I_q(k_1,k_3,k_2).
\label{eq:app_zhh_delta_Q}
\end{align}
Equation\eqref{eq:app_zhh_delta_Q} is the single symmetrization
formula used below. Under this interchange, $r_2\leftrightarrow r_3$.

Substitution of Eqs.\eqref{eq:app_zhh_delta_I_explicit} and
\eqref{eq:app_zhh_delta_Q} into Eq.\eqref{eq:z_hh_final_nlo_amplitude}
completes the scalar--tensor--tensor calculation; the full amplitude is
not written a second time.

\section{First-order slow-variation corrections
to the three-tensor bispectrum}
\label{app:h_h_h_slow_variation}

In this appendix, we evaluate the two corrected shape functions
$\delta\mathcal Q_{\mathrm{new}}$ and $\delta\mathcal Q_{\mathrm{GR}}$
that enter the complete amplitude in
Eq.\eqref{eq:hhh_final_nlo_amplitude}. The correction to the pivot value
of the additional interaction coefficient is already separated there.

As in the previous appendices, within the time integrals we expand the
coefficients of the cubic action themselves. For the new vertex, this
coefficient is $\mu_5(t)$, whereas for the universal GR vertex it is
$F_T(t)$.

\paragraph{Common frequency variables and leading temporal functions.}

All three species are tensor modes. We use
Eqs.\eqref{eq:modes_common_pivot},
\eqref{eq:modes_frequency_variables}, and \eqref{eq:app_zzh_y}, with
$K_*=K_{T*}=c_{T*}K$ and $r_i=k_i/K$. The polynomials $P_i$ and $D_i$
are those of Eqs.\eqref{eq:app_zzh_Pi} and \eqref{eq:app_zzh_Di}.

For the additional interaction $h'_{ij}h'_{jk}h'_{ki}$,
all three tensor modes are differentiated, while in conformal time the
vertex contains one power of the scale factor. Therefore, its
dimensionless temporal function is
\begin{align}
\mathcal T_{\mathrm{new}}(y)
=
\frac{D_1D_2D_3}{y}.
\label{eq:app_hhh_Tnew}
\end{align}
For the universal general-relativistic vertex, all three modes entering
the temporal part are undifferentiated, while the vertex contains
$a^2$. The corresponding temporal function is
\begin{align}
\mathcal T_{\mathrm{GR}}(y)
=
\frac{P_1P_2P_3}{y^2}.
\label{eq:app_hhh_TGR}
\end{align}
Thus, no new types of temporal integrals arise in the $hhh$ sector.
In particular,
\begin{align*}
\mathcal T_{\mathrm{GR}}^{(hhh)}
&=
\mathcal T_1^{(\zeta\zeta h)},
\end{align*}
whereas
\begin{align*}
\mathcal T_{\mathrm{new}}^{(hhh)}
&=
\mathcal T_4^{(\zeta hh)}
=
\mathcal T_6^{(\zeta hh)}.
\end{align*}
The moments are defined once in
Eqs.\eqref{eq:temporal_moments} and
\eqref{eq:app_zzh_regulated_projection}. Using
Eq.\eqref{eq:S_2S_3}, their values in the present sector are
\begin{align*}
F_{\mathrm{new}}
&=-2S_3,
&
\ell_{\mathrm{new}}
&=(2\gamma_E-3)S_3,
\\
F_{\mathrm{GR}}
&=1-S_2-S_3,
&
\ell_{\mathrm{GR}}
&=1-S_3-\gamma_EF_{\mathrm{GR}}.
\end{align*}
The first pair is the $\zeta hh$ result in
Eq.\eqref{eq:app_zhh_temporal_correspondence} for the all-differentiated
case, while the second pair is the $q=1$ scalar--scalar--tensor result.

It is convenient to express the dimensionful leading temporal functions
as
\begin{align*}
I_*^{(r)}
&=
\eta_r\Lambda_rF_r,
\qquad
r=\mathrm{new},\mathrm{GR},
\end{align*}
where
\begin{align*}
\eta_{\mathrm{new}}
=
-1,
\qquad
\Lambda_{\mathrm{new}}
=
\frac{k_1k_2k_3}{2},
\end{align*}
\begin{align*}
\eta_{\mathrm{GR}}
=
-1,
\qquad
\Lambda_{\mathrm{GR}}
=
\frac{K}{16}.
\end{align*}

Substituting $r_i=k_i/K$ into the preceding formulas reproduces the
temporal factors in the leading shapes defined in the main text. Thus,
\begin{align*}
\mathcal S_{\lambda_1\lambda_2\lambda_3,\mathrm{new}}^{(0)}
=
V_{\lambda_1\lambda_2\lambda_3}^{(\mathrm{new})}
I_*^{(\mathrm{new})},
\qquad
\mathcal S_{\lambda_1\lambda_2\lambda_3,\mathrm{GR}}^{(0)}
=
V_{\lambda_1\lambda_2\lambda_3}^{(\mathrm{GR})}
I_*^{(\mathrm{GR})}.
\end{align*}

\paragraph{Corrections from the cubic-vertex coefficients and the scale factor.}

All slow-variation parameters below are understood to be evaluated at
the common pivot $\tau_*$. For brevity, the star is omitted in
intermediate expressions.

The additional coefficient follows the additive convention in
Eq.\eqref{eq:cubic_coefficient_additive_convention}, with
$g=\mu_5$ and logarithmic rate $\sigma_{\mu_5}$. Its pivot-value
correction is contained in the reduced coefficient of
Eq.\eqref{eq:reduced_tensor_coefficient}. For the GR interaction, the
coefficient variation is $F_T(\tau)=F_{T*}[1-f_T\ln y]
+\mathcal O(\varepsilon^2F_T)$, as follows from
Eq.\eqref{eq:tensor_slow_rates}.

The powers of the scale factor are
\begin{align*}
p_{\mathrm{new}}
=
1,
\qquad
p_{\mathrm{GR}}
=
2.
\end{align*}
Using the general expansion
\eqref{eq:z_hh_scale_factor_power_expansion}, we obtain
\begin{align*}
\Delta_{\mathrm{new}}^{[\mu_5+a]}
=
\epsilon F_{\mathrm{new}}
-
\left(
\sigma_{\mu_5}+\epsilon
\right)
\ell_{\mathrm{new}},
\qquad
\Delta_{\mathrm{GR}}^{[F_T+a]}
=
2\epsilon F_{\mathrm{GR}}
-
\left(
f_T+2\epsilon
\right)
\ell_{\mathrm{GR}}.
\end{align*}

\paragraph{Boundary and bulk corrections to the tensor modes.}

The tensor index shift and external correction are given in
Eqs.\eqref{eq:tensor_hankel_index} and
\eqref{eq:tensor_external_mode}. Hence the common external contribution is
\begin{equation}
E_{\mathrm{ext}}^{(hhh)}
=E_T(k_1)+E_T(k_2)+E_T(k_3),
\label{eq:app_hhh_Eext}
\end{equation}
and it multiplies $I_*^{(r)}$ for
$r=\mathrm{new},\mathrm{GR}$.

No new tensor-mode expansions are required for the bulk corrections.
All three lines are tensor lines, so the tensor cases in
Eqs.\eqref{eq:app_zzh_DeltaP_common_regulator} and
\eqref{eq:app_zzh_DeltaD_common_regulator} are used throughout.

For the GR interaction, all three lines are undifferentiated, and its
temporal function is the $q=1$ scalar--scalar--tensor function. Its
polynomial and Hankel-order responses are therefore exactly those of
Eqs.\eqref{eq:app_zzh_XP1i},
\eqref{eq:app_zzh_YP1i}, and
\eqref{eq:app_zzh_H1i_analytic}, with $s_T$ and $\delta_T$.

For the additional interaction, all three lines are differentiated. The
polynomial responses and the Hankel-order response are, respectively,
Eqs.\eqref{eq:app_zhh_DDD_responses} and
\eqref{eq:app_zhh_HDDD}, again with $s_T$ and $\delta_T$. The complete
line corrections are then obtained directly from
Eqs.\eqref{eq:app_zzh_DeltaP_general} and
\eqref{eq:app_zzh_DeltaD_general}; no extra response term is added.

In contrast to some differentiated insertions in the
$\zeta\zeta h$ sector, no new singular Hankel-order structures arise in
the present case. For the new vertex, only
$\mathcal B_1^{(i)}$, which is regular at $\alpha=0$, is required, while
for the GR interaction the combination in
Eq.\eqref{eq:app_zzh_H1i_analytic} is already regular.
Nevertheless, all terms are understood within the common regularization
scheme
\eqref{eq:app_zzh_common_regulator}--
\eqref{eq:app_zzh_regulated_projection}.

\paragraph{Assembly of the corrected temporal functions.}

Define
\begin{align*}
\Delta_{\mathrm{new}}^{[\mathrm{bulk}]}
&=\sum_{i=1}^{3}\Delta_{\mathrm{new},i}^{(D)},
&
\Delta_{\mathrm{GR}}^{[\mathrm{bulk}]}
&=\sum_{i=1}^{3}\Delta_{\mathrm{GR},i}^{(P)}.
\end{align*}
Only one propagator is corrected in each term. The two temporal
corrections are
\begin{align}
\delta I_{\mathrm{new}}
={}&E_{\mathrm{ext}}^{(hhh)}I_*^{(\mathrm{new})}
+\eta_{\mathrm{new}}\Lambda_{\mathrm{new}}
\left[
\epsilon F_{\mathrm{new}}
-(\sigma_{\mu_5}+\epsilon)\ell_{\mathrm{new}}
+\Delta_{\mathrm{new}}^{[\mathrm{bulk}]}
\right],
\nonumber\\
\delta I_{\mathrm{GR}}
={}&E_{\mathrm{ext}}^{(hhh)}I_*^{(\mathrm{GR})}
+\eta_{\mathrm{GR}}\Lambda_{\mathrm{GR}}
\left[
2\epsilon F_{\mathrm{GR}}
-(f_T+2\epsilon)\ell_{\mathrm{GR}}
+\Delta_{\mathrm{GR}}^{[\mathrm{bulk}]}
\right].
\label{eq:app_hhh_delta_I}
\end{align}

The corresponding corrections to the shape functions are
\begin{align}
\delta\mathcal Q_{\mathrm{new}}
=
V_{\lambda_1\lambda_2\lambda_3}^{(\mathrm{new})}
\delta I_{\mathrm{new}},
\qquad
\delta\mathcal Q_{\mathrm{GR}}
=
V_{\lambda_1\lambda_2\lambda_3}^{(\mathrm{GR})}
\delta I_{\mathrm{GR}}.
\label{eq:app_hhh_delta_Q}
\end{align}

Unlike the mixed correlators, no additional symmetrization of the form
$k_i\leftrightarrow k_j$ is required here: the necessary permutations
are already included in the definitions of the tensor structures
$V^{(\mathrm{new})}$ and $V^{(\mathrm{GR})}$.

Equations\eqref{eq:app_hhh_delta_I} and
\eqref{eq:app_hhh_delta_Q}, together with
Eq.\eqref{eq:hhh_final_nlo_amplitude}, complete the three-tensor
calculation. The full amplitude is not written a second time.


\section{Explicit evaluation of the nonlinearity parameters}
\label{app:nonlinearity_evaluation}

This appendix gives the elementary functions used in
Eq.\eqref{eq:nlf_effective_bare_coefficients} and the polarization
contractions needed for the results in
Sec.\ref{sec:nonlinearity_nlo}. The calculation does not introduce
additional interactions or mode corrections, but combines the terms
already evaluated in Appendices\ref{app:zeta_zeta_h_slow_variation},
\ref{app:zeta_h_h_slow_variation}, and
\ref{app:h_h_h_slow_variation} into expressions that can be substituted
directly into the nonlinearity parameters.

\subsection{Temporal functions and complete mode insertions}
\label{app:nlf_temporal}
We use the common dimensionless time and frequency fractions and other useful notations of the
preceding main text and appendices, namely \eqref{eq:modes_common_pivot}, \eqref{eq:modes_frequency_variables}, and \eqref{eq:app_zzh_Di}.
We recall that the factor $P_i$ represents an undifferentiated conjugate mode, while
$D_i$ represents a differentiated mode after its constant normalization
and derivative prefactor have been removed. The real projection
$\mathscr R$ is the common-regulator projection of
Appendix\ref{app:zzh_hankel_integrals}, with all terms of a complete
insertion combined before the finite part and the Bunch--Davies limit
are taken.

Only six types of temporal polynomial are required for the thirteen
mixed operators and the two tensor operators. We denote them by
\begin{align}
 \mathcal T_{\mathsf A}
 &=\frac{P_1P_2P_3}{y^2},
 &\mathcal T_{\mathsf B_d}
 &=\frac{P_aP_bD_d}{y},
 &\mathcal T_{\mathsf C_u}
 &=\frac{P_uD_vD_w}{y^2},
 \nonumber\\
 \mathcal T_{\mathsf D}
 &=P_1P_2P_3,
 &\mathcal T_{\mathsf E}
 &=\frac{D_1D_2D_3}{y},
 &\mathcal T_{\mathsf F_u}
 &=P_uD_vD_w.
 \label{eq:temporal_function_basis}
\end{align}
The indices in $(a,b,d)$ or $(u,v,w)$ are distinct and form a
permutation of $(1,2,3)$; thus $d$ labels the differentiated leg of
$\mathsf B_d$, whereas $u$ labels the undifferentiated leg of
$\mathsf C_u$ or $\mathsf F_u$. These letters label temporal functions
only and do not change the operator numbering of the cubic actions.

For each temporal function, the moments $F$ and $\ell$ are those of
Eq.\eqref{eq:temporal_moments}. Their polynomial integrals are evaluated
from the master formula \eqref{eq:regulated_moment_master} and the values
listed in Eqs.\eqref{eq:app_zzh_Inegative}--\eqref{eq:app_zzh_I4_log}.
Applying these results to the six functions in
Eq.\eqref{eq:temporal_function_basis} yields
\begin{align}
 F_{\mathsf A}&=1-S_2-S_3,
 &\ell_{\mathsf A}
 &=1-S_3-\gamma_EF_{\mathsf A},
 \nonumber\\
 F_{\mathsf B_d}
 &=r_d(1+r_a+r_b+2r_ar_b),
 &\ell_{\mathsf B_d}
 &=-\gamma_EF_{\mathsf B_d}
 +r_d(r_a+r_b+3r_ar_b),
 \nonumber\\
 F_{\mathsf C_u}
 &=r_vr_w(1+r_u),
 &\ell_{\mathsf C_u}
 &=-\gamma_EF_{\mathsf C_u}+S_3,
 \nonumber\\
 F_{\mathsf D}
 &=2(1+S_2+3S_3),
 &\ell_{\mathsf D}
 &=-\gamma_EF_{\mathsf D}+1+3S_2+11S_3,
 \nonumber\\
 F_{\mathsf E}
 &=-2S_3,
 &\ell_{\mathsf E}
 &=(2\gamma_E-3)S_3,
 \nonumber\\
 F_{\mathsf F_u}
 &=-2r_vr_w(1+3r_u),
 &\ell_{\mathsf F_u}
 &=-\gamma_EF_{\mathsf F_u}
 -r_vr_w(3+11r_u).
 \label{eq:nlf_Fell_explicit}
\end{align}
All six $F$ functions are nonzero for positive $r_i$ in the interior of
the frequency simplex, so the ratios used in
Eq.\eqref{eq:nlf_effective_bare_coefficients} are well defined there.

The remaining terms arise from inserting one corrected mode at a time.
It is important to combine the elementary and Hankel-order pieces of a
differentiated mode before projecting, since the apparent inverse power
of its argument cancels in this combination. To make that cancellation explicit, we use the representation in
Eq.\eqref{eq:nlf_hankel_E1} and define
\begin{equation}
W_r(y)\equiv \mathrm e^{2iry}E_1(2iry),
\qquad x=ry.
\end{equation}
Differentiating Eq.\eqref{eq:nlf_hankel_E1}, one obtains
\begin{align*}
 -i\mathrm e^{ix}\mathfrak h^*(x)
 &=2+\frac{i\pi}{2}(1+ix)+(1-ix)W_r(y),
 \\
 \left(\frac1x+i\right)
 -i\mathrm e^{ix}\mathfrak h^{*\prime}(x)
 &=x\left[\frac{i\pi}{2}+W_r(y)\right].
\end{align*}
The $1/x$ terms have disappeared in the second line, and substitution
into Eqs.\eqref{eq:app_zzh_delta_P} and
\eqref{eq:app_zzh_delta_D} gives
\begin{subequations}
\label{eq:nlf_deltaP_combined}
\begin{align}
\delta P_S
={}&-\epsilon P
+s\left[-P+x^2-x^2\ln y\right]
+\delta_S\left[
P\ln y+2+\frac{i\pi}{2}P+(1-ix)W_r(y)
\right],\\
\delta P_T
={}&-\epsilon P
+s_T\left[-P+x^2-x^2\ln y\right]
+\delta_T\left[
P\ln y+2+\frac{i\pi}{2}P+(1-ix)W_r(y)
\right],\\
\delta D_S
={}&-\epsilon D
+s\left[-ix^2-2D\ln y+ix^2\ln y\right]
+\delta_S D\left[
\ln y+\frac{i\pi}{2}+W_r(y)
\right],\\
\delta D_T
={}&-\epsilon D
+s_T\left[-ix^2-2D\ln y+ix^2\ln y\right]
+\delta_T D\left[
\ln y+\frac{i\pi}{2}+W_r(y)
\right].
\end{align}
\end{subequations}
For each complete leading temporal function, the finite-part integral
$\int_0^\infty dy\,\mathrm e^{-iy}\mathcal T$ is purely imaginary, and
therefore the terms proportional to $i\pi\mathcal T/2$ in these
expressions have vanishing real projection.

The remaining exponential-integral transforms reduce to ordinary
logarithms. For a nonnegative integer $n$, the integral representation
of $E_1$ gives
\begin{align*}
 \int_0^\infty dy\,y^n\mathrm e^{-iy}W_r(y)
 &=n!(-i)^{n+1}
 \int_1^\infty
 \frac{du}{u[1-2r+2ru]^{n+1}}
 \\
 &=n!(-i)^{n+1}
 \int_0^1\frac{v^n\,dv}{1-(1-2r)v}
 =n!(-i)^{n+1}L_n(r),
\end{align*}
\begin{align}
 L_n(r)
 =\frac{
 -\ln(2r)-\displaystyle\sum_{m=1}^n(1-2r)^m/m
 }{(1-2r)^{n+1}},
 \qquad n=0,1,2,3.
 \label{eq:nlf_Ln}
\end{align}
The sum is empty for $n=0$, and the apparent singularity at $r=1/2$
is removable, since the integral representation gives
\begin{align*}
 L_n\left(\frac12\right)&=\frac1{n+1}.
\end{align*}
It is convenient to collect the particular combinations that occur in
undifferentiated insertions into four real functions,
\begin{align*}
 H_0(r)&=2+L_0(r)-rL_1(r),
 \\
 H_1(r)&=2+L_1(r)-2rL_2(r),
 \\
 H_2(r)&=4+2L_2(r)-6rL_3(r),
 \\
 H_-(r)&=(1-2r)[1-\ln(2r)]
 +r(1-r)L_0(r).
\end{align*}
For the $\mathsf A$ temporal function, the separate negative-power transforms must
be retained with the same regulator. With $r=r_i$ and $B=r_jr_k$,
the two relevant polynomial ratios are
\begin{align*}
 \frac{\mathcal T_{\mathsf A}}{P_i}
 &=y^{-2}+i(1-r)y^{-1}-B,
 \\
 \frac{\mathcal T_{\mathsf A}}{P_i}(1-iry)
 &=y^{-2}+i(1-2r)y^{-1}
 +r(1-r)-B+irBy.
\end{align*}
Combining their analytically continued integrals before projection
cancels the dilogarithms of the separate negative-power transforms,
leaving the elementary expression
\begin{align}
 2\mathscr R\left[
 \frac{\mathcal T_{\mathsf A}}{P_i}
 \right]
 +\mathscr R\left[
 \frac{\mathcal T_{\mathsf A}}{P_i}
 (1-iry)W_r(y)
 \right]
 &=H_-(r_i)-r_jr_kH_0(r_i).
 \label{eq:nlf_A_hankel_result}
\end{align}
This is the combined real part of the Hankel-order expression already
given in Eq.\eqref{eq:app_zzh_H1i_analytic}, and is not an additional
insertion to be added to it.

For either an undifferentiated or differentiated line, define the
complete bulk response by
\begin{align}
\Delta_{qi}^{\mathrm{bulk}}
&=
\begin{cases}
-\epsilon F_q+s U_{qi}+\delta_S V_{qi},&\text{scalar line},\\
-\epsilon F_q+s_T U_{qi}+\delta_T V_{qi},&\text{tensor line}.
\end{cases}
\label{eq:nlf_UV_definition}
\end{align}
Here $V_{qi}$ is a real temporal function and is unrelated to the
polarization tensor $V_{\lambda_2\lambda_3}^{(q)}$; its two lower
indices identify an operator and an external leg. Equations
\eqref{eq:nlf_deltaP_combined}--\eqref{eq:nlf_A_hankel_result} give the
following fully evaluated functions, where $F$ and $\ell$ on each
right-hand side refer to the temporal function being considered.

For $\mathsf A$, all three legs are undifferentiated; writing
$A_i=r_j+r_k$ and $B_i=r_jr_k$, with $i,j,k$ distinct, the responses are
\begin{align}
 U_i^{\mathsf A}
 &=-F+(1+\gamma_E)r_i^2(1+A_i+2B_i)
 -r_i^2(A_i+3B_i),
 \nonumber\\
 V_i^{\mathsf A}
 &=\ell+H_-(r_i)-B_iH_0(r_i).
 \label{eq:nlf_UV_A}
\end{align}
For $\mathsf B_d$, the two undifferentiated legs are $a,b$, and an
insertion on $i\in\{a,b\}$, with $j$ the other undifferentiated leg,
gives
\begin{align*}
 U_i^{\mathsf B_d}
 &=-F-2(1+\gamma_E)r_i^2r_d(1+3r_j)
 +r_i^2r_d(3+11r_j),
 \\
 V_i^{\mathsf B_d}
 &=\ell+r_d[H_0(r_i)+r_jH_1(r_i)].
\end{align*}
For the differentiated leg of this temporal function, set $A=r_a+r_b$ and
$B=r_ar_b$, so that
\begin{align*}
 U_d^{\mathsf B_d}
 &=-2\ell-(1+\gamma_E)r_d^2(1+2A+6B)
 +r_d^2(1+3A+11B),
 \\
 V_d^{\mathsf B_d}
 &=\ell+r_d[
 L_0(r_d)+AL_1(r_d)+2BL_2(r_d)
 ].
\end{align*}
For $\mathsf C_u$, the undifferentiated leg is $u$, and its response is
\begin{align*}
 U_u^{\mathsf C_u}
 &=-F+(1-2\gamma_E)r_u^2r_vr_w,
 \\
 V_u^{\mathsf C_u}
 &=\ell+r_vr_wH_0(r_u).
\end{align*}
An insertion on a differentiated leg $i\in\{v,w\}$, with $j$ the other
differentiated leg, instead gives
\begin{align*}
 U_i^{\mathsf C_u}
 &=-2\ell+r_i^2r_j[
 r_u-\gamma_E(1+2r_u)
 ],
 \\
 V_i^{\mathsf C_u}
 &=\ell+r_vr_w[
 L_0(r_i)+r_uL_1(r_i)
 ].
\end{align*}
For $\mathsf D$, every leg is undifferentiated, and the result with
$A_i=r_j+r_k$ and $B_i=r_jr_k$ is
\begin{align}
 U_i^{\mathsf D}
 &=-F-2(1+\gamma_E)r_i^2(1+3A_i+12B_i)
 +r_i^2(3+11A_i+50B_i),
 \nonumber\\
 V_i^{\mathsf D}
 &=\ell+H_0(r_i)+A_iH_1(r_i)+B_iH_2(r_i).
 \label{eq:nlf_UV_D}
\end{align}
For $\mathsf E$, every leg is differentiated, and the two responses
reduce to
\begin{align}
 U_i^{\mathsf E}
 &=-2\ell+(6\gamma_E-5)r_iS_3,
 \nonumber\\
 V_i^{\mathsf E}
 &=\ell-2S_3L_2(r_i).
 \label{eq:nlf_UV_E}
\end{align}
Finally, the undifferentiated leg of $\mathsf F_u$ gives
\begin{align*}
 U_u^{\mathsf F_u}
 &=-F+(24\gamma_E-26)r_u^2r_vr_w,
 \\
 V_u^{\mathsf F_u}
 &=\ell-r_vr_wH_2(r_u),
\end{align*}
while a differentiated leg $i\in\{v,w\}$, with $j$ the other
differentiated leg, gives
\begin{align*}
 U_i^{\mathsf F_u}
 &=-2\ell+r_i^2r_j[
 6\gamma_E-5+(24\gamma_E-26)r_u
 ],
 \\
 V_i^{\mathsf F_u}
 &=\ell-2r_vr_w[
 L_2(r_i)+3r_uL_3(r_i)
 ].
\end{align*}
These functions include the full Hankel-order correction of a leg;
in particular, the differentiated $V_i$ already combines the
$Y^{(D)}$ and $\mathfrak H^{(D)}$ terms of the earlier appendices.
No separate Hankel contribution is to be added to
Eq.\eqref{eq:nlf_UV_definition}.

The assignments for the scalar--scalar--tensor operators are
\begin{align}
\begin{array}{c|c|c|c}
q&\text{temporal type}&p_q&\eta_q\Lambda_q\\ \hline
1&\mathsf A&2&-K_*/H^2\\
2&\mathsf B_3&1&\omega_3/H\\
3&\mathsf C_1&2&K_*\omega_2\omega_3/H^2\\
4&\mathsf B_2&1&\omega_2/H\\
5&\mathsf D&0&1/K_*\\
6&\mathsf C_3&2&K_*\omega_1\omega_2/H^2
\end{array}
\qquad
(\omega_1,\omega_2,\omega_3)=(c_Sk_1,c_Sk_2,c_Tk_3).
\label{eq:nlf_ssh_temporal_assignment}
\end{align}
For the scalar--tensor--tensor operators,
\begin{align}
\begin{array}{c|c|c|c}
q&\text{temporal type}&p_q&\eta_q\Lambda_q\\ \hline
1&\mathsf C_1&2&K_*\omega_2\omega_3/H^2\\
2&\mathsf A&2&-K_*/H^2\\
3&\mathsf C_3&2&K_*\omega_1\omega_2/H^2\\
4&\mathsf E&1&-\omega_1\omega_2\omega_3/H\\
5&\mathsf F_1&0&-\omega_2\omega_3/K_*\\
6&\mathsf E&1&-\omega_1\omega_2\omega_3/H\\
7&\mathsf F_1&0&-\omega_2\omega_3/K_*
\end{array}
\qquad
(\omega_1,\omega_2,\omega_3)=(c_Sk_1,c_Tk_2,c_Tk_3).
\label{eq:nlf_shh_temporal_assignment}
\end{align}
The species parameters multiplying the leg responses are those in
Eqs.\eqref{eq:nlf_ssh_assignment} and
\eqref{eq:nlf_shh_assignment}, rather than a common scalar or tensor
rate for all three legs.

To verify the assembly, expand the product of the bare coefficient and
the scale factor as
\begin{align*}
 a^{p_q}g_q
 =a_0^{p_q}\Bigg[
 g_{q*}^{\mathrm{lead}}+\delta g_{q*}
 +p_q\epsilon g_{q*}^{\mathrm{lead}}
 -\left(
 \frac{\dot g_{q*}^{\mathrm{lead}}}{H_*}
 +p_q\epsilon g_{q*}^{\mathrm{lead}}
 \right)\ln y
 \Bigg]+\mathcal O(\varepsilon^2g_{q,\mathrm{char}}).
\end{align*}
The sum of the three bulk insertions is
\begin{align*}
\sum_{i=1}^3\Delta_{qi}^{\mathrm{bulk}}
&=-3\epsilon F_q+\Sigma_{q,\mathrm{mode}}^{(A)},
\end{align*}
where the two mixed-sector sums are given in
Eq.\eqref{eq:nlf_species_mode_sum}. This combines with the coefficient
and scale-factor terms above, the
external-mode correction, and the denominator subtraction to reproduce
Eq.\eqref{eq:nlf_effective_bare_coefficients}. In terms of the original
temporal functions, their complete sum satisfies
\begin{align*}
 &\left(
 g_{q*}^{\mathrm{lead}}+\delta g_{q*}
 \right)J_q^{(0)}
 +g_{q*}^{\mathrm{lead}}\delta J_q
 -4E_{X_A}(k_0)g_{q*}^{\mathrm{lead}}J_q^{(0)}
 =\eta_q\Lambda_qF_q\widehat g_q,
\end{align*}
where $J_q^{(0)},\delta J_q$ are replaced by $I_q^{(0)},\delta I_q$
for the scalar--tensor--tensor sector. This identity also makes clear
that the constant reduced normalization multiplying $c_q$ or $b_q$
is not treated as a second time-dependent interaction coefficient.


\subsection{Momentum and helicity contractions}
\label{app:nlf_polarizations}

The numerical coefficients in the equilateral nonlinearity parameters
can be checked without choosing a special propagation speed, since
polarization tensors depend on the spatial momentum triangle rather
than on the acoustic frequencies. Choose a unit normal $\bm n$ to the
momentum plane and define
\begin{align*}
 \bm t_a&=\bm n\times\widehat{\bm k}_a,
 &e_{ij}^{(\lambda_a)*}(\bm k_a)
 &=\frac12
 (t_{a,i}-i\chi_an_i)(t_{a,j}-i\chi_an_j),
 &\chi_a&=\frac{\lambda_a}{2}.
\end{align*}
This choice has the norm specified in the main text, and all helicity
projections below use the conjugate polarization as required by
Eq.\eqref{eq:helicity_extraction_convention}. A different polarization
phase changes the phase of a helicity amplitude, but not the
underlying tensor correlator.

Introduce the triangle polynomial
\begin{align*}
 \Delta_\triangle
 &=(k_1+k_2+k_3)(-k_1+k_2+k_3)
 (k_1-k_2+k_3)(k_1+k_2-k_3).
\end{align*}
For the scalar--scalar--tensor sector, transversality and momentum
conservation give
\begin{equation}
V_\lambda^{(1)}
=e_{ij}^{(\lambda)*}(\bm k_3)k_{1i}k_{2j}
=-\frac{\Delta_\triangle}{8k_3^2}.
\end{equation}
The remaining five structures follow from
Eq.\eqref{eq:zzh_momentum_structures}. At equilateral momenta, their
substitution into Eq.\eqref{eq:fnl_zzh_general} gives the weights in
Eq.\eqref{eq:nlf_ssh_weights}; the angular and temporal integrals have
already been evaluated.

For the scalar--tensor--tensor sector, define
\begin{align}
 M_{\lambda_2\lambda_3}
 &=k_1^2-(\chi_2k_2+\chi_3k_3)^2,
 \nonumber\\
 V_{\lambda_2\lambda_3}^{(1)}
 &=\frac{M_{\lambda_2\lambda_3}^2}
 {16k_2^2k_3^2},\\
 V_{\lambda_2\lambda_3}^{(6)}
 &=-\frac{
 \Delta_\triangle M_{\lambda_2\lambda_3}
 }{32k_1^2k_2^2k_3^2}.
\nonumber
\end{align}
The other contractions follow from
Eq.\eqref{eq:z_hh_tensor_momentum_structures}; in particular, at
equilateral momenta they are
\begin{align*}
 V^{(1)}&=V^{(4)}=v_\sigma,
 &V^{(2)}&=-\frac{k^2}{2}v_\sigma,
 &V^{(3)}&=-\frac12v_\sigma,
 \\
 V^{(5)}&=k^2v_\sigma,
 &V^{(6)}&=w_\sigma,
 &V^{(7)}&=k^2w_\sigma.
\end{align*}
Adding the second tensor ordering and the dimensional factors of
Eq.\eqref{eq:nlf_shh_temporal_assignment} to
Eq.\eqref{eq:fnl_zhh_general} gives the weights in
Eq.\eqref{eq:nlf_shh_weights}. The negative signs in the second line of operators combine with the
negative leading functions of $\mathsf E$ and $\mathsf F_1$, yielding
the positive weights $W_4^{(b)},\ldots,W_7^{(b)}$ displayed in
Eq.\eqref{eq:nlf_shh_weights}.

For the three-tensor sector, the additional-vertex contraction can be
written as
\begin{align*}
 V_{\lambda_1\lambda_2\lambda_3}^{(\mathrm{new})}
 &=\operatorname{tr}\left(e_1^*e_2^*e_3^*\right)
 =\frac18
 \prod_{(ij)=(12),(23),(31)}
 \left(
 \widehat{\bm k}_i\cdot\widehat{\bm k}_j-\chi_i\chi_j
 \right),
 \\
 V_{\lambda_1\lambda_2\lambda_3}^{(\mathrm{GR})}
 &=\frac12
 (\chi_1k_1+\chi_2k_2+\chi_3k_3)^2
 V_{\lambda_1\lambda_2\lambda_3}^{(\mathrm{new})}.
\end{align*}
At equilateral momenta these expressions give
\begin{align*}
 V_{+++}^{(\mathrm{new})}&=-\frac{27}{64},
 &V_{++-}^{(\mathrm{new})}&=-\frac3{64},
 \\
 V_{+++}^{(\mathrm{GR})}&=-\frac{243}{128}k^2,
 &V_{++-}^{(\mathrm{GR})}&=-\frac3{128}k^2,
 \\
 I_{\mathrm{new}}^{(0)}&=\frac{k^3}{27},
 &I_{\mathrm{GR}}^{(0)}&=-\frac{17k}{144}.
\end{align*}
Together with the temporal factors displayed above, these contractions
reproduce the leading normalizations in
Eqs.\eqref{eq:nlf_tensor_all_equal} and
\eqref{eq:nlf_tensor_one_opposite}; the elementary multiplication is not
shown separately.


\subsection{Reduction of the tensor corrections at equilateral momenta}
\label{app:nlf_tensor_reduction}

The tensor temporal functions are $\mathsf A$ for the universal vertex
and $\mathsf E$ for the additional vertex, so the general
momentum-dependent relative correction is
\begin{align}
 &\rho_r
 =\overline E_{hhh}+(p_r-3)\epsilon
 -(\sigma_r+p_r\epsilon)\frac{\ell_r}{F_r}
 +\frac{s_T}{F_r}\sum_{i=1}^3U_{ri}
 +\frac{\delta_T}{F_r}\sum_{i=1}^3V_{ri},
 \nonumber\\
 &(p_{\mathrm{GR}},\sigma_{\mathrm{GR}})
 =(2,f_T),\nonumber
 \\&(p_{\mathrm{new}},\sigma_{\mathrm{new}})
 =(1,\sigma_{\mu_5}).
 \label{eq:nlf_tensor_general_rho}
\end{align}
Every function on the right is given by
Eqs.\eqref{eq:nlf_Fell_explicit}, \eqref{eq:nlf_UV_A}, and
\eqref{eq:nlf_UV_E}, evaluated at $r_i=k_i/K$, so this expression
contains no unevaluated time integrals or Hankel derivatives.

At $r_i=1/3$, the elementary functions have the values
\begin{align*}
 L_0\left(\frac13\right)&=3L_{
eq},
 &L_1\left(\frac13\right)&=9L_{
eq}-3,
 \\
 L_2\left(\frac13\right)&=27L_{
eq}-\frac{21}{2},
 &L_3\left(\frac13\right)&=81L_{
eq}-\frac{65}{2},
 \\
 H_0\left(\frac13\right)&=3,
 &H_-\left(\frac13\right)&=\frac13+L_{
eq},
 &L_{
eq}&=\ln\frac32.
\end{align*}
Since all external legs have the same momentum and $k_0=k$, the
normalized external correction is $\overline E_{hhh}=-E_T(k)$, where
\begin{align*}
 E_T(k)
 &=-(\epsilon+s_T)
 -\delta_T(\gamma_E-2-L_{
m eq}).
\end{align*}
For the universal vertex, direct substitution into the temporal
functions gives
\begin{align*}
 F_{\mathrm{GR}}&=\frac{17}{27},
 \\
\frac{\ell_{\mathrm{GR}}}{F_{\mathrm{GR}}}
 &=\frac{26}{17}-\gamma_E,\\
 \frac1{F_{\mathrm{GR}}}
 \sum_iU_{\mathrm{GR},i}
 &=\gamma_E-\frac{43}{17},
 \\
 \frac1{F_{\mathrm{GR}}}
 \sum_iV_{\mathrm{GR},i}
 &=3\left(\frac{26}{17}-\gamma_E\right)
 +\frac{81}{17}L_{
m eq}.
\end{align*}
Substitution into Eq.\eqref{eq:nlf_tensor_general_rho}, together with
$2\delta_T=2\epsilon+f_T+s_T$, gives the compact result
Eq.\eqref{eq:nlf_tensor_GR_rho}. This form makes explicit that the
Euler--Mascheroni constant and the separate speed-running contribution
cancel after normalization, although they occur in the individual
terms.

For the additional vertex, the corresponding values are
\begin{align*}
 F_{\mathrm{new}}&=-\frac2{27},
 \\
 \frac{\ell_{\mathrm{new}}}{F_{\mathrm{new}}}
 &=\frac32-\gamma_E,\\
 \frac1{F_{\mathrm{new}}}
 \sum_iU_{\mathrm{new},i}
 &=3\gamma_E-\frac{13}{2},
 \\
 \frac1{F_{\mathrm{new}}}
 \sum_iV_{\mathrm{new},i}
 &=81L_{
m eq}-3\gamma_E-27.
\end{align*}
Substitution of these values into
Eq.\eqref{eq:nlf_tensor_general_rho} gives
Eq.\eqref{eq:nlf_tensor_new_rho}. In contrast to the
universal contribution, the additional vertex retains the independent
rate of the bare cubic coefficient, because $\mu_5(t)$ is not fixed by
the normalization of the tensor quadratic action.

\end{appendices}

\end{document}